\documentclass[aps,prd,reprint,longbibliography,showpacs,floatfix,nofootinbib,superscriptaddress,amsmath,amssymb]{revtex4-1}
\usepackage{graphicx} \usepackage{mathrsfs} \usepackage{longtable}
\usepackage{hyperref} \usepackage{dsfont}
\newcommand{\MS}{\overline{\mathrm{MS}}}

\begin{document}
\title{Direct determinations of the nucleon and pion $\sigma$ terms
  at nearly physical quark masses}
\author{Gunnar S.~Bali} \affiliation{Institut f\"ur Theoretische Physik,
  Universit\"at Regensburg, 93040 Regensburg, Germany}\affiliation{Tata
  Institute of Fundamental Research, Homi Bhabha Road, Mumbai 400005,
  India} 
\author{Sara~Collins}\email{sara.collins@ur.de} \affiliation{Institut
  f\"ur Theoretische Physik, Universit\"at Regensburg, 93040
  Regensburg, Germany}
\author{Daniel~Richtmann}  
\affiliation{Institut f\"ur Theoretische Physik, Universit\"at Regensburg,
              93040 Regensburg, Germany}   
\author{Andreas~Sch\"{a}fer}  
\affiliation{Institut f\"ur Theoretische Physik, Universit\"at Regensburg,
              93040 Regensburg, Germany}   
\author{Wolfgang~S\"oldner}  
\affiliation{Institut f\"ur Theoretische Physik, Universit\"at Regensburg,
              93040 Regensburg, Germany}   
\author{Andr\'e~Sternbeck}  
\affiliation{Theoretisch-Physikalisches Institut,                                          
Friedrich-Schiller-Universit\"at Jena, 07743 Jena, Germany}
\collaboration{RQCD Collaboration}
\date{\today}
\begin{abstract}
We present a high statistics study of the pion and nucleon light and
strange quark sigma terms using $N_f=2$ dynamical non-perturbatively
improved clover fermions with a range of pion masses down to
$m_\pi\sim 150$~MeV and several volumes, $Lm_\pi=3.4$ up to $6.7$, and
lattice spacings, $a=0.06-0.08$~fm, enabling a study of finite volume
and discretisation effects for $m_\pi\gtrsim 260$~MeV.  Systematics
are found to be reasonably under control. For the nucleon we obtain
$\sigma_{\pi N}=35(6)$~MeV and $\sigma_s=35(12)$~MeV, or equivalently
in terms of the quark fractions, $f_{T_u}=0.021(4)$,
$f_{T_d}=0.016(4)$ and $f_{T_s}=0.037(13)$, where the errors include
estimates of both the systematic and statistical uncertainties. These
values, together with perturbative matching in the heavy quark limit,
lead to $f_{T_c}=0.075(4)$, $f_{T_b}=0.072(2)$ and $f_{T_t}=0.070(1)$.
In addition, through the use of the (inverse) Feynman-Hellmann theorem
our results for $\sigma_{\pi N}$ are shown to be consistent with the
nucleon masses determined in the analysis.  For the pion we implement
a method which greatly reduces excited state contamination to the
scalar matrix elements from states travelling across the temporal
boundary. This enables us to demonstrate the Gell-Mann-Oakes-Renner
expectation $\sigma_\pi=m_\pi/2$ over our range of pion masses.
\end{abstract}

\maketitle

\section{Introduction}
\label{intro}
How the quark and gluon constituents of matter account for the
properties of hadronic bound states is of fundamental interest. The
decomposition for one of the most basic properties, the hadron mass,
has been understood for some
time~\cite{Shifman:1978zn,Ji:1994av,Ji:1995sv}, however, the magnitude
of each contribution is as yet only approximately known due to its
non-perturbative nature.  Of particular importance are quark scalar
matrix elements that form the quark contribution to the hadron
mass. For the case of the nucleon, these matrix elements are also
needed for determining the size of dark matter-nucleon scattering
cross-sections for direct detection experiments~(see, for example,
Refs.~\cite{Cline:2013gha,Hietanen:2013fya,Hill:2014yxa,Hill:2014yka,Ellis:2015rya,Abdallah:2015ter}). A
variety of approaches have been used to determine the scalar matrix
elements or sigma terms from pion-nucleon scattering
data~\cite{Gasser:1990ce,Pavan:2001wz,Alarcon:2012kn,Hoferichter:2015dsa,Hoferichter:2015hva}. Lattice
calculations, as a first principles approach, are now gaining
prominence, not least due to the refinement of techniques and increase
in computational power available which now allows for the direct
evaluation of the sigma terms at the physical
point~\cite{Yang:2015uis,Abdel-Rehim:2016won}. In this work, we
present results for the pion and nucleon scalar matrix elements close
to the physical point, but also investigate the quark mass dependence
up to $m_\pi\lesssim 500$~MeV and the lattice systematics including
lattice spacing and volume dependence. On a technical note, our
analysis includes a method for reducing excited state contamination to
pion three-point functions by isolating the forward propagating pion
for lattices with anti-periodic fermionic boundary conditions in time.

By way of introduction, we review the decomposition of hadron masses
into the quark and gluon contributions and the scalar matrix elements
of interest. The starting point is the
energy momentum tensor of QCD~\cite{Freedman:1974gs,Freedman:1974ze,Caracciolo:1989pt}
\begin{equation}
T_{\mu\nu}  = \frac{1}{4}\sum_q
\bar{q}\gamma_{(\mu}\overleftrightarrow{D}_{\nu)}q+F_{\mu\alpha}F_{\nu\alpha}-\frac{1}{4}\delta_{\mu\nu}F^2\end{equation}
and its anomalous trace~\cite{Coleman:1970je,Chanowitz:1972da,Crewther:1972kn,Nielsen:1977sy,Adler:1976zt}: 
\begin{equation} 
T_{\mu\mu}=(\gamma_m(\alpha)-1)\sum_q m_q\bar{q}q +
\frac{\beta(\alpha)}{4\alpha}F^2.\end{equation} For our conventions see
Appendix~\ref{app:definitions}.
We define the expectation value for the hadron state,
$|H\rangle$:
\begin{align}
\langle T_{\mu\nu}\rangle_H & = \langle H|T_{\mu\nu}|H\rangle-\langle 0|T_{\mu\nu}|0\rangle\nonumber\\
& = \frac{\left\langle H| \int d^3x\,
T_{\mu\nu}(x)|H\right\rangle}{\langle H|H\rangle} - \left\langle 0\left| \int\! d^3\!x\,
T_{\mu\nu}(x)\right|0\right\rangle.\label{eq:defmatt}
\end{align}
Since $-T_{44}$ is the Hamiltonian density, in the rest frame of the hadron $H$,
$\langle T_{44}\rangle_H=-M_H$ gives the mass while $\langle T_{ij}\rangle_H=\langle T_{4i}\rangle_H=0$.
This means that in any Lorentz frame:
\begin{align}
m_H & = -\langle T_{\mu\mu}\rangle_H\nonumber\\
& =\sum_q m_q \langle\bar{q}q \rangle_H - \gamma_m(\alpha) \sum_q m_q
\langle \bar{q}q \rangle_H - \frac{\beta(\alpha)}{4\alpha}\langle F^2\rangle_H.
\label{eq:anomaly}
\end{align}
For zero momentum this is the same as~\cite{Ji:1994av},
\begin{widetext}
\begin{align}
m_H = -\langle T_{44}\rangle_H = \underbrace{\sum_q m_q \langle\bar{q}q \rangle_H}_{\langle H_m\rangle_H} 
\underbrace{+\frac{1}{2}\langle \mathbf{B}^2-\mathbf{E}^2\rangle_H+\sum_q \langle \bar{q}{\bf D}\cdot\boldsymbol{\gamma}q\rangle_H}_{\langle H_{\rm kin}\rangle_H = 3 \langle H_a\rangle_H}
\underbrace{-\frac{1}{4}\left[\gamma_m \sum_q m_q \langle \bar{q}q \rangle_H +
  \frac{\beta}{4\alpha}\langle \mathbf{E}^2+\mathbf{B}^2\rangle_H\right]}_{\langle H_a\rangle_H =\frac{1}{4}\left( m_H-\langle H_m\rangle_H\right)}.\label{eq:massd}
\end{align}
\end{widetext}
The terms are grouped into scale invariant
combinations~\cite{Tarrach:1981bi}: $\langle H_m\rangle_H$, the quark
mass contribution, $\langle H_{\rm kin}\rangle_H$, arising from the
quark and gluon kinetic energies and $\langle H_a\rangle_H$ from the
trace anomaly.  Comparison with Eq.~(\ref{eq:anomaly}) demonstrates
that knowledge of the sigma terms $\sigma^H_q = m_q\langle
\bar{q}q\rangle_H$ and $m_H$ is sufficient to determine all three
components. We remark that the individual~(scale dependent) quark and
gluon kinetic energies can be computed on the lattice, however, this
is not attempted here. In a theory with only two light quarks,
$q=u,d$, then $\langle H_m\rangle_N=\sigma^N_u+\sigma^N_d=\sigma_{\pi
  N}$ for the nucleon.  We define
$\sigma^\pi_u+\sigma^\pi_d=\sigma_\pi$ for the light quark pion sigma
term. Early estimates, employing Eq.~(\ref{eq:anomaly}) and SU(3)
flavour symmetry breaking of baryon octet masses~\cite{Cheng:1988im}
suggested $\sigma_{\pi N}\sim 26$~MeV, while $\sigma_\pi\sim m_\pi/2$
can be inferred from the Gell-Mann-Oakes-Renner~(GMOR) relation and
the Feynman-Hellmann theorem, $\sigma_q^H = m_q\partial m_H/\partial
m_q$~(this was noted in
Refs.~\cite{Donoghue:1990xh,Ananthanarayan:2004xy,Yang:2014xsa} and
confirmed in Ref.~\cite{Yang:2014xsa} at $m_\pi=281$~MeV).  Using
$\sigma_{\pi N}\sim 35\,\mathrm{MeV}\,\sim 0.04\, m_N$, which is close
to the result presented later in this paper, we have the
decompositions for $N_f=2$ QCD:
\begin{align}
m_N &\approx \left(0.04\, m_N\right)_{m} + \left(0.72\, m_N\right)_{\rm kin} + \left(0.24\, m_N\right)_{\rm a}, \label{eq:massdecomp1} \\
m_\pi &\approx \left(\frac 12 m_{\pi}\right)_{m} + \left(\frac 38 m_\pi\right)_{\rm kin} + \left(\frac 18 m_\pi\right)_{\rm a},\label{eq:massdecomp2}
\end{align}
reflecting the different impact of spontaneous chiral symmetry
breaking in the two cases, i.e. $m_N>0$ for $m_q=0$. These
decompositions will be modified in the presence of the sea quarks
$s,c,b,t$. While the sigma terms for the light and strange quarks,
$q=u,d,s$, must be determined via non-perturbative methods one can
appeal to the heavy quark limit in order to evaluate
$\sigma_{c,b,t}$. Following Ref.~\cite{Shifman:1978zn}, in the
effective theory the heavy quark~($h$) term $m_h \bar{h}h$, to leading
order in the QCD coupling $\alpha$, transforms as
$-(2/3)(\alpha/(8\pi)) F^2+O\left(\Lambda^2/m_h^2\right)$, where
$\Lambda$ is the typical QCD scale. One can then
use Eq.~(\ref{eq:anomaly}) to express $\sigma_h$ in terms of the sum
of the sigma terms\footnote{Below we will suppress the super- and
  subscripts indicating $H=\pi, N$, whenever this is clear from the
  context.} for which $m_q\ll m_h$. To leading order in $1/m_h$ and
$\alpha$ one obtains:
\begin{align}
\sigma_{h=c,b,t} = \frac{2}{27}\left(m_H-\sum_{q=u,d,s}\sigma_q\right).
\label{eq:sigheavy}
\end{align}
See Refs.~\cite{Vecchi:2013iza,Hill:2014yxa} for radiative
corrections. Alternatively, in terms of the quark mass fractions,
$f_{T_q}=\sigma_q/m_{H}$,
\begin{align}
m_H f_{T_h} = \frac{2}{27} m_H \left(1-\sum_{q=u,d,s} f_{T_q}\right) \equiv \frac{2}{27} m_H f_{T_G}.\label{eq:heavyftq}
\end{align}
For the nucleon, these quark fractions are needed to determine the
coupling to the Standard Model Higgs boson or to other scalar
particles, for example, in dark matter-nucleon
scattering~\cite{Cline:2013gha,Hietanen:2013fya,Hill:2014yxa,Hill:2014yka,Ellis:2015rya,Abdallah:2015ter}. The
cross-section is proportional to $|f_N|^2$, where~(using
Eq.~(\ref{eq:heavyftq}))
\begin{equation}
\label{eq:fn}
\frac{f_{\mathrm{N}}}{m_{\mathrm{N}}}\approx\sum_{q=u,d,s} f_{T_q}\frac{\alpha_q}{m_q}
+\frac{2}{27}f_{T_G}\sum_{q=c,b,t}\!\!\frac{\alpha_q}{m_q}\,,\end{equation}
with the couplings $\alpha_q\propto m_q/m_W$ in the Higgs case.

For the light and strange sigma terms one can go further and
disentangle the contributions from the valence and sea quarks through
the ratios,
\begin{align}
r^{\mathrm{sea}}
=\frac{\langle\bar{u}u+\bar{d}d\rangle^{\mathrm{sea}}}
{\langle \bar{u}u+\bar{d}d\rangle}\hspace{0.5cm} \mathrm{and}\hspace{0.5cm}
y
&=\frac{2\langle \bar{s}s\rangle}{\langle \bar{u}u+\bar{d}d\rangle},
\label{eq:rvalsea}
\end{align}
while the SU(3) flavour symmetry of the sea is probed
with the ratio
\begin{equation}
a^{\mathrm{sea}}=
\frac{2\langle\bar{s}s\rangle}
{\langle \bar{u}u+\bar{d}d\rangle^{\mathrm{sea}}}.
\end{equation}
Other quantities of interest are the non-singlet
sigma term, $\sigma_0=\frac{1}{2}(m_u+m_d)\langle \bar{u}u +\bar{d}d-2\bar{s}s\rangle$ and the isospin asymmetry ratio
\begin{equation}
z = \frac{\langle \bar{u}u-\bar{s}s\rangle}{\langle \bar{d}d-\bar{s}s\rangle}.
\end{equation}
In a naive picture of the proton with only valence quarks
$z=1.5$. Using Gell-Mann-Okubo mass relations,
Ref.~\cite{Cheng:1988im} estimated this to be only slightly modified
to 1.49 in the presence of sea quarks.  The individual light and
strange quark sigma terms can be obtained from different combinations
of $\sigma_0$, $z$ and $\sigma_{\pi N}$, see, for example,
Ref.~\cite{Cline:2013gha}.

\begin{table*}
\caption{\label{tab:sim} Details of the ensembles used in the analysis
  including the lattice spacing $a$, the light quark mass parameter
  $\kappa_{\ell}$, the lattice volume $V$, the pion mass $m_\pi$ and the
  spatial lattice extent $L$ in units of $m_\pi$. The finite volume
  pion masses were determined in Ref.~\cite{Bali:2014gha} and the
  errors include an estimate of both the systematic and statistical
  uncertainty. The number of configurations $n^{\rm conf}$ employed is
  given along with the number of measurements of the three-point
  functions on each configuration for the connected $n^{\rm 3pt,
    conn}_{N}$ and disconnected $n^{\rm 3pt,dis}_{N}$
  contributions for the nucleon and similarly for the pion. The number
  of Wuppertal smearing iterations $n_{\mathrm{sm}}$ applied to the
  light quark appearing in the pion and nucleon interpolators is also
  shown. }
\begin{center}
\begin{ruledtabular}
\begin{tabular}{ccccccccccccc}
Ensemble& $\beta$ &  $a$ [fm] & $\kappa_{\ell}$     &   $V$   &  $m_\pi$ [GeV] &  $Lm_\pi$  &   $n^{\rm conf}$ & $n^{\rm 3pt, conn}_{N}$ & $n^{\rm 3pt,dis}_{N}$ &  $n^{\rm 3pt, conn}_{\pi}$ & $n^{\rm 3pt,dis}_{\pi}$ &  $n_{\mathrm{sm}}$ \\
\hline
I    &5.20 &  0.081& 0.13596 & $32^3\times 64$& 0.2795(18)& 3.69& $1986$ & $4$&8 &  & &  300 \\\hline
II   &5.29 &  0.071& 0.13620 & $24^3\times 48$& 0.4264(20)& 3.71& $1999$ & $2$&8 &  & &  300 \\
III  &     &       & 0.13620 & $32^3\times 64$& 0.4222(13)& 4.90& $1998$ & $2$&8 &2 & 8  &  300 \\
IV   &     &       & 0.13632 & $32^3\times 64$& 0.2946(14)& 3.42& $2023$ & $2$&8 &2 & 8 &  400 \\
V    &     &       &         & $40^3\times 64$& 0.2888(11)& 4.19& $2025$ & $2$&8 &2 & 8 &  400 \\
VI   &     &       &         & $64^3\times 64$& 0.2895(07)& 6.71& $1232$ & $2$&8 &  & &  400 \\
VIII &     &       & 0.13640 & $64^3\times 64$& 0.1497(13)& 3.47& $1593$ & $3$&8 &3 & 8  &  400 \\\hline
IX   &5.40 & 0.060 & 0.13640 & $32^3\times 64$& 0.4897(17)& 4.81& $1123$ & $2$&8  & & &  400 \\
X    &     &       & 0.13647 & $32^3\times 64$& 0.4262(20)& 4.18& $1999$ & $2$&8  & & &  450 \\       
XI   &     &       & 0.13660 & $48^3\times 64$& 0.2595(09)& 3.82& $2177$ & $2$&8  & & &  600  
\end{tabular}
\end{ruledtabular}
\end{center}
\end{table*}

This paper is organized as follows: in the next section we provide
details of the simulation including the lattice set-up and the
construction of the connected and disconnected quark line diagrams
needed for the computation of the pion and nucleon scalar matrix
elements. These matrix elements typically suffer from significant
excited state contamination. The fitting procedures employed to ensure
the ground states are extracted reliably is discussed in
Sections~\ref{sec:pionfits} and~\ref{sec:nucfits}, for the pion and
the nucleon, respectively.  Some of the quantities given above require
renormalization due to the explicit breaking of chiral symmetry for
our lattice fermion action. The relevant renormalization
factors are detailed in Section~\ref{renorm}. Our final results for
the sigma terms, including mass and volume dependence are presented in
Section~\ref{pionsig} for the pion and, also including lattice spacing
effects, for the nucleon in Section~\ref{nucleonsig}. For the latter a
comparison is made with other recent lattice determinations by direct
and indirect~(via the Feynman-Hellmann theorem) methods and also other
theoretical results in Section~\ref{comparison}. We conclude in
Section~\ref{conc}. For the sake of brevity, our conventions for the
definition of the energy-momentum tensor are collected in
Appendix~\ref{app:definitions}. For the pion, in order to reduce
excited state contamination, we construct the relevant two- and
three-point functions from quark propagators with both periodic and
anti-periodic boundary conditions in time. This approach is discussed
in Appendix~\ref{app:spectral}. Finally, the finite volume chiral
perturbation theory expressions we use when investigating finite size
effects on the sigma terms and nucleon mass are given in
Appendix~\ref{app:chiralsigma}.

\section{Simulation details}
\subsection{Lattice set-up and methods}
\label{latsetup}

\begin{figure}
\centerline{
\includegraphics[width=.48\textwidth,clip=]{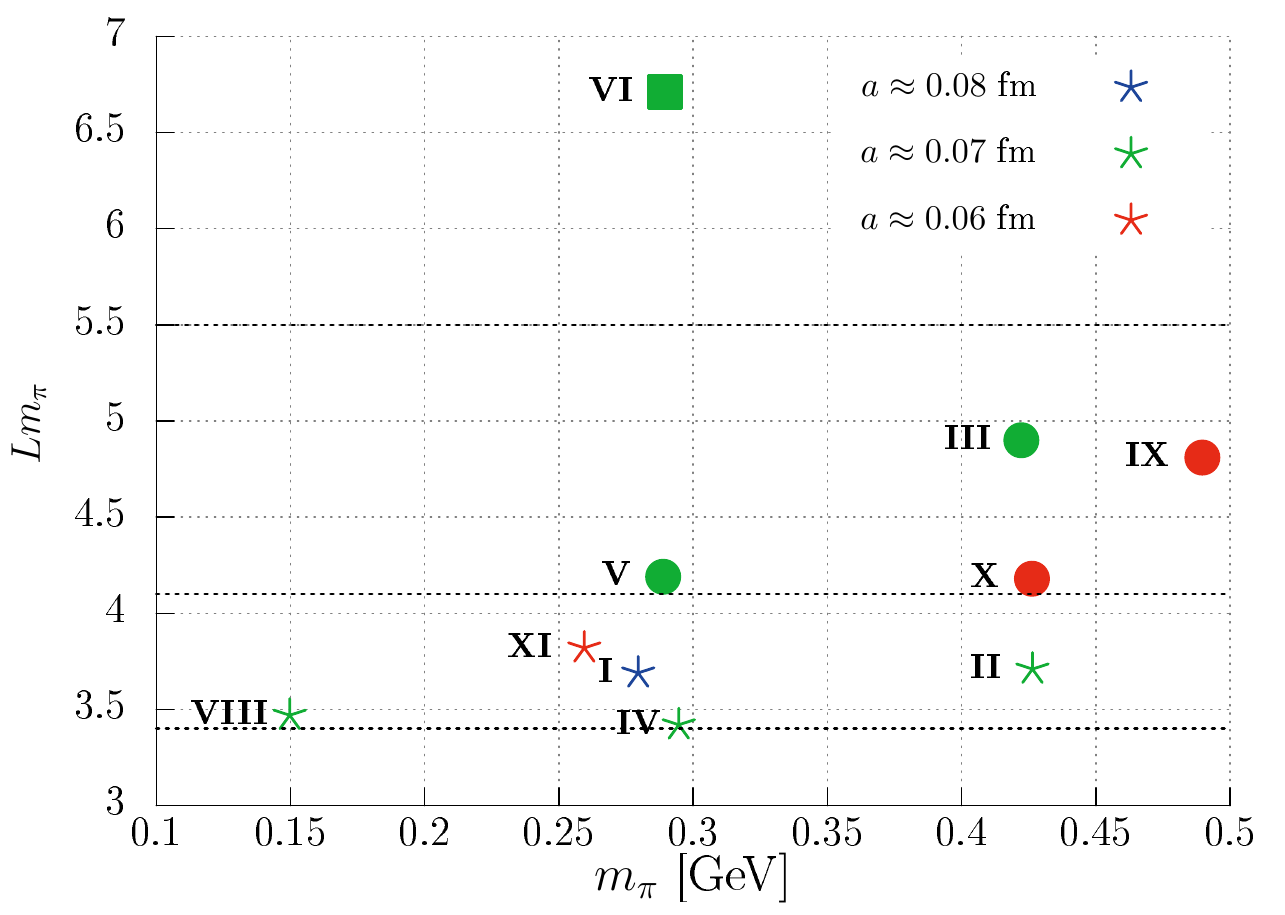}
}
\caption{Overview of the ensembles listed in
  Table~\protect\ref{tab:sim}.  Colours indicate the lattice spacings
  and symbols the lattice extents.  This labelling will be used in the
  figures presented in
  Secs.~\ref{pionsig}--\ref{comparison}. Different volume ranges are
  indicated by the horizontal lines. }
\label{fig:ens}
\end{figure}

The analysis was performed on $N_f=2$ ensembles using the Wilson gauge
action with non-perturbatively improved clover fermions generated by
QCDSF and the Regensburg lattice QCD group~(RQCD). A wide range of
pion masses~($m_\pi=490$~MeV down to $150$~MeV) and spatial lattice
extents~($Lm_\pi=3.4$ up to $6.7$) were realized over a limited range
of lattice spacings~($a=0.08$~fm to $0.06$~fm).  The scale
was set using the value $r_0 \approx 0.5$~fm at vanishing quark mass,
obtained by extrapolating the nucleon mass to the physical
point~(within our range of $a$)~\cite{Bali:2012qs}.
Table~\ref{tab:sim} gives details of the ensembles and
Fig.~\ref{fig:ens} illustrates the range of volumes available for each
pion mass.

The full set of ensembles was used in the determination of the
nucleon scalar matrix elements enabling a constrained approach
to the physical point and a thorough investigation of finite volume
effects using three spatial extents with $m_\pi\sim 290$~MeV at fixed
lattice spacing. 
Discretisation effects are
$\mathcal{O}(a^2)$ for some non-singlet combinations of scalar
currents and $\mathcal{O}(a)$ for others~(see
Section~\ref{renorm}). The latter being due to mixing with the gluonic
operator $aF^2$~\cite{Bhattacharya:2005rb} or $\mathcal{O}(am_q)$
terms.  Note that mixing with $aF^2$ is present also for other actions such
as the twisted mass (including maximal twist) and overlap actions. No clear
indication of significant discretisation effects is seen in our
results, however, $a$~($a^2$) only varies by a factor 1.3~(1.8) in our
simulations and, hence, this cannot be checked decisively.

A further source of systematic uncertainty is excited state
contamination.  As in our studies of nucleon isovector
quantities~\cite{Bali:2014gha,Bali:2014nma} a careful investigation
of excited state contributions is performed, see
Sections~\ref{sec:pionfits} and~\ref{sec:nucfits} for the pion and
nucleon, respectively.  This is an important issue for our analysis of
pion scalar matrix elements, since terms arising from multi-pion
states which propagate around the temporal boundary can dominate the
three-point function if the temporal extent of the lattice is not
large.  In particular, for the near physical point ensemble the
temporal extent of the lattice is only $T\sim
4.58$~fm~$\approx 3.5/m_\pi$. Our method for reducing this contribution and
ensuring ground state dominance, detailed in
Appendix~\ref{app:spectral}, was applied to four ensembles~(labelled
III, IV, V and VIII in Table~\ref{tab:sim}) at one lattice spacing
$a=0.07$~fm. The pion mass is varied between $420$~MeV and $150$~MeV
and a limited study of finite size effects is possible through the use
of two volumes with $Lm_\pi=3.4$ and $4.2$ for $m_\pi=290$~MeV.

\begin{figure}
\centerline{
\includegraphics[width=.3\textwidth,clip=]{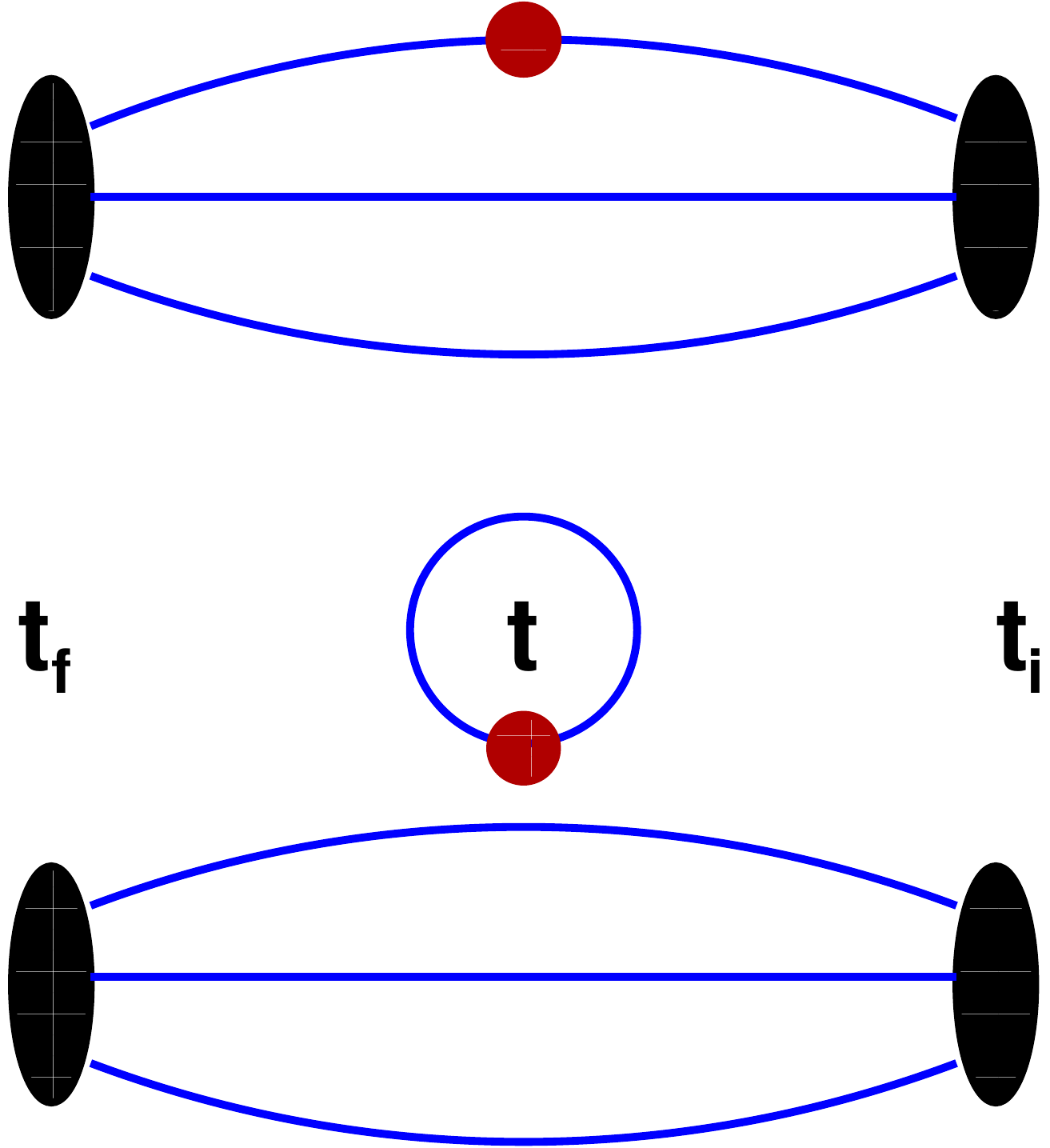}}
\caption{Quark line connected ($C^{\mathrm{conn}}_{{\rm
      3pt}}(t_{\mathrm{f}},t,t_{\mathrm{i}})$, top) and disconnected
  ($C^{\mathrm{dis}}_{{\rm 3pt}}(t_{\mathrm{f}},t,t_{\mathrm{i}})$,
  bottom) three-point functions for the nucleon. The quark
  contractions give a relative minus sign between the diagrams.  Note
  that for scalar matrix elements, the vacuum expectation value of the
  current insertion needs to be subtracted
  ($\bar{q}q\mapsto\bar{q}q-\langle\bar{q}q\rangle$), see
  Eqs.~(\ref{eq:defmatt}) and~(\ref{eq:threept0}). The blue lines
  represent light quarks.  The disconnected loop is evaluated
  for both light and strange quarks.
\label{fig:condiscon}
}
\end{figure}

High statistics was achieved in all cases and the signals of the
required two-point and three-point functions were further improved by
performing multiple measurements per configuration using different
source positions. This is necessary in particular for scalar matrix
elements since the intrinsic gauge noise can be substantial.  The
isoscalar three-point functions contain both connected and
disconnected quark line contributions, as shown in
Fig.~\ref{fig:condiscon}, with the latter dominating the noise. Eight
measurements of the disconnected diagrams were performed on each
configuration compared to two measurements for the connected part.
Signal to noise ratios are worse for coarser lattice spacings and for
smaller pion masses and the number of determinations of the connected
terms was increased to $4$ and $3$ for ensembles I~($m_\pi=280$~MeV
with $a\sim 0.08$~fm) and VIII~($150$~MeV, $a=0.07$~fm), respectively.
For the nucleon the connected terms were generated as part of a
previous study of isovector charges~\cite{Bali:2014nma}. The number
of disconnected measurements was not increased due to the
computational cost and the limited reduction in error due to
correlations within the data.  Measurements performed on the same
configuration are averaged and binning over configurations was applied
to a level consistent with four times the integrated autocorrelation
time.

\begin{table*}
\caption{\label{tab:res} Results for the hadron masses and axial Ward
  identity masses~($\tilde{m}$) in lattice units, the source-sink
  separation for the connected three-point functions, $t_{\rm f}^{\rm
    conn}$~($t^{\rm conn}_{\rm i} = 0$), and the minimum
  source-operator insertion separation for the disconnected
  three-point function, $\Delta t_{\rm min}$. For the nucleon, the
  statistical errors of $C^{\mathrm{conn}}_{{\rm 3pt}}$ decrease for
  smaller values of $t_{\rm f}$, such that for ensembles IV and VIII
  it is sufficient to perform a smaller number of measurements per
  configuration~(shown in brackets) than indicated in
  Table~\ref{tab:sim}.  In all cases the errors include both
  statistical and systematic uncertainties.  For the axial Ward
  identity masses the asterisk~($^*$) indicates which values are used
  to determine the ratio of non-singlet to singlet renormalization
  factors presented in Section~\ref{renorm}.}
\begin{center}
\begin{ruledtabular}
\begin{tabular}{ccccccc}
Ensemble&  $am_\pi$ & $am_N$& $a\tilde{m}$ & $(t_{\mathrm{f}}/a)^{\rm conn}_{\rm nuc}$&  $(t_{\mathrm{f}}/a)^{\rm conn}_{\rm pion}$ & $\Delta t_{\mathrm{min}}/a$  \\
\hline
I   & 0.11516(73)&  0.4480(31) &  0.003676(39) &   13              &   &4 \\\hline
II  & 0.15449(74)&  0.4641(53) & 0.007987(44) &   15              &   &4 \\
III & 0.15298(46)&  0.4486(30) & 0.007964(34)$^*$  &  15,17           &32 &5 \\
IV  & 0.10675(51)&  0.3855(46) & 0.003794(28) &  7(1),9(1),11(1),&32 &4 \\
    &            &             & &  13,15,17        &   &    \\
V   & 0.10465(38)&  0.3881(35) & 0.003734(21) &  15              &32 &4 \\
VI  & 0.10487(24)&  0.3856(19) & 0.003749(18)$^*$ &  15              &   &5 \\
VIII& 0.05425(49)&  0.3398(63) & 0.000985(19)$^*$ &  9(1),12(2),15 &32 &5 \\\hline
IX  & 0.15020(53)&  0.3962(34) & 0.009323(25)$^*$ &  17              &   &4 \\
X   & 0.13073(61)&  0.3836(32) & 0.007005(23)$^*$ &  17              &   &6 \\       
XI  & 0.07959(27)&  0.3070(50) & 0.002633(14)$^*$ &  17              &   &5 
\end{tabular}
\end{ruledtabular}
\end{center}
\end{table*}

In the two flavour theory the strange quark is quenched and the size
of the corresponding systematic uncertainty is difficult to
quantify~(note that the dominant strange quark contribution can still be
computed, c.f., for example, Eq.~(\ref{eq:rvalsea})). This source of
uncertainty will be removed in future work on $N_f=2+1$ configurations
generated as part of the CLS effort~\cite{Bruno:2014jqa}. We fix the
valence strange quark mass parameter, $\kappa_s$, by tuning the
hypothetical strange-antistrange pseudoscalar meson mass to the value
$(m_{K^{\pm}}^2+m_{K^0}^2-m_{\pi^{\pm}}^2)^{1/2}\approx 686.9$~MeV 
within statistical errors, where experimental values are used for the kaon and
pion masses.

The two-point and three-point functions, needed to extract the
scalar matrix elements, have the form\footnote{Note that for the nucleon we apply the parity projection
  operator $\frac{1}{2}[\mathds{1}+\mathrm{sign}(t_{\rm f}-t_i)\gamma_4]$.}
\begin{eqnarray}
C_{{\rm 2pt}}(t_{\mathrm{f}},t_{\mathrm{i}}) &= & \sum_{\vec{x}}\langle  {\cal H}(\vec{x},t_{\mathrm{f}})\overline{ {\cal H}}(\vec{0},t_{\mathrm{i}})\rangle ,\label{eq:twopt0} \\
C_{{\rm 3pt}}(t_{\mathrm{f}},t,t_{\mathrm{i}}) & = &\sum_{\vec{x},\vec{y}}\langle {\cal H}(\vec{x},t_{\mathrm{f}}) S(\vec{y},t)\overline{ {\cal H}}(\vec{0},t_{\mathrm{i}})\rangle \nonumber\\
& &- \sum_{\vec{x},\vec{y}}\langle S(\vec{y},t)\rangle \langle {\cal H}(\vec{x},t_{\mathrm{f}}) \overline{ {\cal H}}(\vec{0},t_i)\rangle ,\label{eq:threept0}
\end{eqnarray}
for a hadron, ${\cal H}$, at rest created at a time $t_{\mathrm i}$, destroyed
at a time $t_{\mathrm{f}}$ and with the operator $S=\bar{q}q$ inserted
at a time $t$. The interpolators ${\cal H}=(u^TC\gamma_5 d)u$ and
$\bar{u}\gamma_5 d$ for the proton and pion, respectively, create
both ground and excited states with contributions which fall off
exponentially with the energy of the state in Euclidean time.  To
improve the overlap with the ground state, spatially extended
interpolators were constructed using Wuppertal
smeared~\cite{Gusken:1989ad,Gusken:1989qx} light quarks with spatially
APE smoothed gauge transporters~\cite{Falcioni:1984ei}. The number of
Wuppertal smearing iterations applied, $n_{\rm sm}$, shown in
Table~\ref{tab:sim}, was optimized for each ensemble such that ground
state dominance was achieved at similar physical times for different
light quark masses and lattice spacings, see Ref.~\cite{Bali:2014nma}
for more details.

Wick contractions for the three-point function lead to the connected
$C^{\mathrm{conn}}_{{\rm 3pt}}(t_{\mathrm{f}},t,t_{\mathrm{i}})$ and
disconnected contributions $C^{\mathrm{dis}}_{{\rm
    3pt}}(t_{\mathrm{f}},t,t_{\mathrm{i}})$, shown in
Fig.~\ref{fig:condiscon} for the nucleon.  The standard sequential
source method is employed to determine the connected diagram. This
provides the three-point function at all $t\in
[t_{\mathrm{i}}+2a,t_{\mathrm f}-2a]$ for fixed $t_{\mathrm{f}}$,
where the minimal distance $2a$ from source and sink is due to the use
of clover fermions. Table~\ref{tab:res} details the values of
$t_{\mathrm{f}}$ chosen~(relative to a source at the origin,
i.e. $t_{\mathrm{i}}=0$). For the nucleon the relative statistical
errors of $C^{\mathrm{conn}}_{{\rm 3pt}}$ increase rapidly with
increasing $t_{\mathrm{f}}-t_{\mathrm{i}}$, motivating small
source-sink separations. However, several $t_{\mathrm{f}}$ values are
needed to check for excited state contributions which, as we will see in
Sections~\ref{sec:pionfits} and~\ref{sec:nucfits}, are significant for
scalar matrix elements, even with optimized spatially extended
interpolators. 
In the pion case, the signal does not decay rapidly with
$t_{\mathrm{f}}$ and we choose $t_{\mathrm{f}}=T/2$.  This has the
advantage that $C_{{\rm 3pt}}(t_{\mathrm{f}},t,t_{\mathrm{i}})$ can be
averaged over the regions with $t_{\mathrm{i}}=0<t<t_{\mathrm{f}}$ and
$t_{\mathrm{f}}<t<t_{\mathrm{i}}=T$. Excited state contributions are
controlled using our method discussed in Appendix~\ref{app:spectral}.

\begin{figure}
\centerline{\includegraphics[width=.48\textwidth]{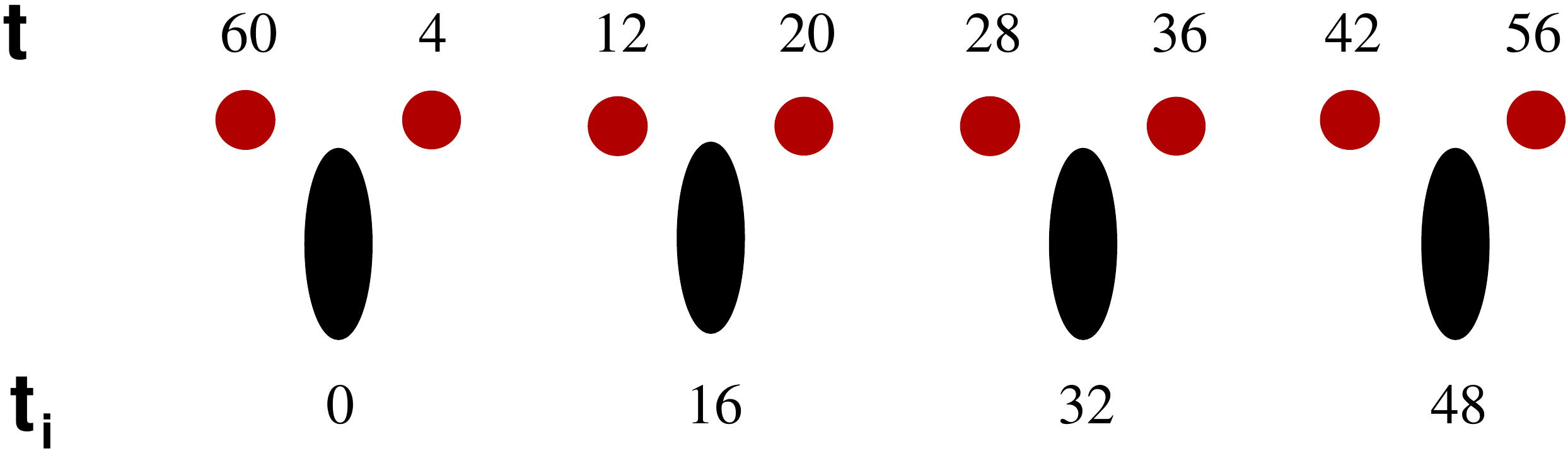}}
\caption{The time positions $t$ of the 8 disconnected loops and the
  source positions $t_{\mathrm{i}}$ of the 4 two-point functions  which
  are used to construct the disconnected three-point function in
  Eq.~(\ref{eq:dis}), for an ensemble with $T=64a$ and minimum
  $|t-t_{\mathrm{i}}|=4a$.\label{fig:relpos}}
\end{figure}

\begin{figure*}
\centerline{
\includegraphics[width=.48\textwidth,clip=]{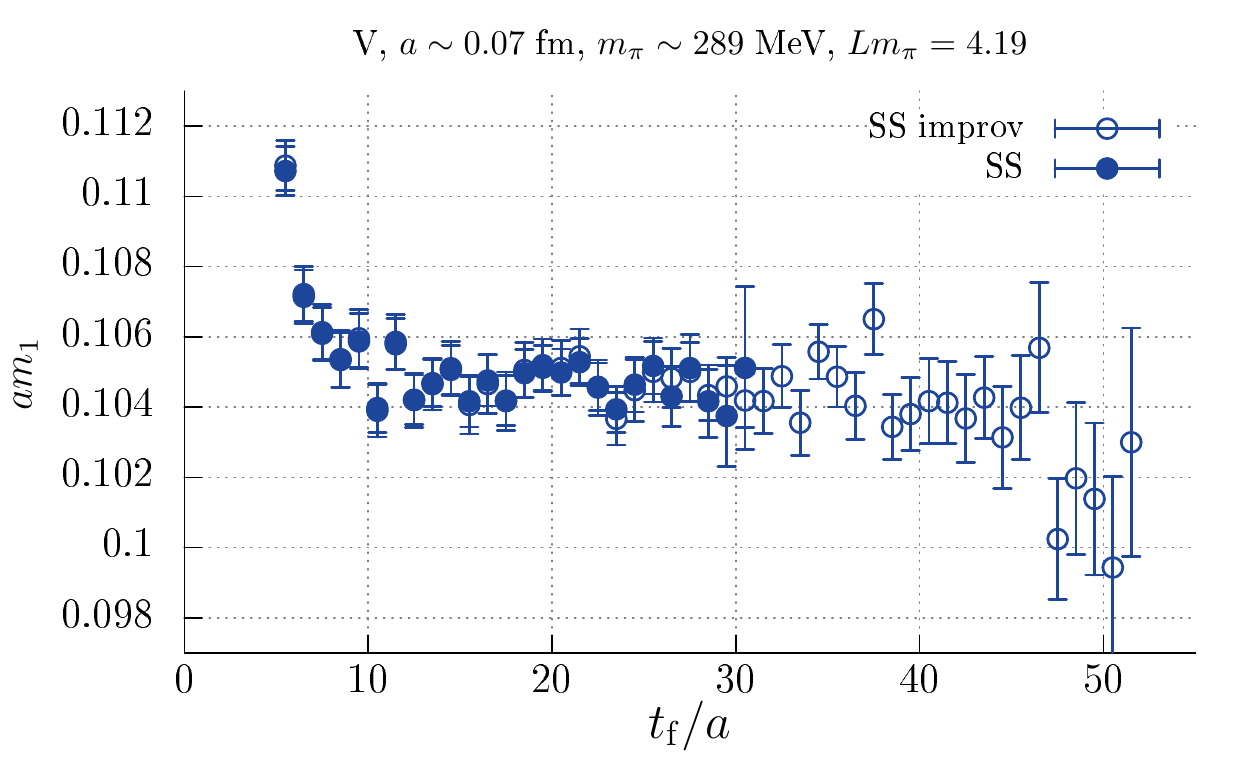}
\includegraphics[width=.48\textwidth,clip=]{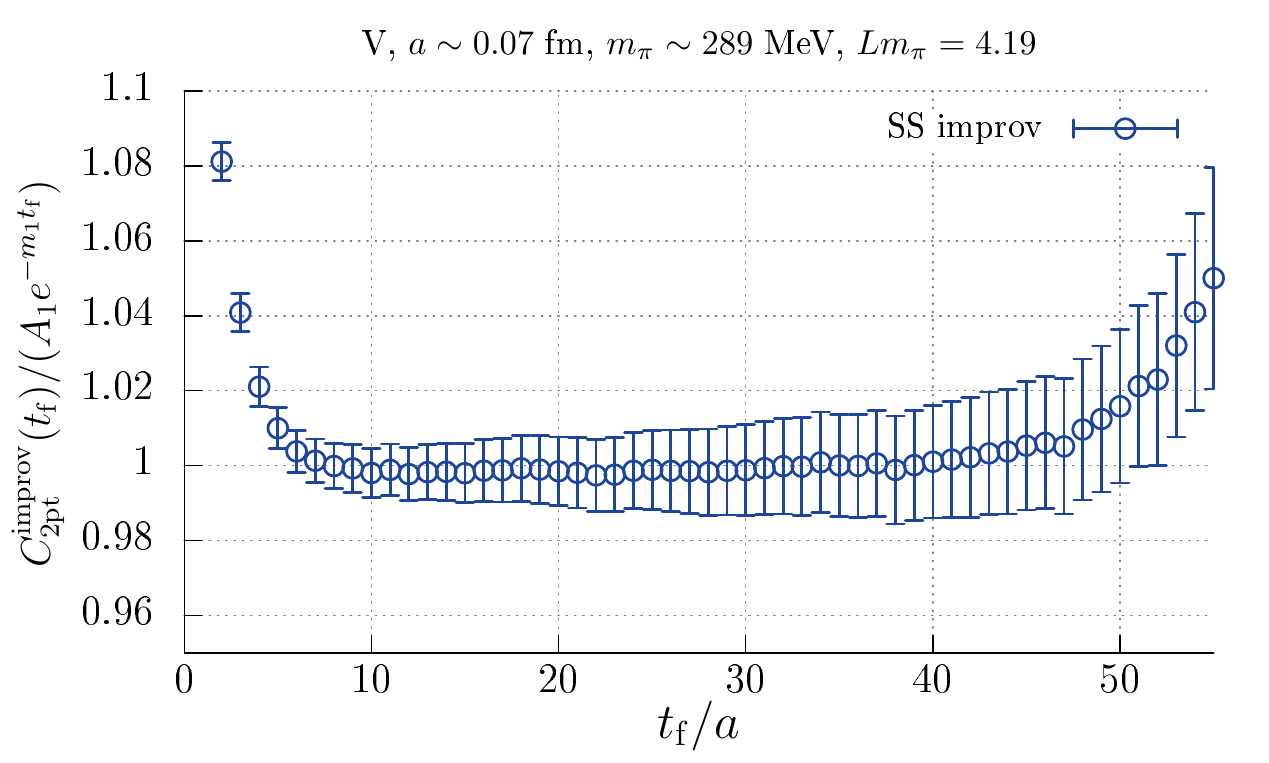}
}
\centerline{
\includegraphics[width=.48\textwidth,clip=]{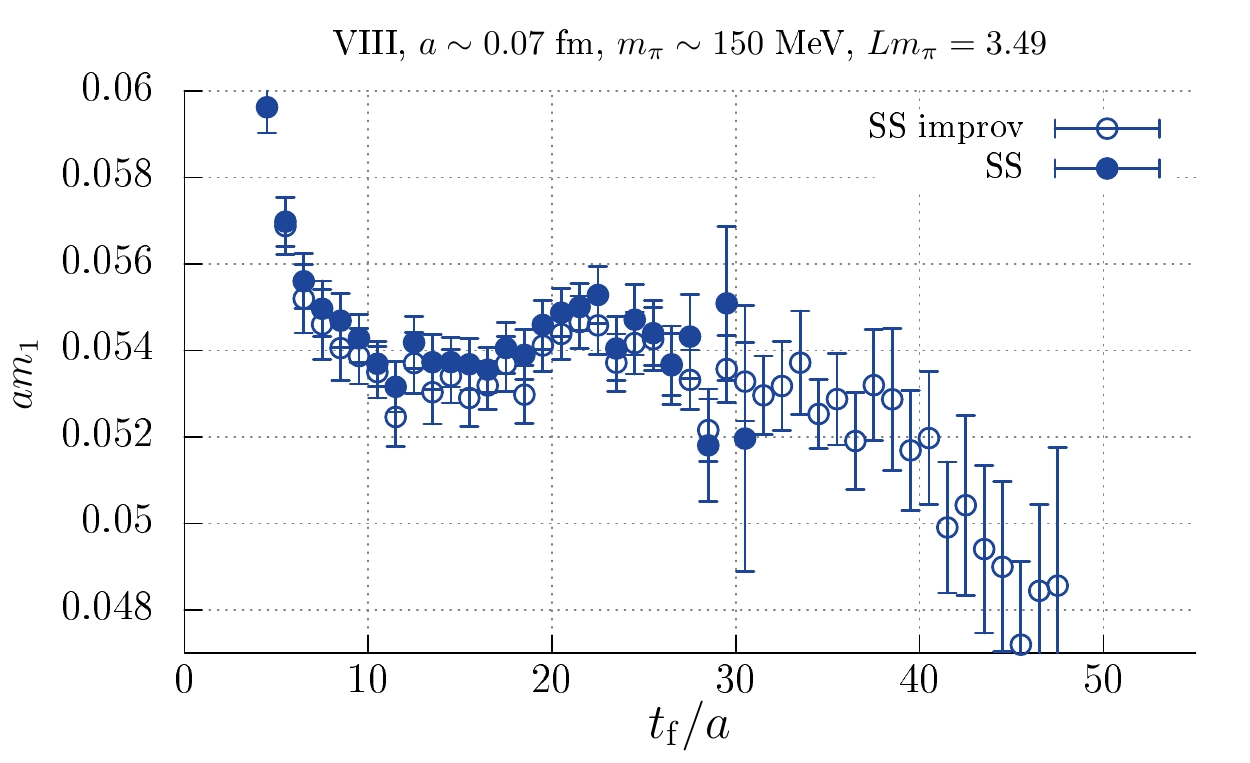}
\includegraphics[width=.48\textwidth,clip=]{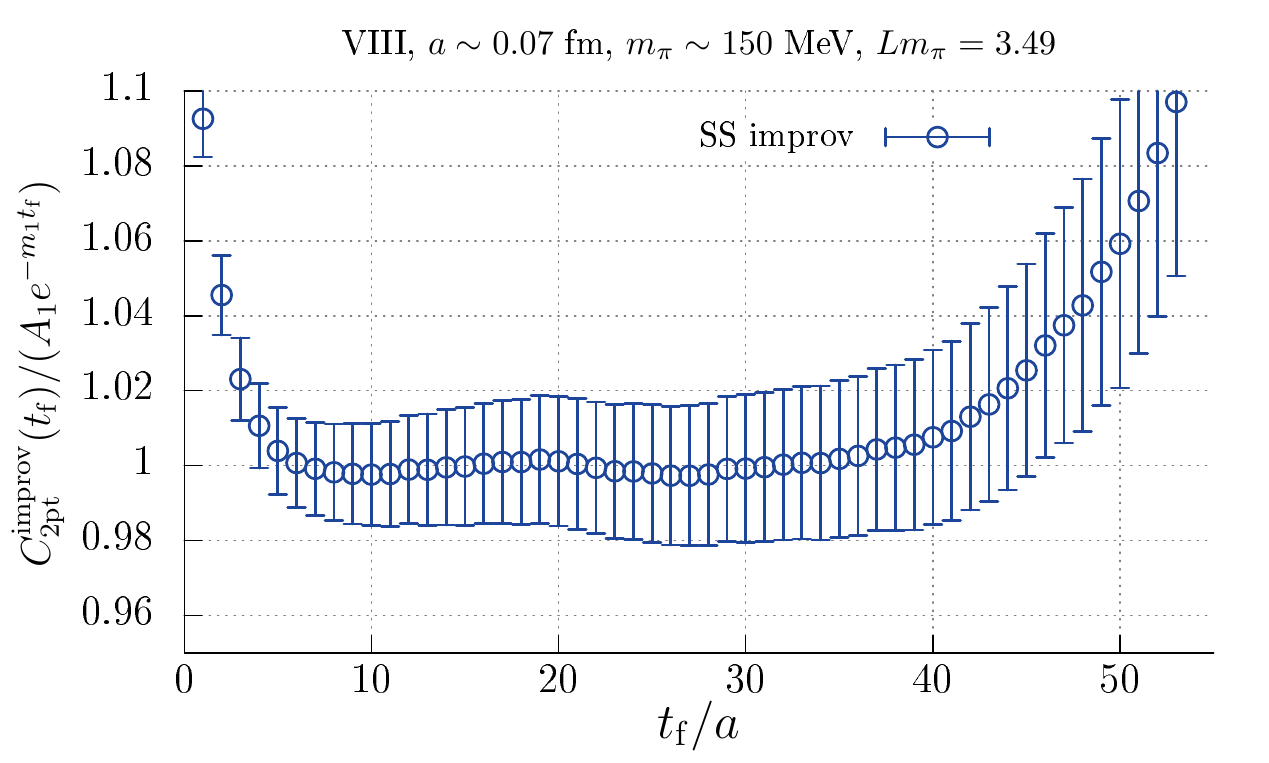}
}
\caption{(Left) A comparison of the effective masses of the
  pseudoscalar $C_{\rm 2pt}$ and $C_{\rm 2pt}^{\rm improv}$ for
  ensembles V~(top) and VIII~(bottom) with pion masses $m_\pi\sim
  289$~MeV and $\sim 150$~MeV, respectively. The correlators are
  smeared at both the source and the sink~(SS). Note that for $C_{\rm
    2pt}$ the inverse cosh effective mass is shown. (Right) $C_{\rm
    2pt}^{\rm improv}$ divided by the corresponding forward
  propagating ground state contribution, $A_1e^{-m_\pi t_{\rm
      f}}$~($A_1=|Z_{01}|^2$), extracted from a fit, for the same
  ensembles.}
\label{fig:pioneffmass}
\end{figure*}

The disconnected term is
constructed from a disconnected ``loop'' $L(t)$ and a two-point function
computed on each configuration:
\begin{equation}
C^{\mathrm{dis}}_{{\rm 3pt}}(t_{\mathrm{f}},t,t_{\mathrm{i}}) = \langle C^c_{{\rm 2pt}}(t_{\rm f},t_{\mathrm{i}}) L^c(t) \rangle_c - \langle C^c_{{\rm 2pt}}(t_{\rm f},t_{\mathrm{i}})\rangle_c\langle L^{c}(t) \rangle_{c}, \label{eq:dis}
\end{equation}
where $L^c(t) = \sum_{\vec{x}} \mathrm{Tr}[M^{-1}(x,x)\mathds{1}]$ on
configuration $c$, $\langle\cdot \rangle_c$ makes the configuration
average explicit and $x=(\vec{x},t)$. The quark propagator
$M^{-1}(x,x)$ is estimated stochastically using 25 complex
$\mathds{Z}_2$ random source vectors that are non-zero on 8
timeslices.\footnote{In the stochastic estimation of
the trace, terms off-diagonal in space or
  time average to zero, see Ref.~\cite{Bali:2009hu} for details.} This
number of stochastic estimates and level of time partitioning for
$L(t)$ ensured the additional random noise introduced to
$C^{\mathrm{dis}}_{{\rm 3pt}}$ was below the level of the intrinsic
gauge noise while also allowing for 8 measurements of the three-point
function per configuration.  The latter requires four different source
times $t^n_{\mathrm{i}}=nT/4$, $n=0,1,2,3$ for the two-point functions
appearing in Eq.~(\ref{eq:dis}). The disconnected loops are positioned
at timeslices $t^n_{\mathrm{i}}\pm \Delta t_{\mathrm{min}}$, where
$\Delta t_{\mathrm{min}}$ is a fixed minimum value of
$|t-t_{\mathrm{i}}|$ for each ensemble, see
Table~\ref{tab:res}. Figure~\ref{fig:relpos} illustrates the relative
positions of the disconnected loops and two-point function source
times for the example of an ensemble with $T=64a$ and $\Delta
t_{\mathrm{min}}=4a$. By correlating a forward~(backward) propagating
two-point function with source position $t^n_{\mathrm{i}}$ with a loop
at $t_{\rm i}^n+\Delta t$~($t_{\rm i}^n-\Delta t$), for each $n$ one
obtains the 8 estimates of the disconnected three-point function,
\begin{widetext}
\begin{align}
C^{\mathrm{dis}}_{{\rm 3pt}}(\Delta t_{\mathrm{f}},\Delta t) \equiv &
\frac{1}{8}\sum_{n=0}^3 \left\{\left[\langle C^c_{{\rm 2pt}}(t_i^n+\Delta t_{\rm f},t^n_{\mathrm{i}})
L^c(t^n_{\mathrm{i}} + \Delta t) \rangle_c - \langle C^c_{{\rm 2pt}}(t_i^n+\Delta t_{\rm f},t^n_{\mathrm{i}})\rangle_c\langle L^c(t^n_{\mathrm{i}} + \Delta t) \rangle_c \right]_{t_{\mathrm{f}}>t>t_{\mathrm{i}}}\right. \nonumber\\
& \left. + \left[\langle C^c_{{\rm 2pt}}(t_i^n-\Delta t_{\rm f},t^n_{\mathrm{i}})
L^c(t^n_{\mathrm{i}} - \Delta t) \rangle_c- 
\langle C^c_{{\rm 2pt}}(t_i^n-\Delta t_{\rm f},t^n_{\mathrm{i}})\rangle_c\langle
L^c(t^n_{\mathrm{i}} - \Delta t) \rangle_c\right]_{t_{\mathrm{i}}>t>t_{\mathrm{f}}} \right\}, \label{eq:dis2}
\end{align}
\end{widetext}
for multiple time separations $\Delta t = |t - t_{\rm i}| = \Delta
t_{\mathrm{min}}, T/4 - \Delta t_{\mathrm{min}} , T/4 + \Delta
t_{\mathrm{min}}, \ldots < | t_{\mathrm{f}} - t_{\rm i} |\equiv \Delta
t_{\mathrm{f}}$.  In order to determine both the light and the strange
quark content of the nucleon and pion, the loop $L^c(t)$ is evaluated
both for $\kappa_{\ell}$ and $\kappa_s$ and contracted with the two-point
function constructed from light quark propagators. For the nucleon, as we will
see in Section~\ref{sec:nucfits}, the statistical noise
increases rapidly with increasing source-operator insertion
separations and only $\Delta t = \Delta t_{\mathrm{min}}$ and $\Delta
t= T/4 - \Delta t_{\mathrm{min}}$ provide useful signals.  For the
pion, several $\Delta t$ give meaningful results.

The connected and disconnected three-point functions are analysed
separately to extract the corresponding contributions to the scalar
matrix elements, $g_S^{\rm latt,conn}=\langle
H|\bar{q}q|H\rangle^{\rm latt}_{\rm conn}$ and $g_S^{\rm latt,
  dis}=\langle H|\bar{q}q|H\rangle^{\rm latt}_{\rm dis}-\langle
0|\bar{q}q|0\rangle^{\rm latt}_{\rm dis}$, respectively, where
$|H\rangle\propto\overline{{\cal H}}|0\rangle$ is a nucleon or pion
state and we used the normalization $\langle H|H\rangle=1$, see below.
This procedure is described in the next two sections.

\subsection{Pion three-point function fits}
\label{sec:pionfits}

\begin{figure*}
\centerline{
\includegraphics[width=.48\textwidth,clip=]{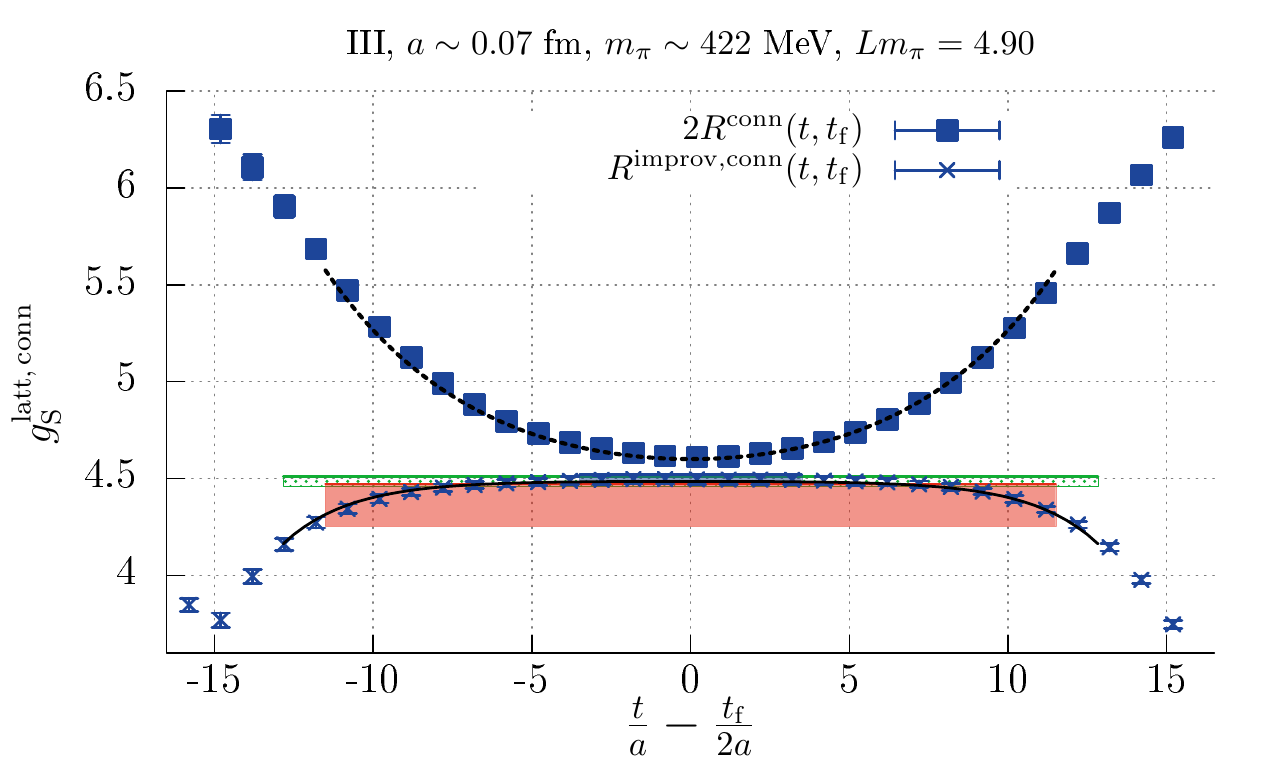}
\includegraphics[width=.48\textwidth,clip=]{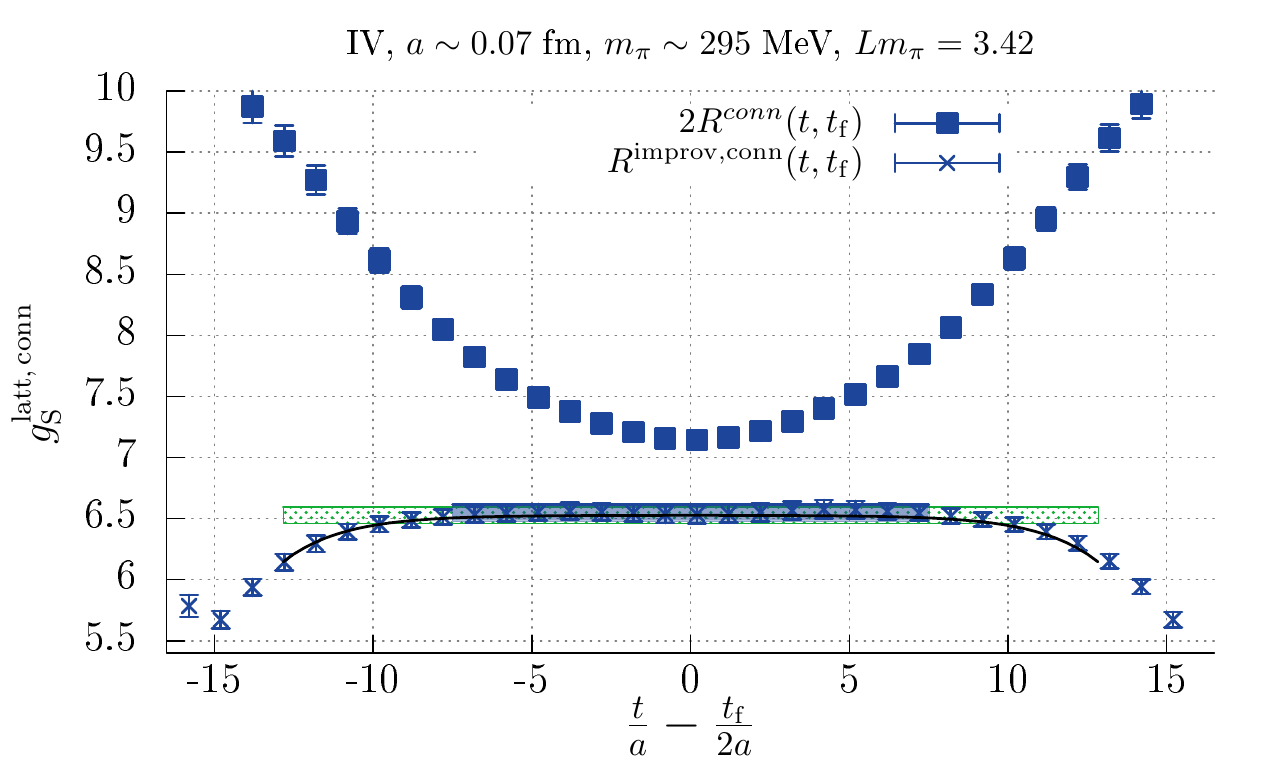}
}
\centerline{
\includegraphics[width=.48\textwidth,clip=]{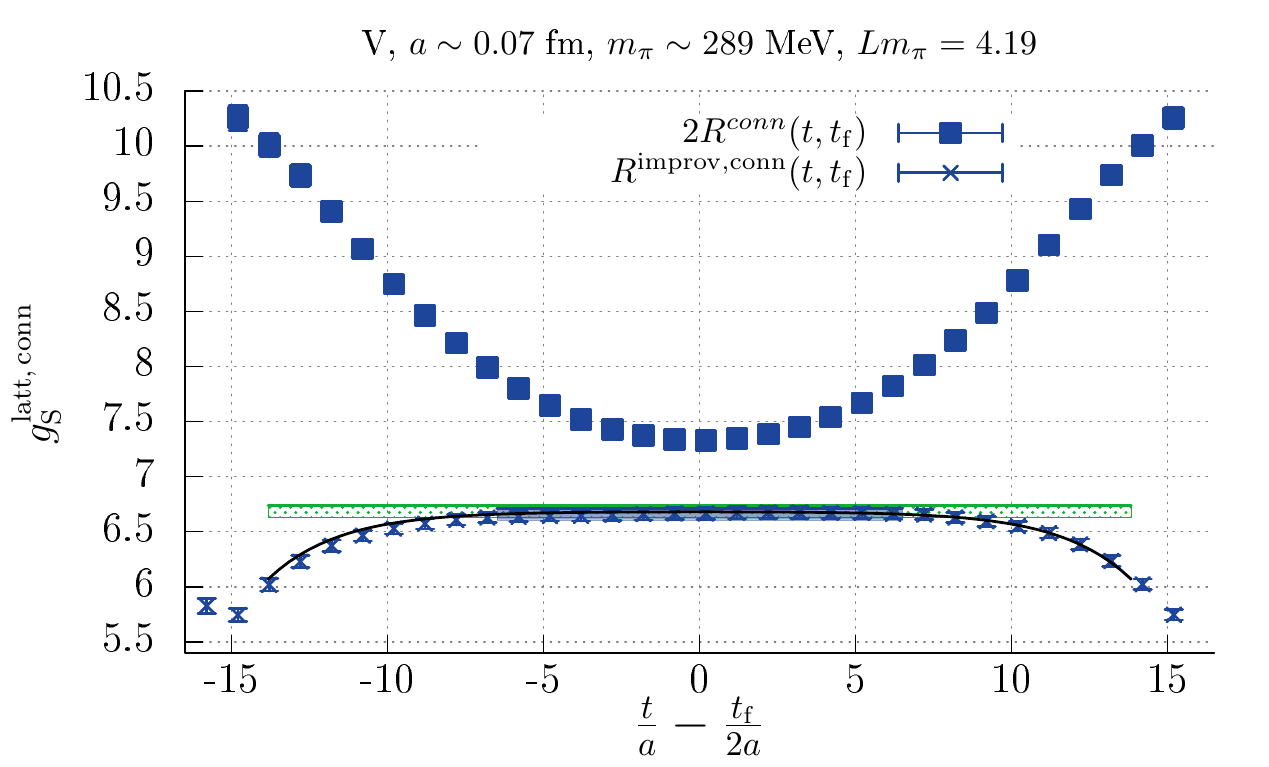}
\includegraphics[width=.48\textwidth,clip=]{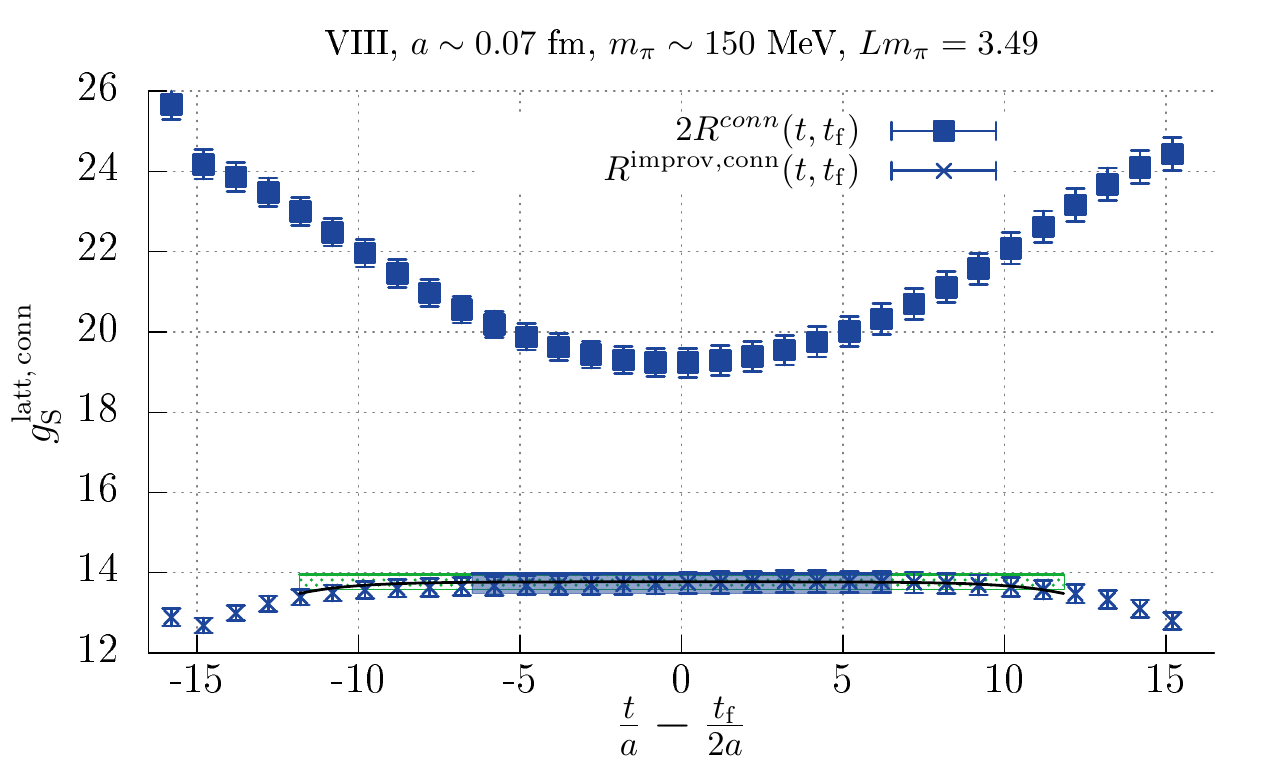}
}
\caption{The ratio of the pion connected three-point to two-point
  functions for both the standard~($R^{\rm conn}$) and
  improved~($R^{\rm improv,conn}$, Eq.~(\ref{eq:ratimprov})) cases,
  for ensembles III, IV, V and VIII. The blue and green shaded regions
  show the results for $g_S^{\rm latt,conn}$ obtained from a constant
  fit to $R^{\rm improv,conn}$ and a simultaneous fit to $C_{\rm
    3pt}^{\rm improv,conn}$ and $C_{\rm 2pt}^{\rm improv}$ using
  Eqs.~(\ref{eq:fitform1}) and~(\ref{eq:fitform2}), respectively.  The
  corresponding fits to Eq.~(\ref{eq:fitform2}) divided by the ground
  state two-point contribution, i.e. $B_1 + B_2\left(e^{(t_{\rm
      f}-t)\Delta E} + e^{-t\Delta E}\right)$, are shown as the black
  lines. For ensemble III, a similar fit to the unimproved $C_{\rm
    3pt}^{\rm conn}$ and $C_{\rm 2pt}$ is indicated by the dashed line
  and the resulting value of $g_S^{\rm latt,conn}$ as the pink
  region. In all cases the fitting range is indicated by the width of
  the shaded region and the range shown for the black and dashed
  lines. The fits shown are representative and, as discussed in the
  text, for the final results the variation arising from different
  fitting ranges is taken into account.  }
\label{fig:pionconn}
\end{figure*}

Pion three-point functions calculated on ensembles with anti-periodic
boundary conditions in time can suffer from large contributions
involving a backward propagating pion~(across the boundary) in
combination with a forward propagating scalar state, if the temporal
extent of the lattice is not large and the pion mass is close to the
physical value, as is the case, for example, for ensemble VIII~(see
Table~\ref{tab:sim}). These contributions and our method for reducing
them are discussed in detail in Appendix~\ref{app:spectral}. We
utilize correlation functions computed using quark propagators with
different temporal boundary conditions~(periodic and anti-periodic) to
isolate the forward propagating negative parity terms.
This is illustrated in Fig.~\ref{fig:pioneffmass} for the two-point
function on two representative ensembles with $m_\pi=289$~MeV and
$150$~MeV.

\begin{figure*}
\centerline{
  \includegraphics[width=.48\textwidth,clip=]{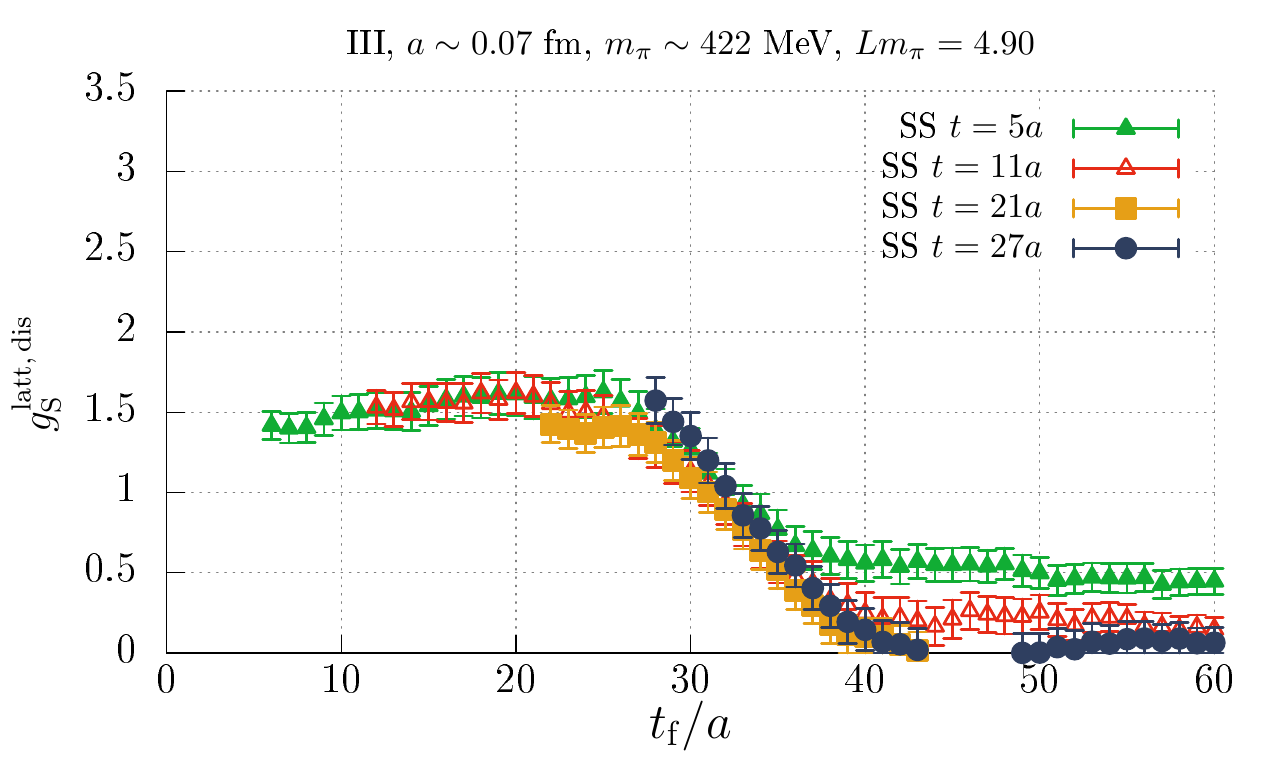}
  \includegraphics[width=.48\textwidth,clip=]{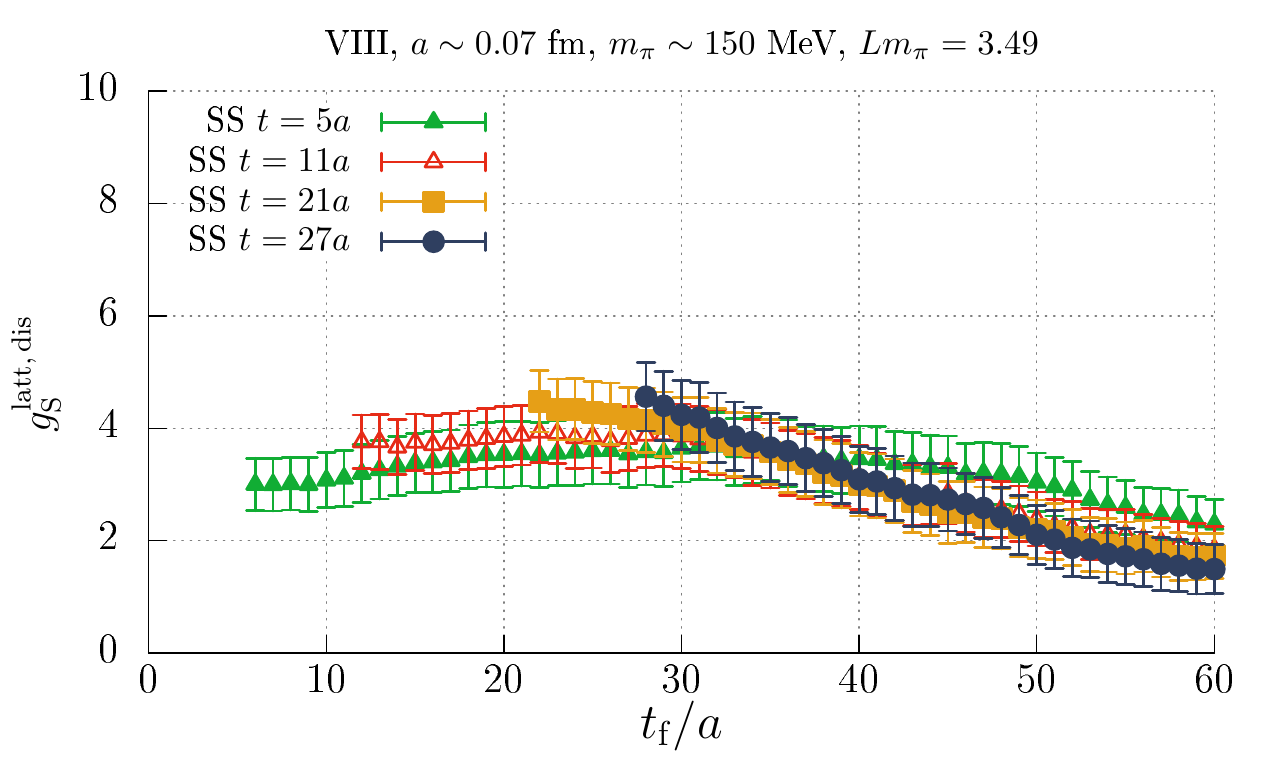}
} \centerline{
  \includegraphics[width=.48\textwidth,clip=]{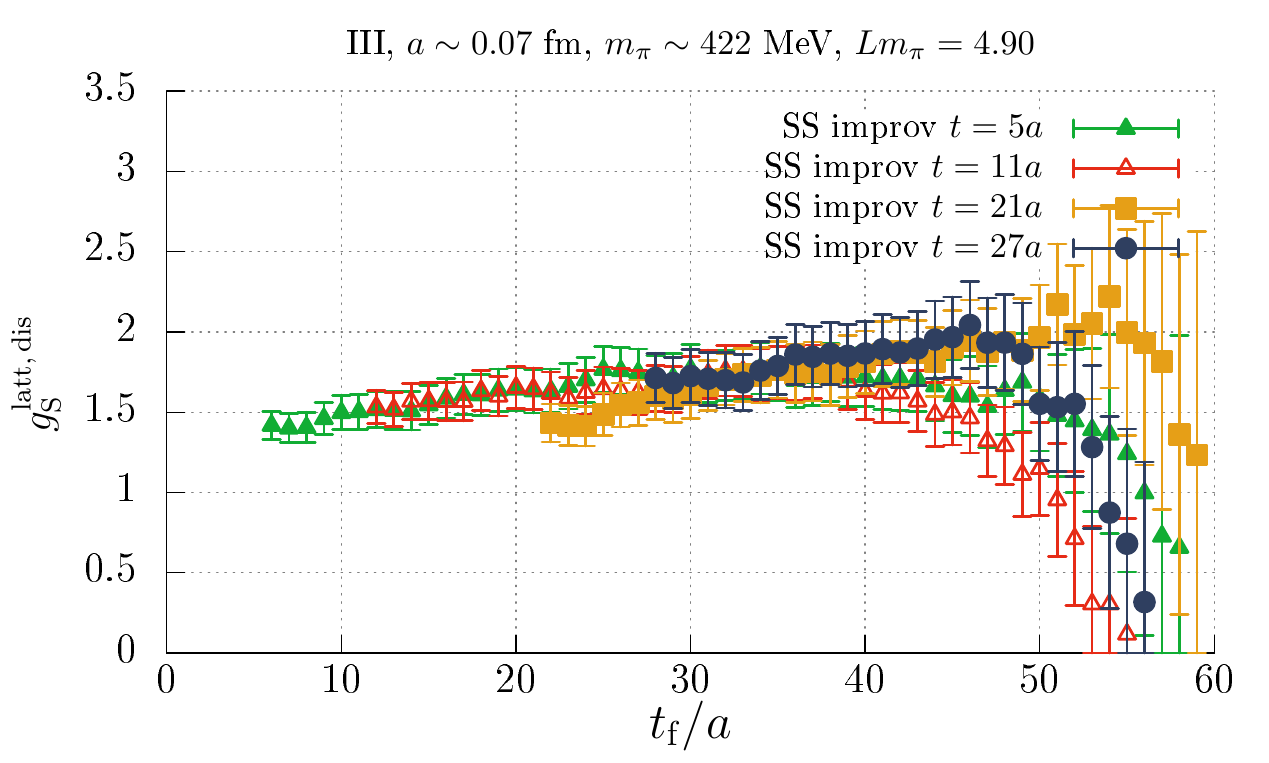}
  \includegraphics[width=.48\textwidth,clip=]{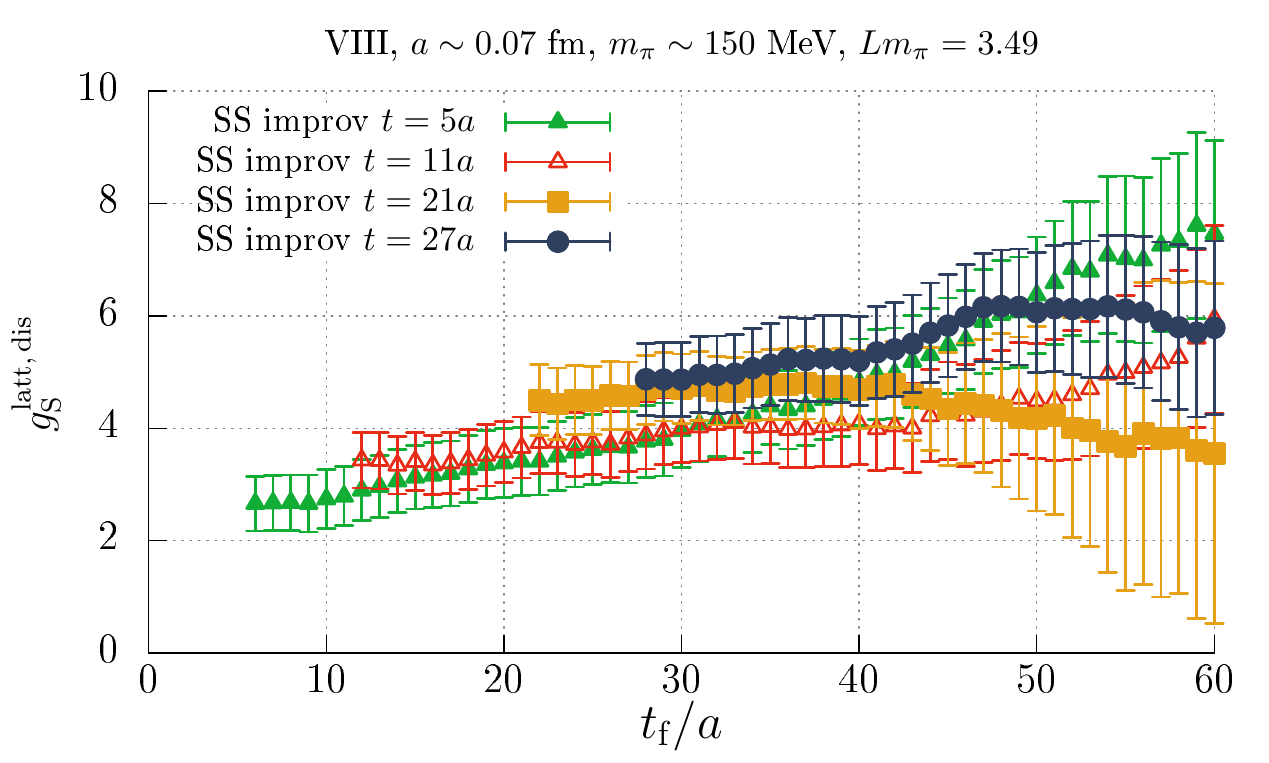}
}
\caption{(Top) The ratio $R^{\rm dis}(t_{\rm f},t,0)=C_{\rm 3pt}^{\rm
    dis}(t_{\rm f},t,0)/C_{\rm 2pt}(t_{\rm f},0)$ for the pion and $S=\bar{s}s$ with
  different values of the current insertion time $t$ on two ensembles
   with (left) $m_\pi=422$~MeV and (right)
  $m_{\pi}=150$~MeV. The correlation functions are smeared at both the source
  and the sink. The expectation value of the disconnected loop
  $\langle 0|S|0\rangle$ has been subtracted. (Bottom) The same ratios
  evaluated using improved correlation functions. }
\label{fig:piondiscon0}
\end{figure*}

The expected time dependence of the ``improved'' two-point
function, smeared at the source and sink~(SS), is given by
Eq.~(\ref{eq:improv3}),
\begin{align}
C^{\rm improv}_{{\rm 2pt}}(t_{\mathrm{f}},0) =    &
|Z_{01}|^2 e^{-t_{\rm f} E_1}\left[\vphantom{\frac{|Z_{12}|^2}{|Z_{01}|^2}} 1 + \frac{|Z_{12}|^2}{|Z_{01}|^2} e^{-(T-t_{\rm f})E_2}\right.\nonumber\\
& \left. + \frac{|Z_{03}|^2}{|Z_{01}|^2}e^{-t_{\rm f} (E_3-E_1)} + \ldots\right],\label{eq:main2}
\end{align}
where $t_{\rm i}=0$, $Z_{nm}= Z_{mn}^*=\langle n|{\cal
  \overline{H}}|m\rangle$. For simplicity we use the normalization
convention $\langle m|n\rangle=\delta_{m,n}$, rather than the
customary convention, $\langle 0|0\rangle=1$ and $\langle
n|n\rangle=2E_n$. Note that $E_1=m_\pi$. The ``$\ldots$'' indicate the
neglected higher excitations. We label the negative parity states by
odd numbers where $E_1$ and $E_3$ are the masses of the ground state
pion and a ``three-pion''~(or excited pion) state, respectively. The
latter association is made since in nature the excited state pion is
much heavier in mass.  Similarly, the positive parity states are
represented by even numbers and $E_2$ is the mass of a~(scalar)
``two-pion'' state.  

Figure~\ref{fig:pioneffmass} demonstrates that the contributions from
the three-pion state~(and higher negative parity states) die off
rapidly due to the optimized smearing and that ground state dominance
sets in around $t_{\rm f}=10a\approx 0.7$~fm for both ensembles,
independent of the pion mass, and continues until $t_{\rm f}\sim
40a\approx 2.8$~fm. As one would expect, up to $t_{\rm f}=T/2$ there is
no significant difference between the effective masses for the
improved and unimproved two-point functions. The terms arising from
scalar states propagating across the boundary become visible in the
improved case for large $t_{\rm f}$ values. This can be seen in the
combination $C^{\rm improv}_{{\rm 2pt}}(t_{\mathrm{f}},0)$, divided by
the ground state contribution, $|Z_{01}|^2e^{-m_\pi t_{\rm f}}$, as
determined from a fit, shown on the right in
Fig.~\ref{fig:pioneffmass}. For $t_{\rm f}/a \gtrsim 40$ this
ratio~(equal to the expression within the square brackets in
Eq.~(\ref{eq:main2})) increases from 1, with the deviation becoming
more significant for smaller $m_\pi$. This motivates us to restrict
$t_{\rm f}/a\le 40$ in order to avoid similar terms when fitting to
the pion three-point functions.

The connected three-point function is shown in Fig.~\ref{fig:pionconn}
as a ratio with the two-point function for the four ensembles used in
the pion analysis, with $m_\pi=422$~MeV down to $150$~MeV. In the
mass-degenerate $N_f=2$ theory $\langle 1|\bar{u}u|1\rangle^{\rm
  conn}= \langle 1|\bar{d}d|1\rangle^{\rm conn}$ and only a single
three-point function needs to be considered. Smeared sources and sinks
are implemented and the sink time $t_{\rm f}$ is fixed to $T/2$.
Using standard correlation functions, the ratio has the functional
form,\footnote{This can be seen from Eqs.~(\ref{eq:spectwopt})
  and~(\ref{eq:threept1}) in Appendix~\ref{app:spectral}, where for
  the connected three-point function the terms arising from the
  subtraction of $\langle 0| S|0\rangle$ are omitted and $\langle
  n|S|n\rangle_{\rm sub}$ is replaced by $\langle n|S|n\rangle$ for
  $n=1,2$.} $0\le t\le t_{\rm f}$,
\begin{widetext}
\begin{align}
 R^{\rm conn}(t_{\mathrm{f}},t,0)  = \frac{C_{\rm 3pt}^{\rm conn}(t_{\rm f},t,0)}{C_{\rm 2pt}(t_{\rm f},0)} &=
\left[ \langle 1|S|1\rangle^{\rm conn} +\frac{|Z_{12}|^2}{|Z_{01}|^2}\left( \langle 1|S|1\rangle^{\rm conn} e^{-(T-t_{\rm f})E_2}+\langle 2|S|2\rangle^{\rm conn} e^{-(T-2t_{\rm f})m_\pi}e^{-t_{\rm f}E_2}\right)
\right. \nonumber\\&  \left.
+\frac{Z^*_{01}Z_{21}}{|Z_{01}|^2}\langle 0|S|2\rangle^{\rm conn} e^{-(T-2t_{\rm f})m_\pi}\left( e^{-tE_2}+ e^{-(t_{\rm f}-t)E_2}\right)\right] \nonumber\\ &
\left[1+e^{-(T-2t_{\rm f}) m_\pi} + \frac{|Z_{12}|^2}{|Z_{01}|^2} \left(e^{-(T-t_{\rm f})E_2} + e^{-t_{\rm f} E_2}e^{-(T-2t_{\rm f}) m_\pi} \right)\right]^{-1}, \label{eq:rat0}
\end{align}
\end{widetext}
up to terms involving a three-pion state.

\begin{figure}
\centerline{
\includegraphics[width=.48\textwidth,clip=]{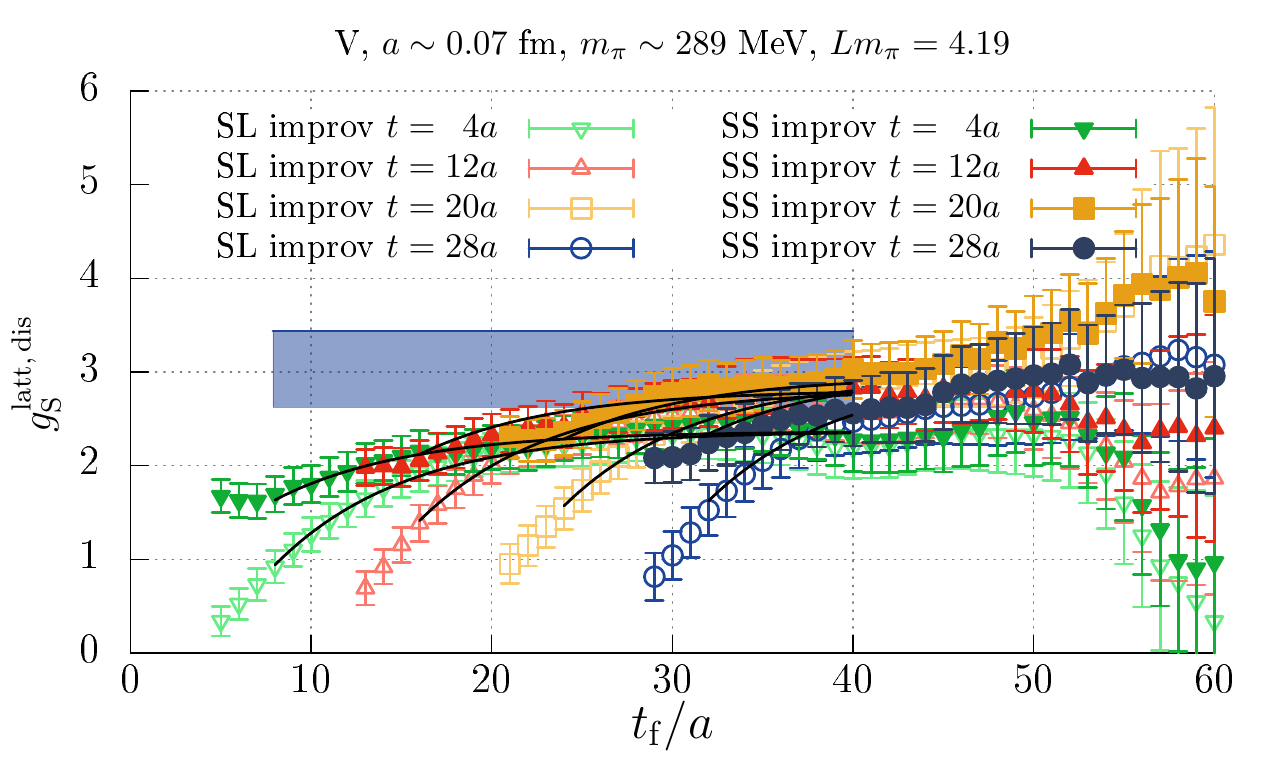}}
\centerline{
\includegraphics[width=.48\textwidth,clip=]{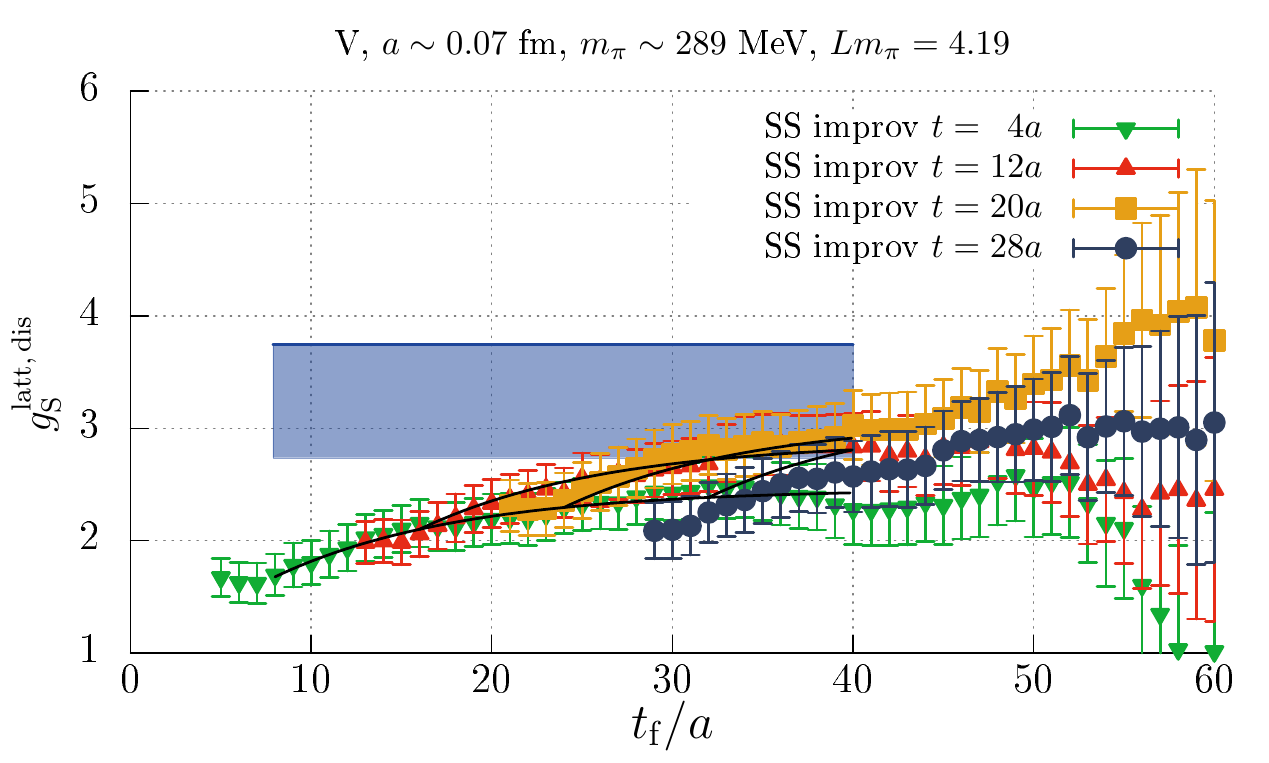}}
\centerline{
\includegraphics[width=.48\textwidth,clip=]{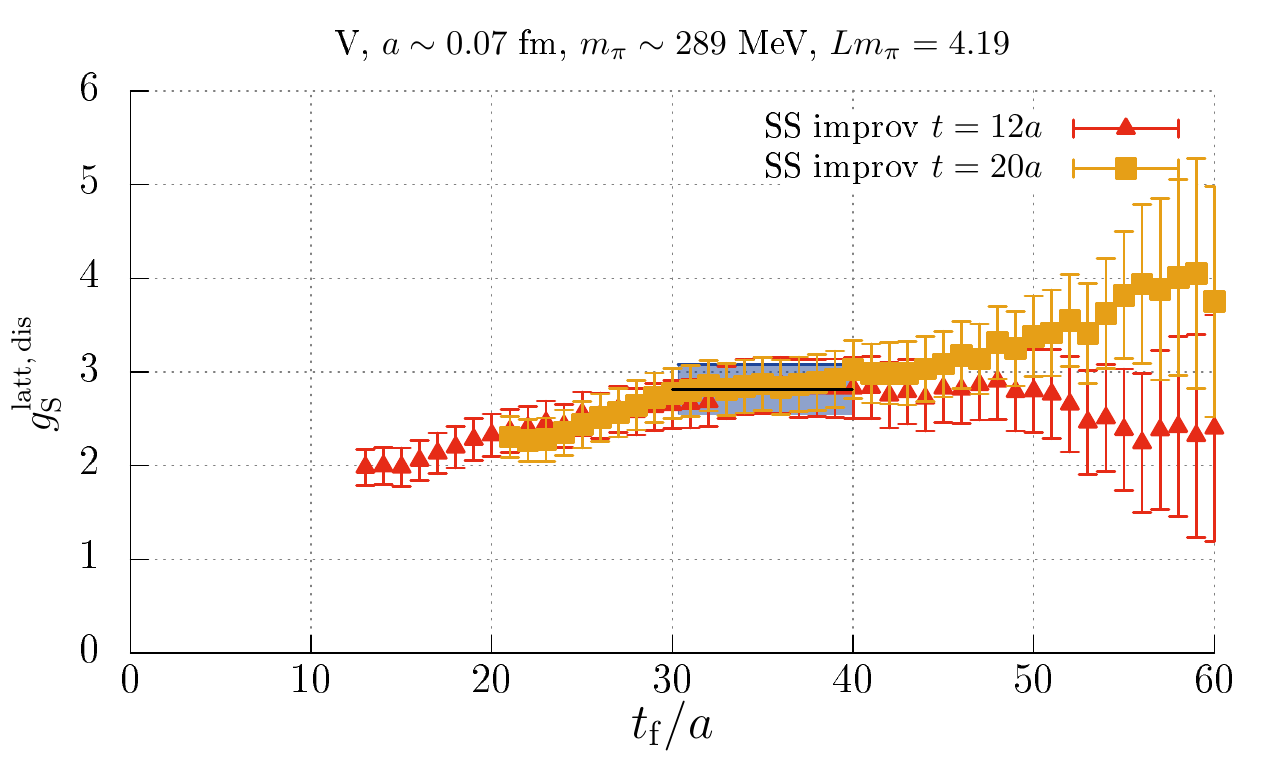}
}
\caption{Examples of fits performed to $R^{\rm improv,dis}(t_{\rm
    f},t,0)$ for the pion with a light quark loop on ensemble
  V~($m_\pi=289$~MeV). The black lines indicate the fit and the blue
  shaded region the value of $g_S^{\rm latt,dis}$ extracted in each
  case. The range shown for the black lines indicates the fit
  range. (Top) a simultaneous fit to the SS and SL ratios for
  multiple operator insertion times $t$, employing the
  parametrization given in Eq.~(\ref{eq:ratimprov3}). (Middle) a
  similar fit to SS ratios, (bottom) a constant fit to SS ratios.  }
\label{fig:piondiscon1}
\end{figure}

Employing our improved
correlation functions, contributions arising from the backward propagating
pion~(those involving factors of $e^{-(T-2t_{\rm f})m_\pi}=1$ for
$t_{\rm f}=T/2$) are removed and
\begin{align}
R^{\rm improv,conn}(t_{\mathrm{f}},t,0) & = 
\langle 1|S|1\rangle^{\rm conn}.\label{eq:ratimprov}
\end{align}
The considerable size of these contributions can be seen by comparing
the improved and unimproved ratios, as shown in
Fig.~\ref{fig:pionconn}. The difference between the two cases becomes
even more dramatic as $m_\pi$ decreases from $422$~MeV down to
$150$~MeV.  For $R^{\rm improv,conn}$ one can extract $\langle 1| S|
1\rangle^{\rm conn}$ by fitting to a constant~($B_1$) for small
$|t_{\rm f}/2-t|$. Examples of such fits are indicated by the blue regions
in Fig.~\ref{fig:pionconn}. However, the fitting range can be extended
by including the next order terms arising from a forward propagating
three-pion state. Equivalently, we perform simultaneous fits to
$C_{\rm 3pt}^{\rm improv,conn}$ and $C_{\rm 2pt}^{\rm improv}$ using
the functional form~(see Eqs.~(\ref{eq:improv4}) and~(\ref{eq:threept3}) in
Appendix~\ref{app:spectral}):
\begin{align}
C_{\rm 2pt}^{\rm improv}(t_{\rm f},0) & = A_1 e^{-m_\pi t_{\rm f}}\left[1 + A_2 e^{-\Delta E t_{\rm f}} \right],\label{eq:fitform1}\\
C_{\rm 3pt}^{\rm improv,conn}(t_{\rm f},t,0) & = \nonumber\\
&\hspace{-2cm} A_1 e^{-m_\pi t_{\rm f}}\left[B_1 + 
  B_2\left(e^{-(t_{\rm f}-t)\Delta E} + e^{-t\Delta E}\right)\right],\label{eq:fitform2}
\end{align}
where $\Delta E = E_3-m_\pi$ and $B_1\sim \langle 1|S|1\rangle^{\rm
  conn}$. For both these fits and the constant fits to $R^{\rm improv,
  conn}$ we have to assume that contributions to $B_1$ containing factors
$e^{-t_{\rm f}\Delta E}$ and $e^{-(T-t_{\rm f}) E_2}$ are small for $t_{\rm
  f} = T/2$. If $E_n=n m_\pi$ then $e^{-t_{\rm f}\Delta E}=
e^{-(T-t_{\rm f}) E_2}\sim 0.03$ for the lightest pion mass ensemble,
suggesting this assumption is reasonable. However, data with different
$t_{\rm f}$ would be needed to confirm this.

Final values for $\langle 1|S|1\rangle^{\rm conn}$ are obtained taking
into account the variation in the results due to the type of fit
used and the fitting range
chosen. For the latter all ranges with correlated $\chi^2/d.o.f. <2$
are included. If the covariance matrix for the fit is ill determined
due to insufficient statistics the fit result can be biased. To avoid
this problem the values for the scalar matrix elements are extracted
for the different fitting ranges using uncorrelated fits.

We remark that for the $m_\pi=422$~MeV ensemble fitting to the
unimproved $C_{\rm 3pt}^{\rm conn}$ and $C_{\rm 2pt}$ leads to a value
for $\langle 1|S|1\rangle^{\rm conn}$ consistent with the improved
result, albeit with larger statistical errors, see
Fig.~\ref{fig:pionconn}. The terms appearing in the numerator of
Eq.~(\ref{eq:rat0}) will dominate and one can see that $C_{\rm
  3pt}^{\rm conn}$ will have the same $t$
dependence as in Eq.~(\ref{eq:fitform2}), replacing $\Delta E$ by
$E_2$~(for fixed $t_{\rm f}=T/2$).  If we assume that these effects only depend on $T m_\pi $,
which is approximately $9.8$ for this ensemble, then $T\lesssim
180a\sim 12.8$~fm would be required at $m_\pi=150$~MeV in order to
ensure $\langle 1| S| 1\rangle^{\rm conn}$ can be reliably extracted
using standard correlators at $t_{\rm f}=T/2$. Having three-point
functions with multiple $t_{\rm f}$ can help, however, at least one
$t_{\rm f}$ value must be large enough that the unwanted terms are
significantly suppressed.

\begin{figure*}
\centerline{
  \includegraphics[width=.48\textwidth,clip=]{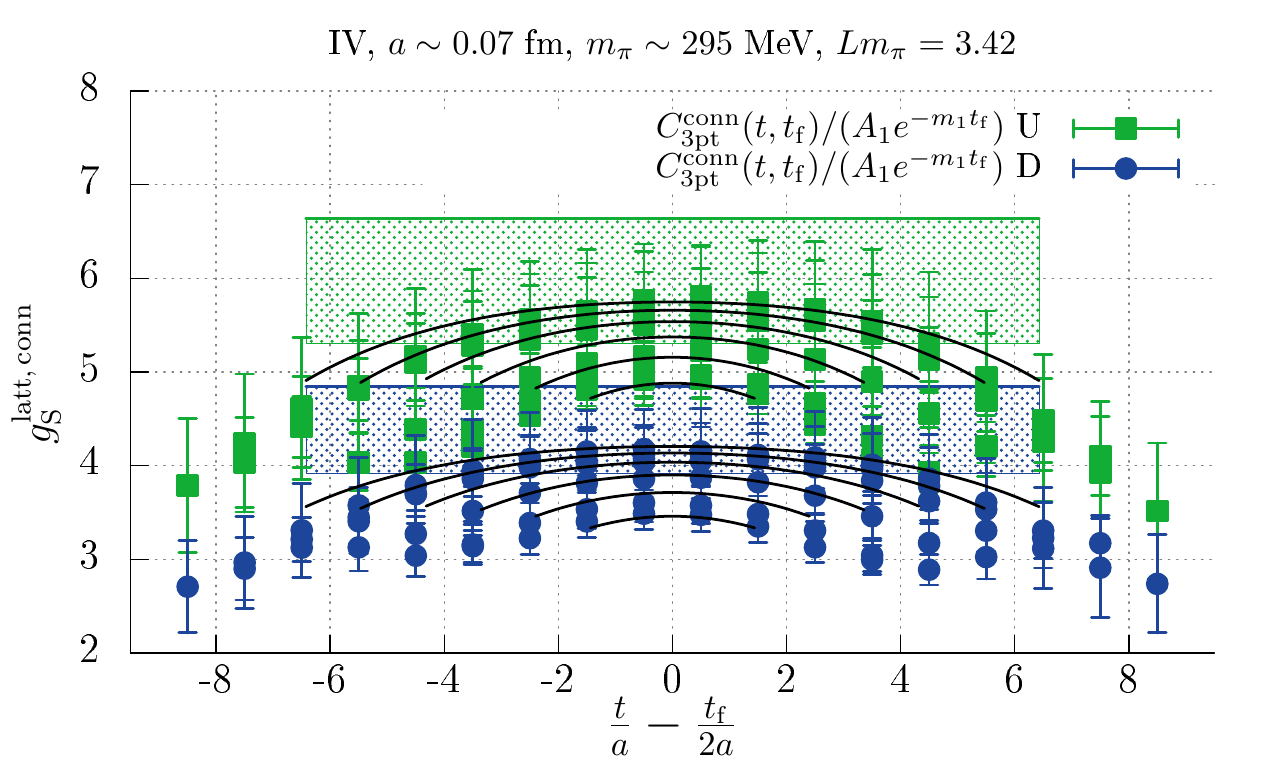}
  \includegraphics[width=.48\textwidth,clip=]{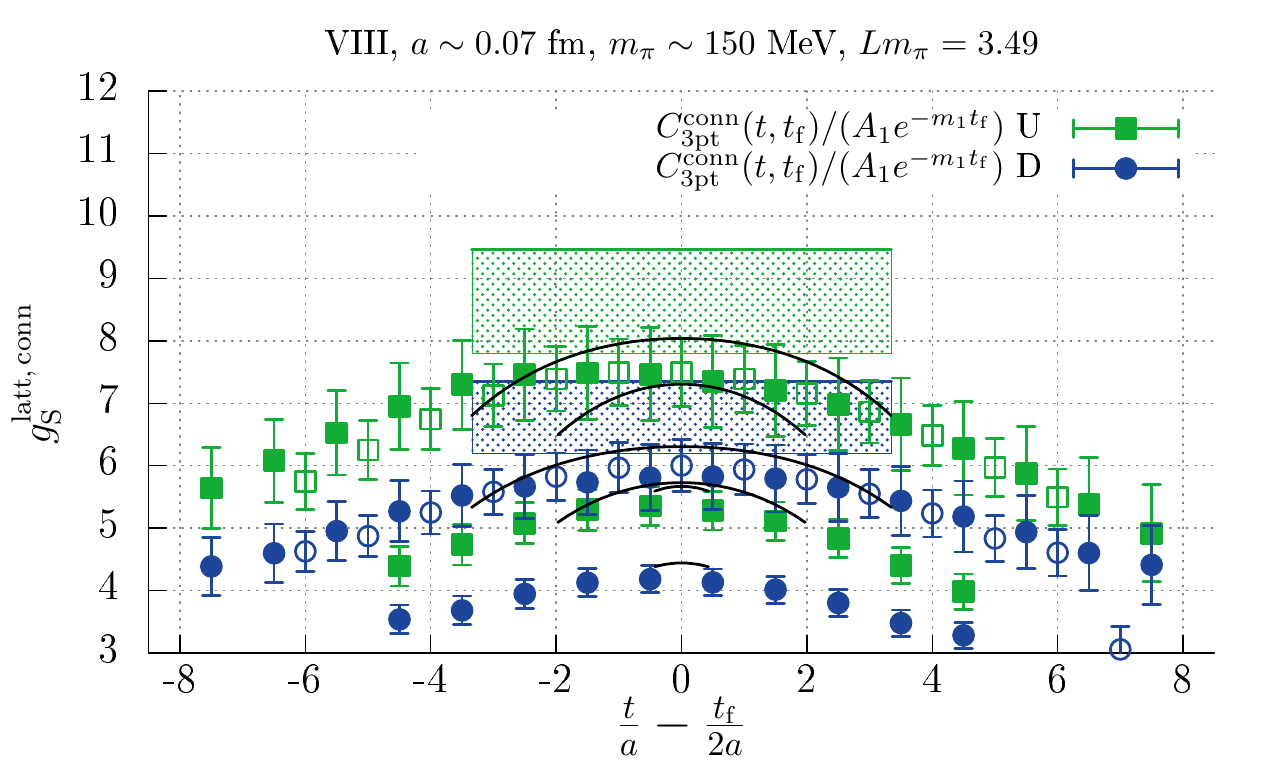}
}
\caption{Examples of simultaneous fits to proton SS $C_{{\rm
      2pt}}(t_{\mathrm{f}},0)$ and the two SS connected $C_{{\rm
      3pt}}^{{\rm conn}}(t_{\mathrm{f}},t)$ corresponding to
  $S=\bar{u}u$ and $\bar{d}d$, including multiple $t_{\rm f}$ values
  on ensembles IV and VIII~($m_\pi=295$ and $150$~MeV,
  respectively). The fit form is given in Eqs.~(\ref{eq:twonuc})
  and~(\ref{eq:threenuc}). The data points are obtained by dividing
  the three-point function by the ground state contribution, $A_1
  e^{-m_1 t_{\rm f}}$, where $A_1=|Z_{01}|^2$, as determined from the
  fit. The black lines indicate the fits while the green~(blue) shaded
  region gives the resulting value for $\langle
  N_1|\bar{u}u|N_1\rangle^{\rm conn}$~($\langle
  N_1|\bar{d}d|N_1\rangle^{\rm conn}$) and the width shows the fitting
  range chosen. For clarity the data points for $t_{\rm f}=12a$ on
  ensemble VIII are shown with open symbols. }
\label{fig:nucconna}
\end{figure*}

\begin{figure}
\centerline{
  \includegraphics[width=.48\textwidth,clip=]{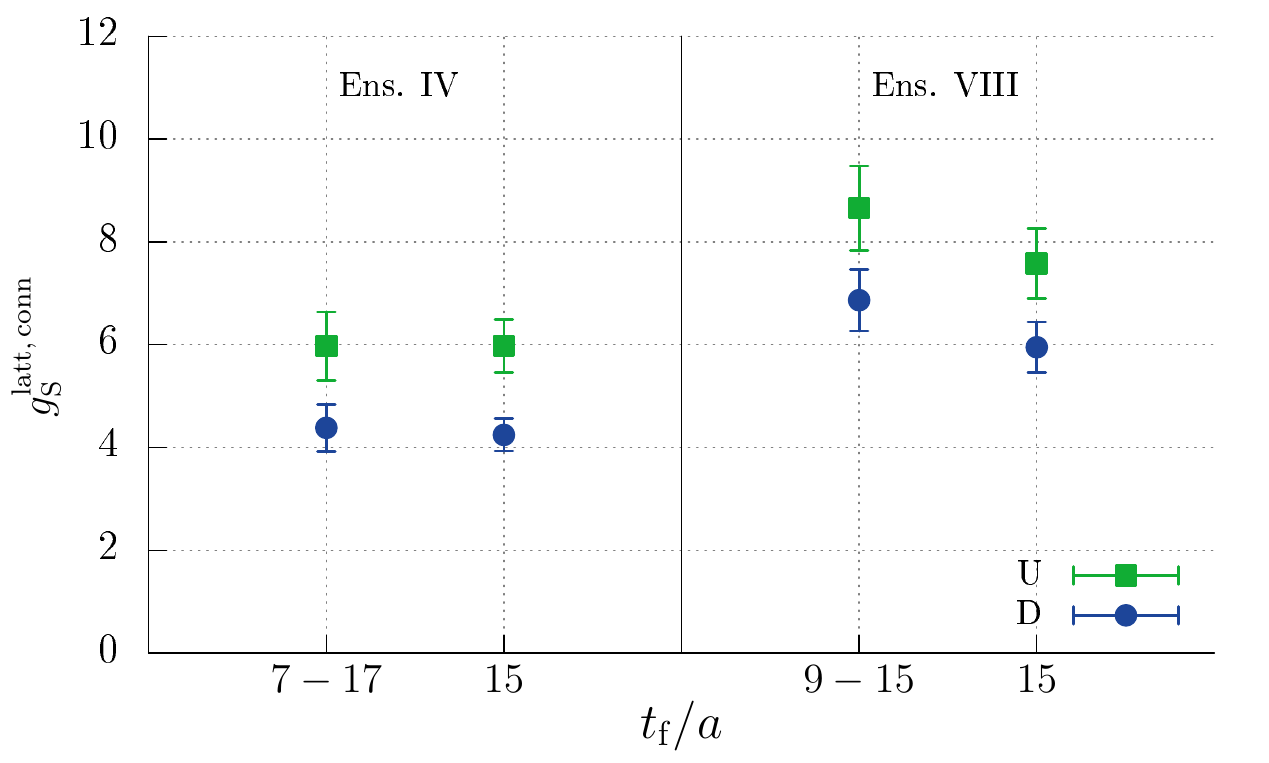}
}
\caption{ A comparison of the results for the scalar matrix elements
  obtained from fits with multiple $t_{\rm f}$, as shown in
  Fig.~\ref{fig:nucconna}, and fits to a single $t_{\rm f}=15a$ with
  the same fit range.  }
\label{fig:nucconnb}
\end{figure}

For the analysis of the disconnected contribution the three-point
function has been computed for both $S=\bar{u}u$~($\bar{d}d$) and
$\bar{s}s$ at multiple current insertion times and for all sink
times. In Fig.~\ref{fig:piondiscon0} we again consider the ratio with the
two-point function for the improved and unimproved cases with a
strange quark loop for ensembles III and VIII~($m_\pi=422$~MeV and
$150$~MeV, respectively). Note that the correlators are smeared
at the source and the sink. Including only the vacuum and pion state in
the spectral decompositions of the correlators, the unimproved ratio
has the time dependence,
\begin{align}
R^{\rm dis}(t_{\rm f},t,0) & = \frac{C_{\rm 3pt}^{\rm dis}(t_{\rm f},t,0)}{C_{\rm 2pt}(t_{\rm f},0)} \nonumber\\
&= \left(\langle 1|S|1\rangle^{\rm dis}-\langle 0|S|0\rangle^{\rm dis}\right) \frac{1}{1+e^{-m_\pi (T-2t_{\rm f})}},\label{eq:01rat}
\end{align}
independent of $t$.  From Eq.~(\ref{eq:01rat}), for small $t_{\rm
  f}>t$, one expects $R^{\rm dis}\sim \langle 1|S|1\rangle^{\rm
  dis}-\langle 0|S|0\rangle^{\rm dis}$. As $t_{\rm f}$ increases the ratio
should drop to half of its value at $t_{\rm f}=T/2$ and continue to
tend to zero, if contributions from states
$|2\rangle, |3\rangle, \ldots$ are small. This behaviour is seen in
Fig.~\ref{fig:piondiscon0}~(top left) for the $m_\pi=422$~MeV ensemble
in the limit of large $t<t_{\rm f}\le T/2$. However, when the pion
mass is decreased the terms arising from the backward propagating pion
across the boundary~(with forward propagating scalar $|2\rangle$),
given in Eqs.~(\ref{eq:spectwopt}) and~(\ref{eq:threept1}), become
very large and $R^{\rm dis}$ does not drop off significantly, as shown
in Fig.~\ref{fig:piondiscon0}~(top right) for $m_\pi=150$~MeV.
Applying our improvement procedure, these terms are removed and one
expects simply $R^{\rm improv,dis}(t_{\rm f},t,0)= \langle
1|S|1\rangle^{\rm dis}-\langle 0|S|0\rangle^{\rm dis}$ for $t_{\rm
  f}\ll T$. As observed in Fig.~\ref{fig:piondiscon0}~(bottom), the
improved ratio is constant for different current insertion times up to
$t_{\rm f}\sim 40a$ on both ensembles. For larger $t_{\rm f}$ values, terms
involving a backward propagating scalar particle cannot be ignored
anymore.

In order to extract $\langle 1|S|1\rangle^{\rm dis}-\langle
0|S|0\rangle^{\rm dis}$ we perform three types of fits to $R^{\rm
  improv,dis}$: for SS correlators we fit the ratio to a constant and,
whenever the next order terms can be resolved, also to a functional form
which includes a three-pion state~(Eq.~(\ref{eq:ratimprov3})). The
latter is also employed to fit the ratio constructed from correlators
smeared at the source and local at the sink~(SL), together with the SS
ratio.\footnote{As discussed at the end of Appendix~\ref{app:spectral}
  the fit function derived from Eq.~(\ref{eq:ratimprov3}) needs to be
  modified for a ratio of SL correlation functions.} As for the
analysis of the connected part, the fitting range is varied with the
restriction that the correlated $\chi^2/d.o.f.<2$, the final error
taking into account the spread of results from uncorrelated fits due
to different fit types and ranges. Representative examples of
fits are given in Fig.~\ref{fig:piondiscon1} for a disconnected
three-point function with a light quark loop on ensemble V~($m_\pi=289$~MeV).

\subsection{Nucleon three-point function fits}
\label{sec:nucfits}

For the nucleon scalar matrix elements three-point functions have been
computed on all ensembles shown in Table~\ref{tab:sim}. Note that we
are working in the isospin limit but take the nucleon corresponding to
a proton~($uud$). This distinction is only necessary for the connected
part. For the latter excited state contamination is explored using
multiple sink times at three pion masses, $m_\pi=426$~MeV, $289$~MeV
and $150$~MeV, at the lattice spacing $a\approx 0.071$~fm. The
standard fit form for SS correlators including contributions from the
first excited state is derived from the spectral decomposition:
\begin{align}
&\,\,\,\,\,C_{{\rm 2pt}}(t_{\mathrm{f}}) =   |Z_{01}|^2 e^{-m_1 t_{\mathrm{f}}} \left[1+ \frac{|Z_{02}|^2}{|Z_{01}|^2} e^{- \Delta m t_{\mathrm{f}}}+\ldots\right],\label{eq:twonuc}\\
& C_{{\rm 3pt}}(t_{\mathrm{f}},t)  = 
 |Z_{01}|^2e^{-m_1 t_{\mathrm{f}}}\left[\vphantom{\frac{|Z_{12}|^2}{|Z_{01}|^2}} \langle  N_1|S|N_1\rangle \right.\nonumber\\\
& \left. \,\,\,\,\,\, + \frac{Z_{20}^*Z_{10}}{|Z_{01}|^2} \langle N_2|S|N_1\rangle\left( e^{-\Delta m (t_{\mathrm{f}}-t)}  +   e^{-\Delta m t} \right)\right.
\nonumber\\& \left.\,\,\,\,\,\,+ |Z_{02}|^2\langle N_2|S|N_2\rangle e^{-\Delta m t_{\mathrm{f}}}+\ldots\vphantom{\frac 12}\right],
\label{eq:threenuc}
\end{align}
where $Z_{i0}=\langle N_i|{\cal \overline{N}}|0\rangle$ are the
overlaps of the state ${\cal \overline{N}}|0\rangle$, created by a
nucleon interpolator ${\cal \overline{N}}$ with the ground and first
excited nucleon states $|N_1\rangle$ and $|N_2\rangle$, respectively. We
denote the corresponding masses as $m_1$ and $m_2$ and the mass
difference as $\Delta m = m_2-m_1$. For the connected three-point
function there are two contributions arising from the scalar current
$S=\bar{u}u$, inserted on a $u$ quark line, and similarly for
$S=\bar{d}d$, inserted on the $d$ quark line. Both contributions are
fitted simultaneously along with the two-point function to extract
$\langle N_1|\bar{u}u|N_1\rangle^{\rm conn}$ and $\langle
N_1|\bar{d}d|N_1\rangle^{\rm conn}$, respectively. With data at
several values of $t_{\rm f}$, see Table~\ref{tab:res}, the last term
in Eq.~(\ref{eq:threenuc}) can be resolved as well as the dependence
on the current insertion time $t$. 

\begin{figure}
\centerline{
\includegraphics[width=.48\textwidth,clip=]{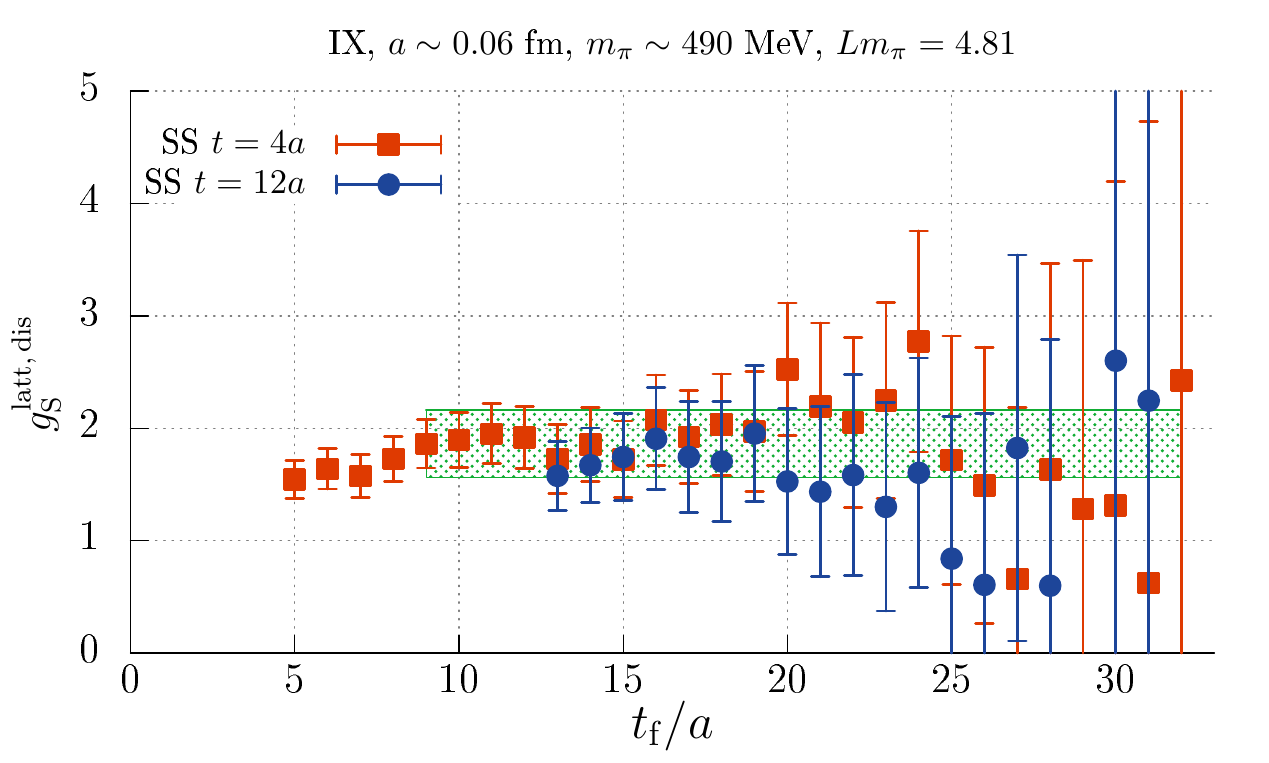}}
\centerline{
\includegraphics[width=.48\textwidth,clip=]{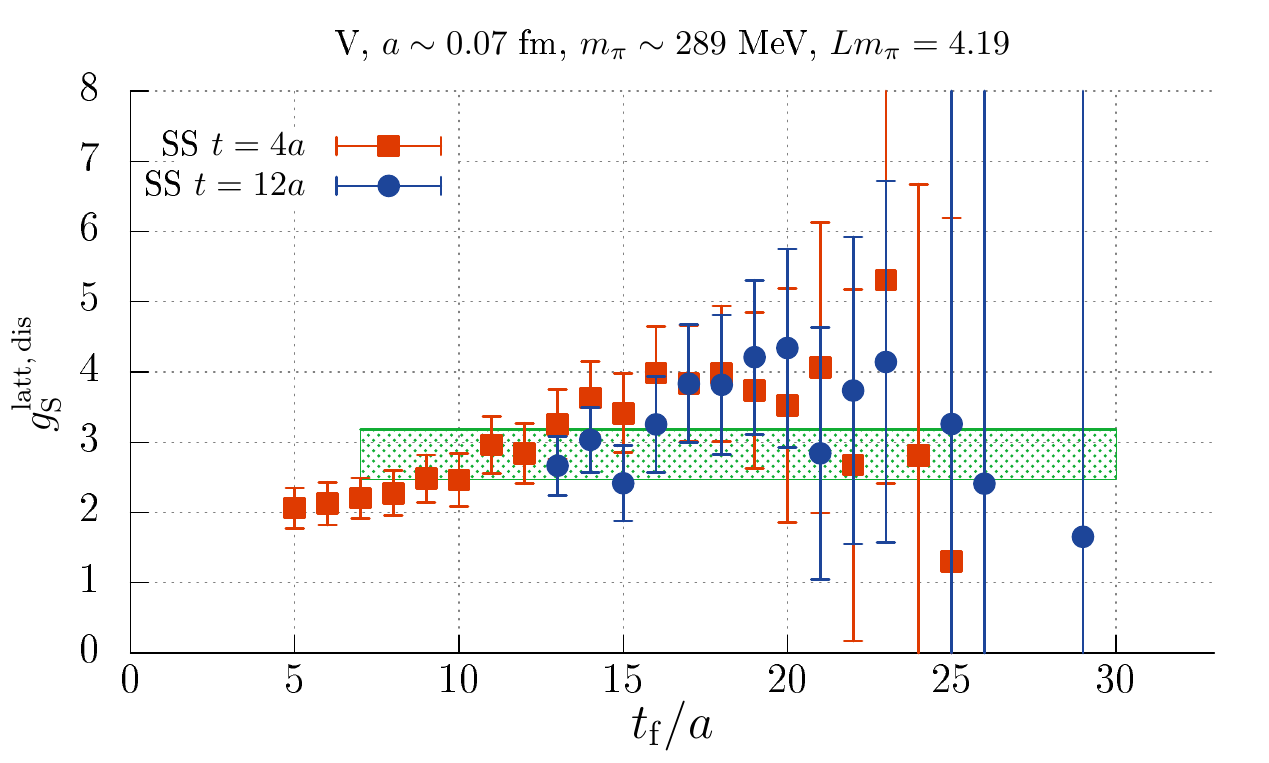}
}
\centerline{
\includegraphics[width=.48\textwidth,clip=]{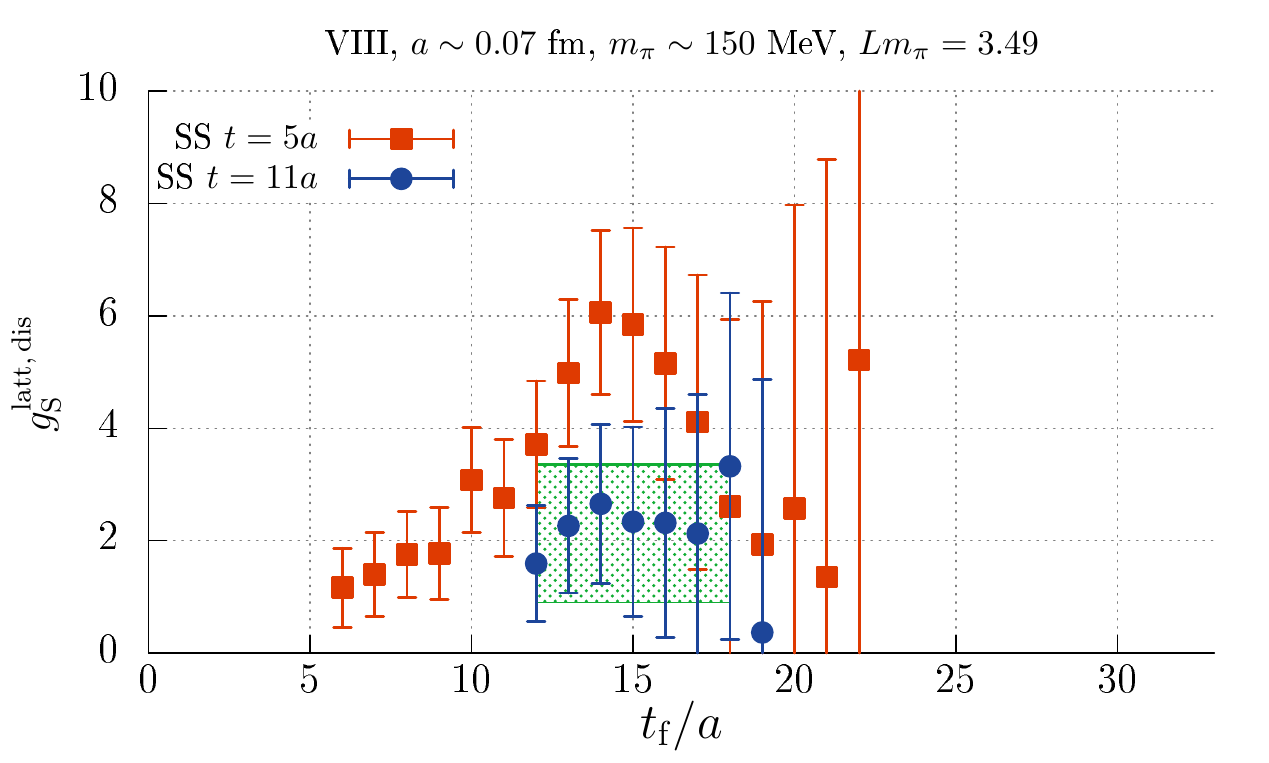}
}
\caption{The ratio $R^{\rm dis}(t_{\rm f},t,0)$ for the nucleon with a
  light quark loop as a function of $t_{\rm f}$ and $t$ on ensembles
  IX, V and VIII~($m_\pi=490$~MeV, $289$~MeV and $150$~MeV,
  respectively). The green shaded regions indicate the value of
  $g_S^{\rm latt,dis}$ extracted in each case.  For ensembles IX and V
  constant fits are performed to the ratios with $t=4a,12a$
  simultaneously with fit ranges $t_{\rm f}/a=9-32,17-32$ and
  $7-30,15-30$, respectively, while for ensemble VIII a constant fit
  is performed to the ratio with $t=11a$ with fit range $t_{\rm
    f}/a=12-18$. }
\label{fig:nucdiscon}
\end{figure}

Typical simultaneous fits to
two-point and multiple three-point functions~(including different
$t_{\rm f}$ values and both the $S=\bar{u}u$ and $\bar{d}d$ contributions)
are illustrated in Fig.~\ref{fig:nucconna} for ensembles IV and VIII
with $m_\pi=295$~MeV and $t_{\rm f}=7a,9a,11a,13a,15a,17a\sim 0.5-1.2$~fm
and $m_\pi=150$~MeV and $t_{\rm f}=9a,12a,15a\sim 0.6-1.1$~fm, respectively.
While contamination from excited states can certainly be resolved, the
last term in Eq.~(\ref{eq:threenuc}) which only depends on the sink
time does not appear to be significant for $t_{\rm f}\gtrsim 13a\sim
0.9$~fm. This can be seen in Fig.~\ref{fig:nucconna} from the
consistency between the data with $t_{\rm f}=13a,15a,17a$ for
$m_\pi=295$~MeV and $t_{\rm f}=13a,15a$ for $m_\pi=150$~MeV.
Performing fits to a single $t_{\rm f}=15a\sim 1.1$~fm~(in this case
the last term cannot be distinguished from the first term in
Eq.~(\ref{eq:threenuc})), we find consistent results for the scalar
matrix elements with the multi-$t_{\rm f}$ fit results, as demonstrated in
Fig.~\ref{fig:nucconnb}.  This gives us confidence that, for the
interpolators employed, excited state contamination can be accounted
for in the analysis of the other ensembles where three-point functions
were generated with a single sink time $t_{\rm f}\gtrsim 1.0$~fm.

The disconnected scalar matrix element was extracted from the ratio of
SS three-point and two-point functions. In contrast to the pion, the
signal deteriorates fairly rapidly for $t_{\rm f}\sim 1.5$~fm, as seen
in Fig.~\ref{fig:nucdiscon}, and only the smallest two values of the
current insertion time $t$ are useful, where $t_{\rm f}>t$. The figure
also shows that excited state contributions are small on the scale
of the statistical errors and indeed
fits employing Eq.~(\ref{eq:ratimprov3}) failed to resolve such
terms. The SL ratios were not included in the analysis as in this case
the excited state contamination was too large to be modelled by
including only the first excited state in the fit function.  For
most ensembles, constant fits were performed to the SS ratios for the
two values of $t$ simultaneously. Statistical noise is larger for
coarser lattice spacings and as the pion mass decreases. For ensemble
I~($a\sim 0.08$~fm) only $t=4a\sim 0.32$~fm provided a reasonable
signal, while for ensemble VIII, $m_\pi=150$~MeV, $t=5a$ and $11a$ are
both noisy, however, we took the conservative choice to fit to $t=11a$.

In the same way as discussed for the analysis of the pion three-point
functions in the previous section, the final results for both the
connected and disconnected matrix elements include an estimate of the
systematic uncertainty arising from the fitting procedure, obtained by
varying the fitting range.

\subsection{Renormalization}
\label{renorm}
The renormalization of the lattice scalar matrix elements in the
$N_f=2$ theory has already been discussed in detail in
Ref.~\cite{Bali:2011ks} and we only repeat the relevant relations here.
In the continuum the combination
$m_q\langle H|\bar{q}q|H\rangle$ is invariant under renormalization
group transformations. However, Wilson fermions explicitly break
chiral symmetry and this enables mixing with other quark flavours. The
renormalization factor that determines the strength of this mixing
is\footnote{In the notation of our previous work~\cite{Bali:2011ks} $r_m=
  1+\alpha_{\mathrm{Z}}$.} $r_m= Z_m^s/Z_m^{ns}$,
the ratio of the singlet~($Z_m^s$) to non-singlet~($Z_m^{ns}$) mass
renormalization factors.  This ratio can be determined
non-perturbatively from the slope of the axial Ward identity quark
mass~($\tilde{m}_q$) as a function of the vector Ward identity
mass~($m_q$), see, e.g, Ref.~\cite{Bhattacharya:2005rb}:
\begin{align}
\tilde{m}_q &=\frac{Z_m^{ns}Z_P^{ns}}{Z_A^{ns}} r_m m_q+O(am^2),\,
\end{align}
where the quark masses are defined as
\begin{align}
m_q &=
\frac{1}{2a}\left(\frac{1}{\kappa_q}-\frac{1}{\kappa_{c,\rm sea}}\right)\hspace{0.1cm}
\mathrm{and}\hspace{0.1cm} \tilde{m}_q = \frac{1}{2}\frac{\partial_t\langle
  A_4^I(t) P^\dagger(0)\rangle}{\langle P(t) P^\dagger(0)\rangle}.\label{eq:mqdefs}
\end{align}
$A_4^I$ and $P$ denote the $\mathcal{O}(a)$ improved axial-vector current and
the pseudoscalar operator, respectively, with corresponding
renormalization factors $Z_A^{ns}$ and $Z_P^{ns}$. $\kappa_{c,\rm sea}$
denotes the critical mass parameter along the isosymmetric line for
which the quark mass is zero.

\begin{figure}
\centerline{
\includegraphics[width=.5\textwidth,clip=]{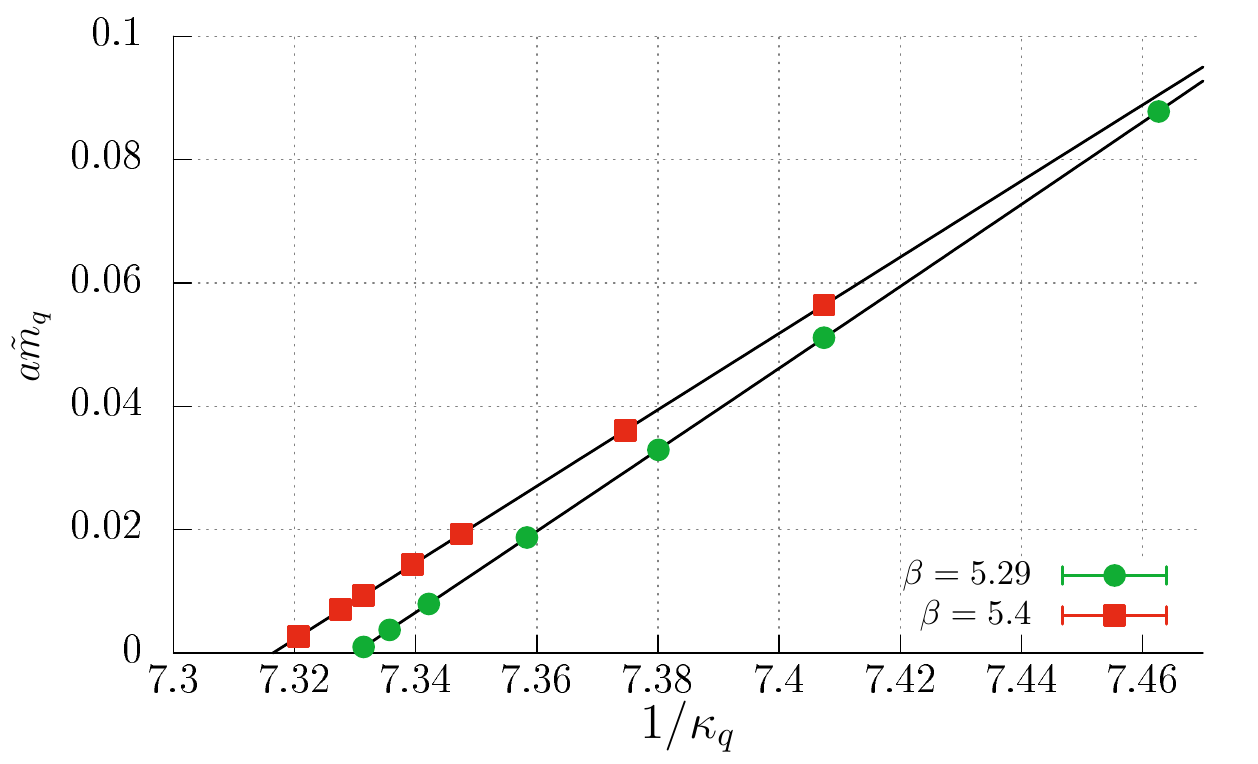}
}
\caption{The axial Ward identity masses as functions of the inverse
  of the sea quark mass parameter $1/\kappa_q$ for
  $\beta=5.29$~($a=0.07$~fm) and $\beta=5.40$~($a=0.06$~fm). A typical
  fit is shown in each case obtained using
  Eq.~(\ref{eq:pcacfitform}).}
\label{fig:fitmpcac}
\end{figure}

We employ the fit form
\begin{align}
\tilde{m}_q & = \frac{Z r_m}{2} \left(\frac{1}{\kappa_q}-\frac{1}{\kappa_{c,\rm sea}}\right)\left(1+\frac{b}{2}\left[\frac{1}{\kappa_q}-\frac{1}{\kappa_{c,\rm sea}}\right]\right),\label{eq:pcacfitform}
\end{align}
accounting for higher order contributions via a quadratic term. The
coefficient $b$ is a combination of improvement coefficients which
include, for instance, $b_g$~\cite{Bhattacharya:2005rb}, which is not
known non-perturbatively. Values for $Z=Z_m^{ns}Z_P^{ns}/Z_A^{ns}$ are
taken from Ref.~\cite{Fritzsch:2010aw}, while $r_m$, $\kappa_{c,\rm sea}$
and $b$ are extracted from fits to a range of masses for $\beta=5.29$
and $\beta=5.40$: those indicated in Table~\ref{tab:res}~(chosen from
ensembles with the largest $Lm_\pi$ for each $\kappa_q$) and at
heavier quark masses produced by QCDSF and
UKQCD~\cite{Gockeler:2006jt,pleitercom}. For $\beta=5.20$ we use $r_m$
and $\kappa_{c,\rm sea}$ as determined in
Ref.~\cite{Fritzsch:2012wq}. Figure~\ref{fig:fitmpcac} shows examples
of typical fits and Table~\ref{tab:renorm} details the results, where
the errors include systematics estimated by varying the fit range and
including and omitting the quadratic term.

\begin{table}
\caption{The critical hopping parameters $\kappa_{\mathrm{crit}}$ and
  ratio of singlet to non-singlet renormalization constants, $r_m$.
  The errors given include systematics.  For $\beta=5.20$ the values
  quoted were determined by the ALPHA
  Collaboration~\cite{Fritzsch:2012wq}.
\label{tab:renorm}}
\begin{ruledtabular}
\begin{tabular}{ccc}
$\beta$ & $\kappa_{\mathrm{c, \rm sea}}$ & $r_m$\\\hline
5.20 &0.1360546(39)&1.549(42)\\
5.29 &0.1364281(12)&1.314(20)\\
5.40 &0.1366793(11)&1.205(14)
\end{tabular}
\end{ruledtabular}
\end{table}

The renormalization pattern for the scalar matrix elements is the same
for the pion and the nucleon. We consider a general hadronic
state $|H\rangle$, with the abbreviation:
\begin{align}
\langle \bar{q}q\rangle^{\rm latt}  = &\langle H | \bar{q}q | H\rangle^{\rm latt}-\langle 0 | \bar{q}q | 0\rangle^{\rm latt} \nonumber\\
 = & \langle H | \bar{q}q| H\rangle^{\rm latt,conn} + \langle H | \bar{q}q | H\rangle^{\rm latt,dis}\nonumber\\ & -\langle 0 | \bar{q}q | 0\rangle^{\rm latt}.
\end{align}
Note that for the strange quark, the connected term is not present.
The dimension three scalar current ($\bar{q}q$ above refers to the
current, integrated over space, and is dimensionless) will receive
contributions $\propto a^{-3}\mathds{1}$, however, these cancel as we
subtract the vacuum expectation value. Flavour singlet and non-singlet
currents not only renormalize differently but are also subject to
different $\mathcal{O}(a)$ improvement terms. In general, terms of
the type $b am_q$ and $a\langle F^2\rangle$ can be added, where the
second term can only affect flavour-singlet combinations. In the
$N_f=2$ theory, the first type of term cancels from combinations
like $(m_u+m_d)\langle
\bar{u}u+\bar{d}d\rangle$~\cite{Bhattacharya:2005rb}.

Following Ref.~\cite{Bali:2011ks}, the light quark
scalar matrix elements are given by
\begin{widetext}
\begin{align}
\sigma_{u} &= \left[m_u\langle \bar{u}u\rangle\right]^{\mathrm{ren}}
= \frac{1}{r_m}\left[m_u^{\mathrm{latt}}+ \frac{(r_m-1)}{2}\left(m_u^{\mathrm{latt}}+m_d^{\mathrm{latt}}\right)\right] \left[\langle \bar{u}u\rangle^{\mathrm{latt}} +\frac{(r_m-1)}{2} \langle \bar{u}u- \bar{d}d\rangle^{\mathrm{latt}}\right],\label{eq:usigma}\\
\sigma_{d} &= \left[m_d\langle \bar{d}d\rangle\right]^{\mathrm{ren}}
= \frac{1}{r_m}\left[m_d^{\mathrm{latt}}+ \frac{(r_m-1)}{2}\left(m_u^{\mathrm{latt}}+m_d^{\mathrm{latt}}\right)\right] \left[\langle \bar{d}d\rangle^{\mathrm{latt}} -\frac{(r_m-1)}{2} \langle \bar{u}u-\bar{d}d\rangle^{\mathrm{latt}}\right],
\end{align}
\end{widetext}
where $m_q^{\rm latt}$ is the $m_q$ of Eq.~(\ref{eq:mqdefs}).
Summing the two sigma terms in the isospin limit, the renormalization factors drop out,
\begin{align}
\sigma_{u}+\sigma_{d} &= \frac{m_u^{\mathrm{latt}}+m_d^{\mathrm{latt}}}{2} \langle \bar{u}u+\bar{d}d\rangle^{\mathrm{latt}},
\end{align}
as expected for the $N_f=2$ theory.  Another combination of interest which does not
require renormalization is the isospin asymmetry ratio,
\begin{align}
z &= \left[\frac{\langle \bar{u}u-\bar{s}s\rangle}{\langle  \bar{d}d-\bar{s}s\rangle}\right]^{\mathrm{ren}} = \frac{\langle \bar{u}u-\bar{s}s\rangle^{\rm latt}}{\langle  \bar{d}d-\bar{s}s\rangle^{\rm latt}}.
\end{align}
The non-singlet sigma term,
\begin{align}
\sigma_0 & = \left[\frac{m_u+m_d}{2}
\langle \bar{u}u+\bar{d}d-2\bar{s}s\rangle\right]^{\mathrm{ren}}\nonumber\\
& = 
r_m \frac{m_u^{\mathrm{latt}}+m_d^{\mathrm{latt}}}{2}
\langle \bar{u}u+\bar{d}d-2\bar{s}s\rangle^{\mathrm{latt}}\,,\label{eq:nonsinglet}
\end{align}
is only multiplicatively renormalized.

For the (quenched) strangeness matrix element we find
\begin{widetext}
\begin{align}
\label{eq:ren4}
\sigma_s &= \left[m_s\langle \bar{s}s\rangle\right]^{\mathrm{ren}}
=\left[m_s^{\mathrm{latt}}+\frac{r_m-1}{2}\left(m_u^{\mathrm{latt}}
+m_d^{\mathrm{latt}}\right)\right]
 \left(
\langle \bar{s}s\rangle^{\mathrm{latt}}-
\frac{(r_m-1)}{2r_m}\langle \bar{u}u+\bar{d}d\rangle^{\mathrm{latt}}
\right)\,.
\end{align}

Large cancellations occur for this quantity at moderate lattice
spacings~($a\gtrsim 0.06$~fm). This can only be mitigated by moving to
finer lattices where $r_m$ is closer to $1$.

Finally, we give the expressions for the ratio of the sea to total light
quark matrix elements\footnote{Here we use $\langle \bar{u}u+\bar{d}d\rangle^{\rm dis}=\langle \bar{u}u+\bar{d}d\rangle-\langle \bar{u}u+\bar{d}d\rangle^{\rm conn}$. The full and connected matrix elements renormalize with $Z_m^{s}$ and $Z_m^{ns}$, respectively.},
\begin{align}
\label{eq:rren}
r^{\mathrm{sea}} &= \left[\frac{\langle \bar{u}u+\bar{d}d\rangle^{\mathrm{dis}}}{\langle \bar{u}u+\bar{d}d\rangle}\right]^{\mathrm{ren}}
= r_m \left(\frac{\langle \bar{u}u+\bar{d}d\rangle^{\mathrm{latt,dis}}}{\langle \bar{u}u+\bar{d}d\rangle^{\mathrm{latt}}}-1\right)+1\,,
\end{align}
the ratio of the strange to~(light) sea contributions,
\begin{align}
a^{\mathrm{sea}} &= \left[\frac{2\langle  \bar{s}s  \rangle}
{\langle  \bar{u}u+\bar{d}d\rangle^{\mathrm{dis}}}\right]^{\mathrm{ren}}
=\frac{2 r_m\langle \bar{s}s\rangle^{\mathrm{latt}}+
(1-r_m)\langle  \bar{u}u+\bar{d}d\rangle^{\mathrm{latt}}}
{r_m\langle  \bar{u}u+\bar{d}d \rangle^{\mathrm{latt,dis}}+
(1-r_m)\langle  \bar{u}u+\bar{d}d  \rangle^{\mathrm{latt}}}\,,
\end{align}
\end{widetext}
and the ratio $y=a^{\mathrm{sea}}/r^{\mathrm{sea}}$,
\begin{align}\label{eq:yrat}
y &=\left[\frac{2\langle \bar{s}s\rangle}{\langle \bar{u}u+\bar{d}d\rangle}\right]^{\mathrm{ren}}=
r_m\left(\frac{2\langle \bar{s}s\rangle^{\mathrm{latt}}}{\langle \bar{u}u+\bar{d}{d}\rangle^{\mathrm{latt}}}-1\right)+1\,.
\end{align}

\begin{table*}[t]
\caption{\label{tab:pion} Final results for the pion sigma terms on a
  subset of ensembles at $\beta=5.29$ with $a=0.071$~fm. The errors
  given include both the systematic and statistical uncertainty, see the text. The
  finite volume corrected pion masses~($m_\pi^\infty$) were determined
  in Ref.~\cite{Bali:2014nma} using Ref.~\cite{Colangelo:2005gd}, while
  $\sigma_\pi^\infty$ was obtained using the finite volume expressions
  for the pion mass~\cite{Gasser:1986vb,Gasser:1987zq} and the
  Feynman-Hellmann theorem, as detailed in
  Appendix~\ref{app:chiralsigma}.  }
\begin{ruledtabular}
\begin{tabular}{cccccccc}
Ensemble&  $Lm_\pi$ & $m_\pi$ [GeV]  & $m^\infty_\pi$ [GeV]& $m^\infty_\pi/2$ [GeV]  & $\sigma_\pi$~[GeV]& $\sigma^\infty_\pi$~[GeV] & $\sigma_s$~[GeV] \\
\hline
III  &  4.90 &  0.4222(13)&  0.4215(13) & 0.2108(7) & 0.2176(86)& 0.2184(86) & 0.014(11) \\
IV   &  3.42 &  0.2946(14)&  0.2895(07) & 0.1448(4) & 0.1336(41)& 0.1348(41) &-0.011(10)\\
V    &  4.19 &  0.2888(11)&  0.2895(07) & 0.1448(4) & 0.1560(75)& 0.1566(75) & 0.031(20)\\
VIII &  3.47 &  0.1497(13)&  0.1495(13) & 0.0748(7) & 0.0780(42)& 0.0782(42) & 0.006(33)\\
\end{tabular}
\end{ruledtabular}
\end{table*}

\begin{figure*}
\centerline{
\includegraphics[width=.48\textwidth,clip=]{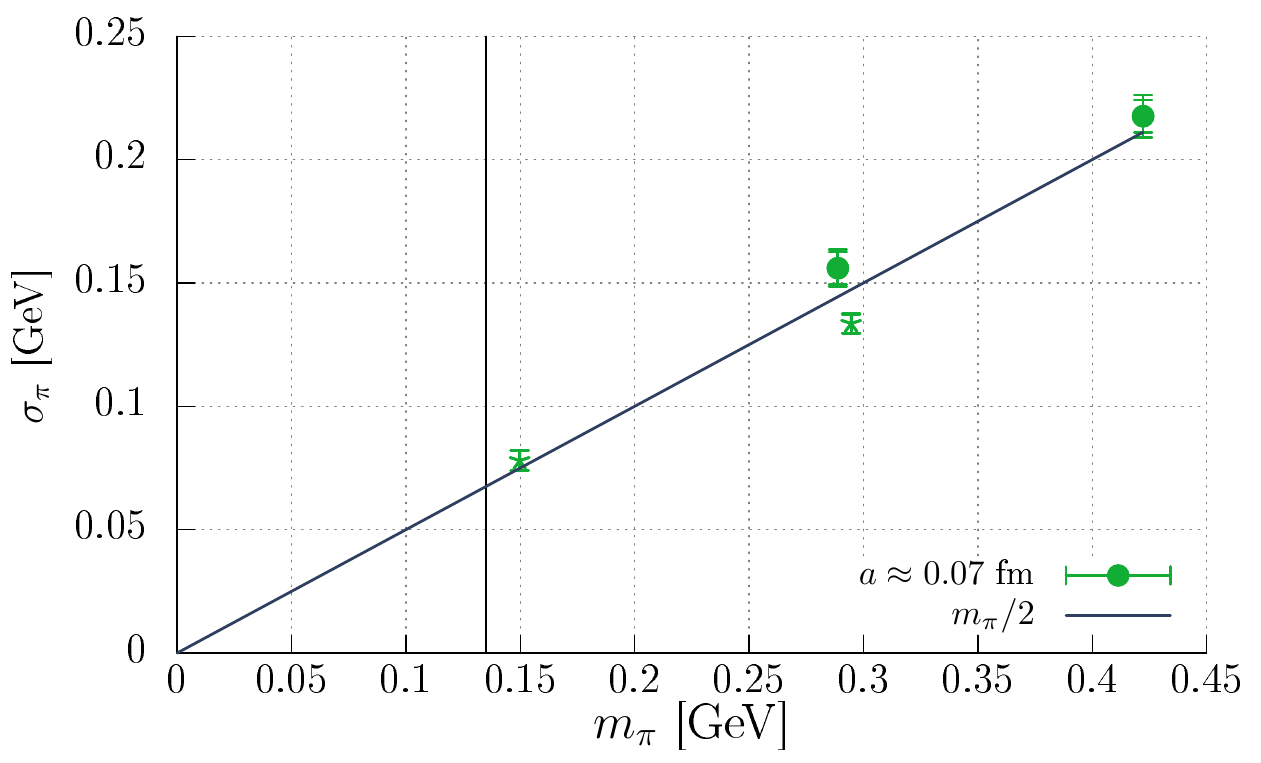}
\includegraphics[width=.48\textwidth,clip=]{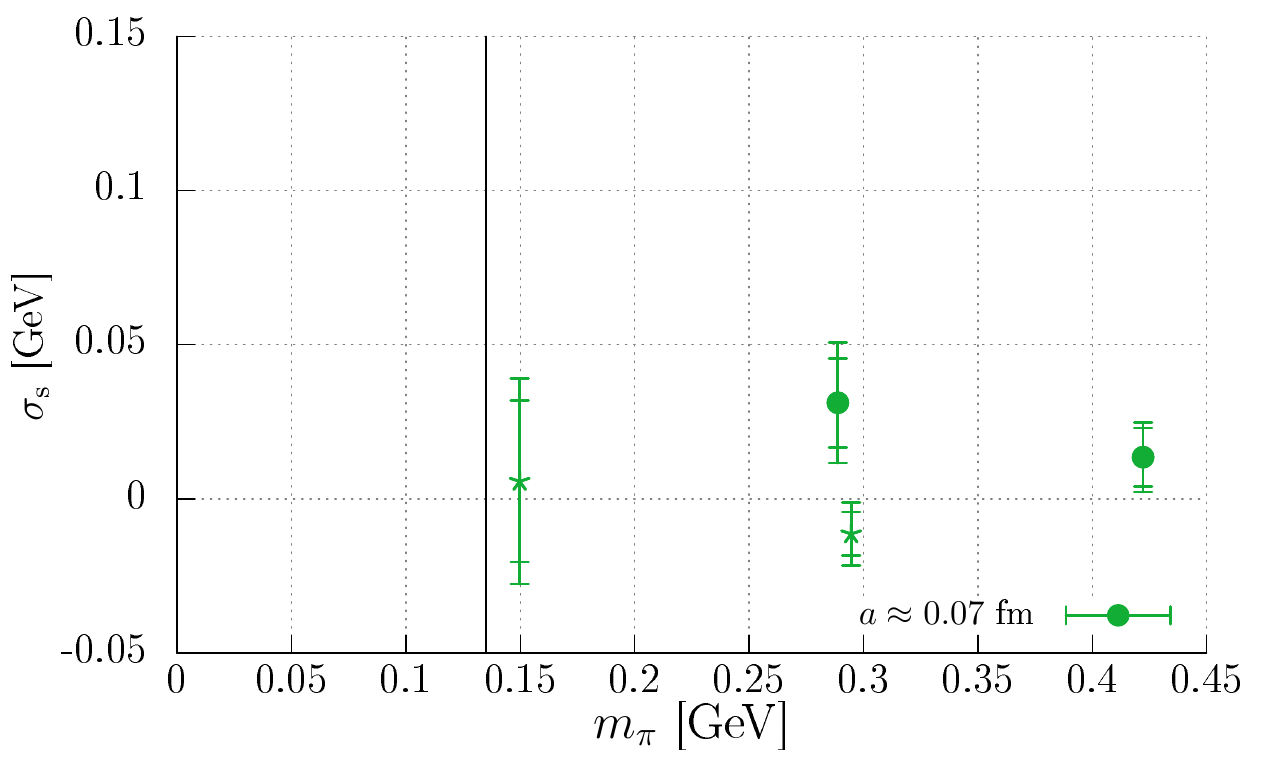}
}
\centerline{
\includegraphics[width=.48\textwidth,clip=]{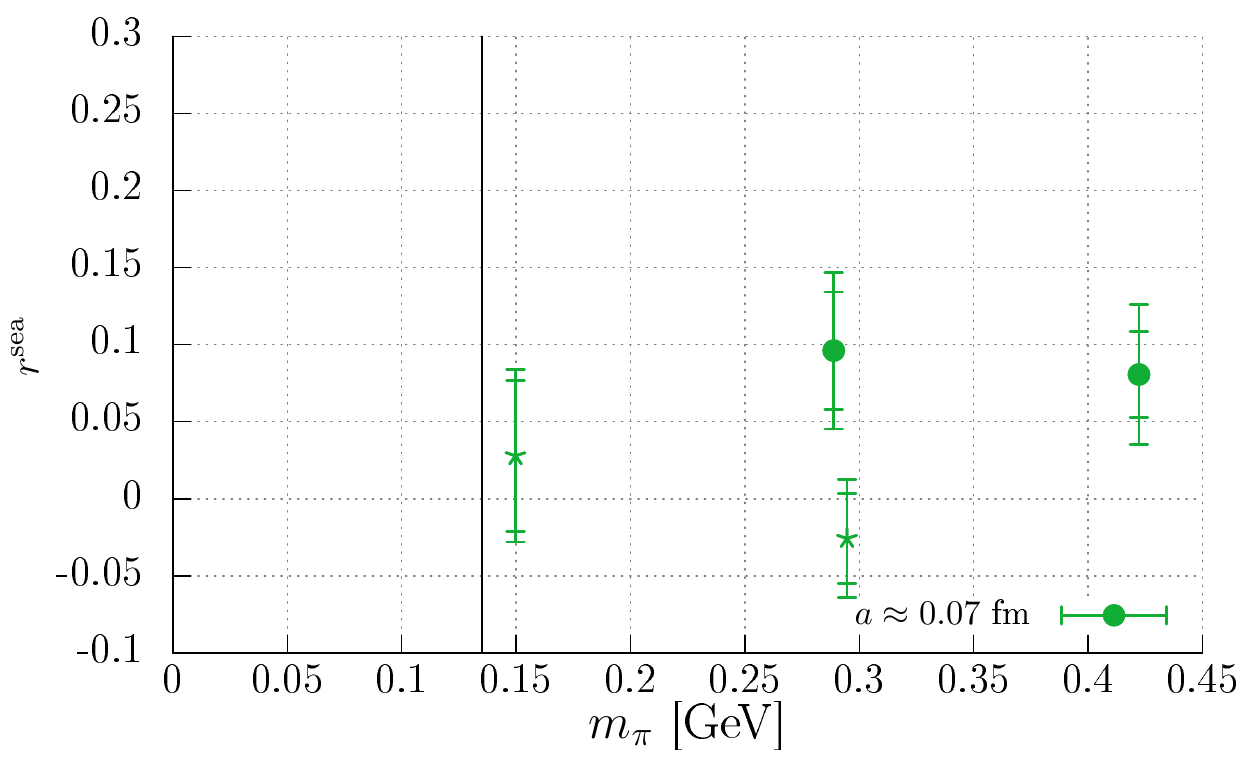}
\includegraphics[width=.48\textwidth,clip=]{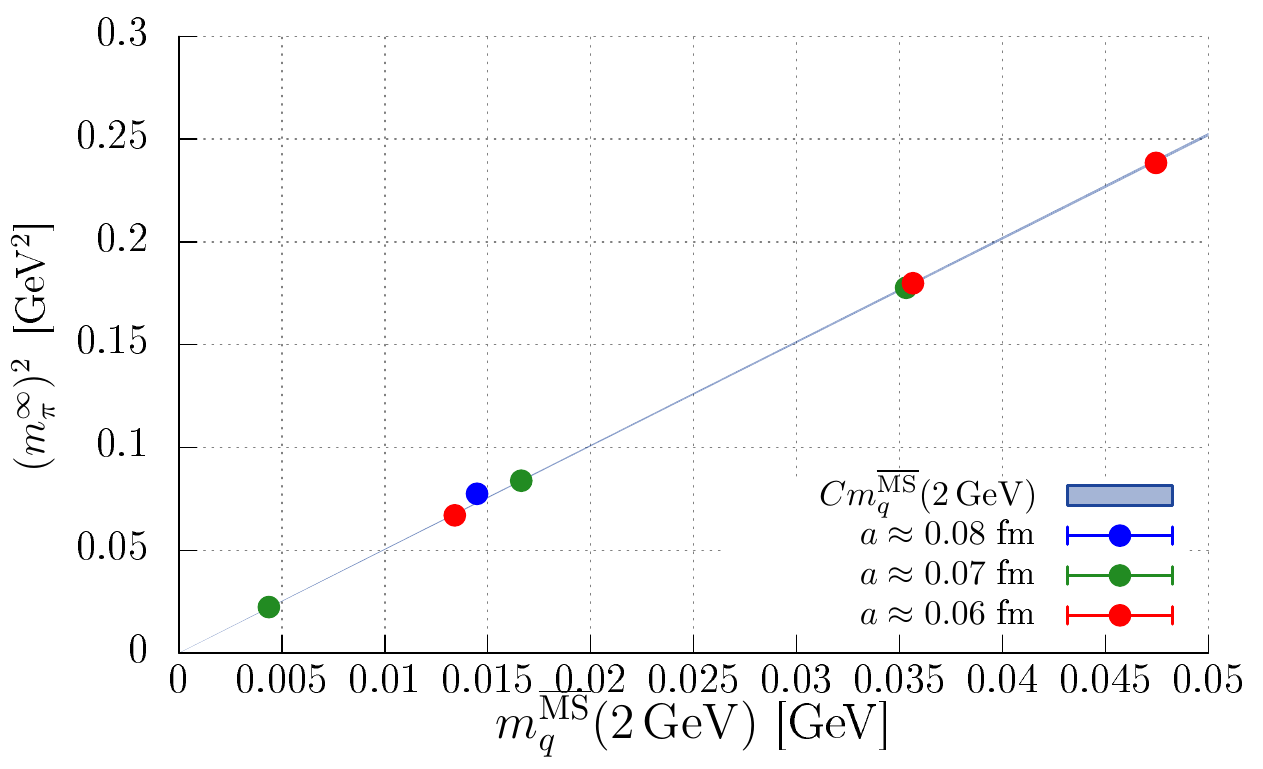}
}
\caption{Results for the finite volume pion light and strange quark sigma terms,
  $\sigma_\pi=\sigma_u+\sigma_d$~(top left) and $\sigma_s$~(top
  right), respectively, and the ratio of the sea to total light quark
  matrix elements, $r^{\mathrm{sea}}$~(bottom left), as a function of
  the pion mass. Both the statistical uncertainty~(inner error bar)
  and the total systematic and statistical uncertainty~(outer error
  bar) are shown. For $\sigma_\pi$, the results are compared to
  $m_\pi/2$ as determined in the simulation~(indicated by a line,
  starting from zero which goes through the central values of
  $m_\pi/2$ for each ensemble). The symbols and colours for the data
  points correspond to those used in Fig.~\ref{fig:ens} for the
  different ensembles. The vertical lines indicate the physical point
  in the isospin limit, $m^{\rm phys}_\pi=135$~MeV. Also
  displayed~(bottom right) is the infinite volume pion mass squared as
  a function of the renormalized quark mass obtained from the axial Ward
  identity, $m_q^{\overline{\rm MS}}(2\,\mathrm{GeV})=Z_A(1+am_q(b_A-b_P))
  \tilde{m}_q/Z_P^{\overline{\rm MS}}(2\,\mathrm{GeV})$ for all
  ensembles, see Table~\ref{tab:res} for the values of $a\tilde{m}$
  and Ref.~\cite{Bali:2014nma} for $b_A-b_P$, $Z_A$ and
  $Z_P^{\overline{\rm MS}}(2\,\mathrm{GeV})$. The shaded region shows
  a fit of the form $C m_q^{\overline{\rm MS}}(2\,\mathrm{GeV})$ to all data points. }
\label{fig:pionsigma}
\end{figure*}

We remark that all the quantities in Eqs.~(\ref{eq:usigma})
to~(\ref{eq:yrat}) do not depend on a renormalization scale.
Considering discretisation effects, only $z$ is
automatically $\mathcal{O}(a)$ improved while all other observables
are subject to $\mathcal{O}(a)$ lattice artifacts.  However, not all
$\mathcal{O}(a)$ terms are likely to be large, for example, the
$a\langle F^2\rangle$ term does not contribute to $\sigma_0$ and if
there is SU(3) flavour symmetry in the sea then for $a^{\rm sea}\approx 1$ 
it cancels.

\section{Pion sigma terms}
\label{pionsig}

\begin{table*}
\caption{\label{tab:resnuc} Final results for the pion-nucleon sigma
  term, $\sigma_{\pi N}$, the individual quark sigma terms of the proton,
  $\sigma_{q=u,d,s}$, the $y$ ratio and the isospin asymmetry ratio
  $z$. The errors include an estimate of both the systematic and
  statistical uncertainty, see the text. The values for the finite volume
  corrected pion mass, $m_\pi^\infty$, were determined as in
  Ref.~\cite{Bali:2014nma}, using Ref.~\cite{Colangelo:2005gd}. For
  $\sigma_{\pi N}^\infty$, we used Ref.~\cite{AliKhan:2003cu}, together
  with the Feynman-Hellmann theorem, as discussed in
  Appendix~\ref{app:chiralsigma}. }
\begin{ruledtabular}
\begin{tabular}{ccccrrrrrrr}
Ensemble&  $m_\pi$ [GeV] &  $Lm_\pi$  &   $m^\infty_\pi$ [GeV]& $\sigma_{\pi N}$~[GeV]& $\sigma^\infty_{\pi N}$~[GeV] & $\sigma_u$~[GeV] & $\sigma_d$~[GeV] & $\sigma_s$~[GeV] & $y$ & $z$\\\hline
I    & 0.2795(18)& 3.69& 0.2783(18)& 0.108(07) & 0.115(07)& 0.0614(36) & 0.0462(33) & 0.025(16) & 0.070(046) & 1.357(52)\\\hline
II   & 0.4264(20)& 3.71& 0.4215(13)& 0.191(14) & 0.214(18)& 0.1086(81) & 0.0829(65) & 0.030(12) & 0.118(046) & 1.361(60)\\
III  & 0.4222(13)& 4.90& 0.4215(13)& 0.230(11) & 0.238(12)& 0.1299(64) & 0.1000(55) & 0.055(14) & 0.178(039) & 1.375(27)\\
IV   & 0.2946(14)& 3.42& 0.2895(07)& 0.125(14) & 0.135(15)& 0.0709(81) & 0.0542(61) & 0.041(18) & 0.108(050) & 1.353(51)\\
V    & 0.2888(11)& 4.19& 0.2895(07)& 0.132(10) & 0.137(11)& 0.0739(54) & 0.0583(51) & 0.048(23) & 0.119(051) & 1.312(31)\\
VI   & 0.2895(07)& 6.71& 0.2895(07)& 0.108(11) & 0.108(11)& 0.0620(56) & 0.0459(53) & -0.009(28) & -0.028(089) & 1.338(29)\\
VIII & 0.1497(13)& 3.47& 0.1495(13)& 0.042(08) & 0.043(08)& 0.0232(42) & 0.0182(35) & -0.036(64) & -0.068(132) &1.258(81)\\\hline
IX   & 0.4897(17)& 4.81& 0.4883(17)&0.275(16) & 0.288(17)&0.1605(83) & 0.1148(81) & 0.053(15) & 0.192(046) & 1.518(23)\\
X    & 0.4262(20)& 4.18& 0.4241(20)&0.226(15) & 0.241(17)&0.1294(86) & 0.0967(67) & 0.062(15) & 0.199(042) & 1.442(39)\\
XI   & 0.2595(09)& 3.82& 0.2588(09)&0.107(07) & 0.112(07)&0.0595(35) & 0.0471(32) & 0.075(18) & 0.191(038) & 1.335(35)\\
\end{tabular}
\end{ruledtabular}
\end{table*}

Our final results for the pion sigma terms for four ensembles at
$a=0.071$~fm are presented in Table~\ref{tab:pion} and
Fig.~\ref{fig:pionsigma}.  The central values are obtained by taking
the average of the maximum and minimum of the sigma terms that result
from independently varying the fit ranges of the connected and
disconnected contributions and, where relevant, the renormalization
factor $r_m\pm\delta r_m$, where $\delta r_m$ is the error given in
Table~\ref{tab:renorm}. The systematic error is then half of the
difference of the maximum and minimum values. This is added in
quadrature to the statistical error arising from typical fits to the
connected and disconnected terms~(computed by combining the jackknife
samples of the individual contributions). For the pion sigma term in
the infinite volume limit, $\sigma^\infty_\pi$, a further systematic arising
from the finite volume correction is added in quadrature corresponding
to half the size of the correction applied.

As discussed in Section~\ref{intro}, we expect $\sigma_\pi=m_\pi/2$
for small pion masses. Fig.~\ref{fig:pionsigma} shows this holds up to
$m_\pi$ of at least $420$~MeV.  The 2.62 $\sigma$ increase going from
$Lm_\pi=3.42$~(ensemble IV) to $4.19$~(ensemble V) for $m_\pi\sim
290$~MeV suggests finite volume effects may be an issue.  Chiral
perturbation theory~(ChPT) provides a framework for evaluating these
effects, as detailed in Appendix~\ref{app:chiralsigma}.  The sigma
term increases in the infinite volume limit, however, the corrections
turn out to be very small, well below the level of statistical
significance. The difference at $m_\pi\sim 290$~MeV is only reduced to
2.55$\sigma$ for $\sigma_\pi^\infty$, see Table~\ref{tab:pion}. If the
next-to-leading order~(NLO) finite volume
formula~(Eq.~(\ref{eq:infvolpi})) is valid down to $Lm_\pi=3.4$ then
the difference in the sigma terms can be ascribed to statistical
variation.  It is worth noting that without the use of our method for
reducing excited state contamination to the pion scalar matrix
element~(see Section~\ref{sec:pionfits} and
Appendix~\ref{app:spectral}) the agreement with the GMOR expectation
would not have been found. In particular, for the near physical point
one may obtain\footnote{This value is obtained by estimating the
  connected scalar matrix element, $g_S^{\rm latt,conn}\sim 9.6$, see
  the unimproved results in the bottom right plot of
  Fig.~\ref{fig:pionconn}, and $g_S^{\rm latt,dis}\sim 4$ for the
  disconnected part.} $\sigma_\pi\sim 97$~MeV.

\begin{figure*}
\centerline{
\includegraphics[width=.48\textwidth,clip=]{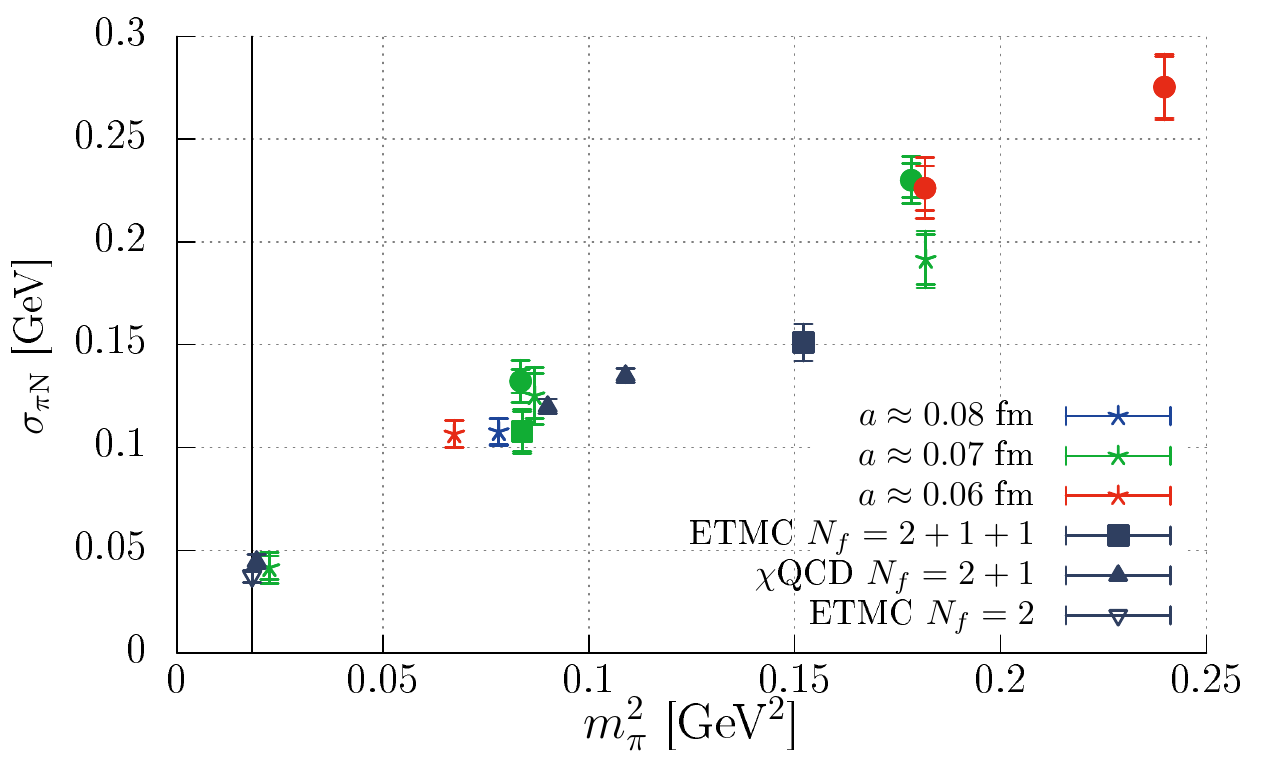}
\includegraphics[width=.48\textwidth,clip=]{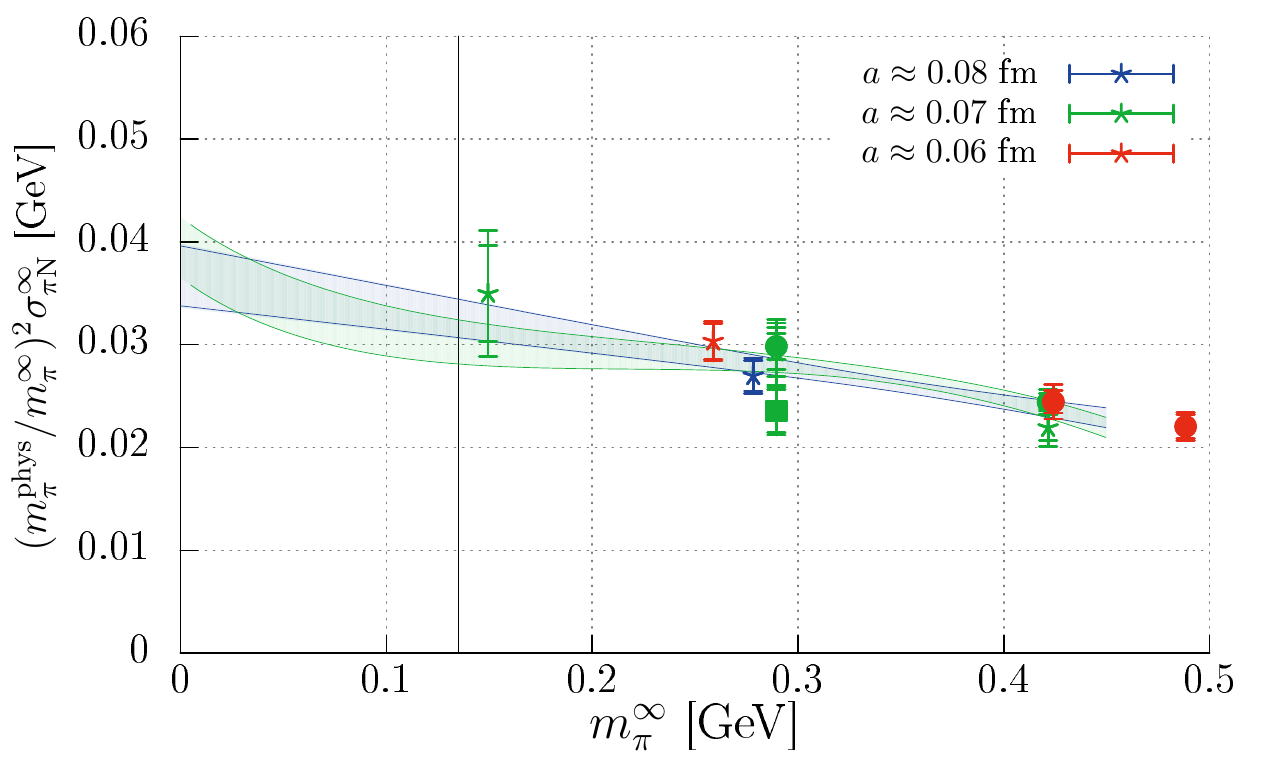}
}
\caption{The pion-nucleon sigma term as a function of the pion
  mass. The symbols and colours of the data points correspond to those
  used in Fig.~\ref{fig:ens}. Both the statistical uncertainty~(inner
  error bar) and the total statistical and systematic uncertainty~(outer
  error bar) are displayed. The vertical lines indicate the physical
  point in the isospin limit, $m^{\rm phys}_\pi=135$~MeV. (Left)
  Recent results from ETMC~\cite{Dinter:2012tt,Abdel-Rehim:2016won}
  and $\chi$QCD~\cite{Yang:2015uis} vs.\ $m_{\pi}^2$ are compared to
  our finite volume values. Note that only the unitary
  values with purely statistical errors have been included from
  Ref.~\cite{Yang:2015uis}. (Right) The finite volume corrected
  results in the combination $(m^{\rm
    phys}_\pi/m_\pi^\infty)^2\sigma^\infty_{\pi N}$ which has a linear
  dependence on $m_\pi^\infty$ at leading
  order~(Eq.~(\ref{eq:nucmpidep})). The blue shaded region indicates a
  fit of the form $a+bm_\pi^\infty$, with $\pm 1\sigma$ error band for
  $m_\pi\lesssim 420$~MeV. Similarly, the green shaded region shows a
  fit using $a+bm_\pi^\infty+c(m_\pi^\infty)^2\ln(m^\infty_\pi/\lambda)$, where the
  linear coefficient is fixed using Eq.~(\ref{eq:nucmpidep}) and our
  determination of
  $g_A/F_\pi=13.88(29)$~GeV$^{-1}$~\cite{Bali:2014nma} and $\lambda=1$~GeV. }
\label{fig:nucsigma}
\end{figure*}

The observed behaviour of $\sigma_\pi$ suggests the GMOR relation is
valid over the same range of pion masses. This is demonstrated in
the bottom right panel of Fig.~\ref{fig:pionsigma} where
$({m_\pi^\infty})^2$ is shown as a function of the renormalized quark
mass for all ensembles. A fit to the simple
form $(m_\pi^\infty)^2= C m^{\overline{\rm MS}}_q(2\,\mathrm{GeV})$
for $m_\pi^\infty \lesssim 500$~MeV gives a $\chi^2/d.o.f.=1.9$ and a
slope $C=-2\Sigma/F^2=5.04(1)(14)$~GeV, where $\Sigma$ is the chiral
condensate and $F$ is the pion decay constant in the chiral limit. The
second error is due to the uncertainty in the non-perturbative
renormalization factors, given in Ref.~\cite{Bali:2014nma}. Additional
uncertainties, such as discretisation effects and, clearly, higher
orders in the quark mass expansion have not been estimated~(although
given that $\chi^2/d.o.f.=1.9$, these terms appear to be small). This
slope compares favourably with $2\Sigma/F^2=-5.3(5)$ obtained using
FLAG estimates~\cite{Aoki:2013ldr}
$-\Sigma=(0.269(8)\,\mathrm{GeV})^3$ and $F_\pi/F=1.0744(67)$ for
the $N_f=2$ theory with $F_\pi=92.2$~MeV.

Recalling the decomposition of the mass of a hadron in
Eq.~(\ref{eq:massd}), the $u$ and $d$ quark sigma term accounts for
half the mass of the pion, i.e. approximately $68$~MeV at the physical
point. From the ratio $r^{\rm sea}$ in Fig.~\ref{fig:pionsigma}, we
find less than $15\%$ of this is due to~(light) sea quarks. While the
disconnected terms are significant, approximately $30\%$ in size of
the connected terms, their contribution is reduced under
renormalization~(Eq.~(\ref{eq:rren})) since $r_m=Z^s_m/Z^{ns}_m>1$.
The strange quark contribution to the pion mass is likely to be small,
however, again due to cancellations under renormalization, the overall
uncertainties are large and we find $\sigma_s < 50$~MeV. Within
errors, $\sigma_s$ is also consistent with zero.

\section{Nucleon $\sigma$ terms}
\label{nucleonsig}
Starting with the pion-nucleon sigma term, our final results on all
ensembles are given in Table~\ref{tab:resnuc} and displayed as a
function of $m_\pi^2$ in Fig.~\ref{fig:nucsigma}~(left). The combined
systematic and statistical errors are calculated as described in the
previous section for the pion. The sigma term tends to zero as
expected as the pion mass is reduced with no significant dependence on
the lattice spacing, but some variation with the volume at heavier
$m_\pi$. Reasonable agreement is seen with other recent direct
determinations from ETMC~\cite{Dinter:2012tt,Abdel-Rehim:2016won} and
$\chi$QCD~\cite{Yang:2015uis}, in particular, close to the physical
point. These other~(near) physical point simulations were performed at
coarser lattice spacings, $a\sim 0.09$~fm and $0.11$~fm, respectively,
and in the case of the ETMC on smaller volumes in terms of
$Lm_\pi=2.97$~\cite{Abdel-Rehim:2015pwa} and, for the $\chi$QCD study,
much lower statistics.  We remark that $\mathcal{O}(a)$ discretisation
errors arise for all fermion actions due to mixing with $aF^2$ and, so
far, these effects have not been removed.

\begin{table*}
\caption{\label{tab:phys} Our determinations of the sigma terms, the
  ratios $y$ and $z$ and the quark mass fractions,
  $f_{T_{q=u,s,d,c,b,t}}$ for the proton, at the physical point. The
  errors encompass all systematics, see the text. Note that
  $\sigma_{\pi N}$, $\sigma_0$, $\sigma_{q=u,d}$,
  $(\sigma_u-\sigma_d)/(\sigma_u+\sigma_d)$, $f_{T_{q=u,d}}$ and $z$
  are obtained using the $m_\pi=150$~MeV results, while $\sigma_s$,
  $f_{T_s}$ and $y$ are derived from fits in the range $m_\pi \lesssim
  420$~MeV.  Finite volume corrections have been applied to the light
  quark sigma terms, see the text. In order to extract $f_{T_{c,b,t}}$
  we used $\sum_{q=u,d,s} f_{T_q}$ together with the perturbative
  relations in Refs.~\cite{Chetyrkin:1997un,Hill:2014yxa}, see the
  text.  }
\begin{ruledtabular}
\begin{tabular}{ccccccc}
$\sigma_{\pi N}$~(MeV) & $\sigma_0$~(MeV) & $\sigma_u$~(MeV) & $\sigma_d$~(MeV) & $\sigma_s$~(MeV)  & $y$ & $z$ \\
\hline
35.0(6.1) & 37.1(7.3) & 19.6(3.4) & 15.4(3.5)  & 34.7(12.2) &  0.104(51) & 1.258(81)\\\hline
$\frac{\sigma_u-\sigma_d}{\sigma_u+\sigma_d}$&$f_{T_u}$& $f_{T_d}$& $f_{T_s}$ & $f_{T_c}$& $f_{T_b}$& $f_{T_t}$\\\hline
0.12(4) & 0.021(4) & 0.016(4)   & 0.037(13)   & 0.075(4) & 0.072(2)  & 0.070(1)\\
\end{tabular}
\end{ruledtabular}
\end{table*}

The leading pion mass dependence of $\sigma_{\pi N}$ is provided by
the application of the Feynman-Hellmann theorem to the NLO baryon 
ChPT expansion of the nucleon
mass~\cite{Steininger:1998ya,Becher:1999he},
\begin{align}
m_N & = m_N^0 -4c_1m_\pi^2-\frac{3 (g^0_A)^2 m_\pi^3}{32\pi F^2}+O\left(m_\pi^4\ln \left(\frac{m_\pi}{\lambda}\right)\right),
\end{align}
which to this order contains the low energy constant, $c_1$, the
renormalization scale $\lambda \sim m_N$ and the chiral limit
nucleon mass, $m_N^0$, axial charge, $g^0_A \sim 1.22$ and pion decay
constant, $F\sim 86$~MeV. At this order $g_A^0/F$ can be replaced
by $g_A/F_\pi$. From $\sigma_{\pi N}=m_\pi^2 \partial m_N/\partial m_\pi^2$ one finds,
\begin{align}
\sigma_{\pi N} & = m_\pi^2\left[-4c_1-\frac{9 (g^0_A)^2 m_\pi}{64\pi
    F^2}+O\left(m_\pi^2\ln
  \left(\frac{m_\pi}{\lambda}\right)\right)\right].
\label{eq:nucmpidep1}
\end{align}
We find it more meaningful to show in Fig.~\ref{fig:nucsigma}~(right)
the combination
\begin{align}
&\frac{\sigma_{\pi N}}{m_\pi^2}(m_\pi^{\rm phys})^2  = \nonumber\\
&(m_\pi^{\rm
  phys})^2\left[-4c_1-\frac{9 (g^0_A)^2 m_\pi}{64\pi
    F^2}+O\left(m_\pi^2\ln
  \left(\frac{m_\pi}{\lambda}\right)\right)\right],
\label{eq:nucmpidep}
\end{align}
which has a milder dependence on the pion mass but also tends to the
physical value as $m_\pi\to m_\pi^{\rm phys}$, where we take $m^{\rm
  phys}_\pi=135$~MeV in the electrically neutral isospin limit. The finite
volume corrections to the sigma term can be derived in a similar way
starting from the corresponding ChPT expressions for the nucleon mass,
see Appendix~\ref{app:chiralsigma}. The size of the corrections, as
shown in Table~\ref{tab:resnuc}, corresponds to 1--2 standard deviations
for the larger pion
mass ensembles~($m_\pi\gtrsim 420$~MeV), becoming much smaller as
$m_\pi$ approaches the physical point. The shift is always to larger
values of the sigma term for $L\to\infty$.  The biggest effect is at
$m_\pi\approx 420$~MeV between ensembles II~($Lm_\pi=3.71$) and
III~($Lm_\pi=4.90$). The difference in $\sigma_{\pi N}$ for these two
ensembles is $39(18)$~MeV, which is reduced to $24(22)$~MeV for
the infinite volume value $\sigma^\infty_{\pi N}$. At $m_\pi\approx 290$~MeV
the results from
the smaller two ensembles, IV~($Lm_\pi=3.42$) and V~($Lm_\pi=4.19$)
become slightly more coincident after finite volume corrections are
applied, while between ensembles V and VI~($Lm_\pi=6.71$) this is less
so.  However, all differences are within the expected range for
statistical variations.

\begin{figure*}
\centerline{
\includegraphics[width=.48\textwidth,clip=]{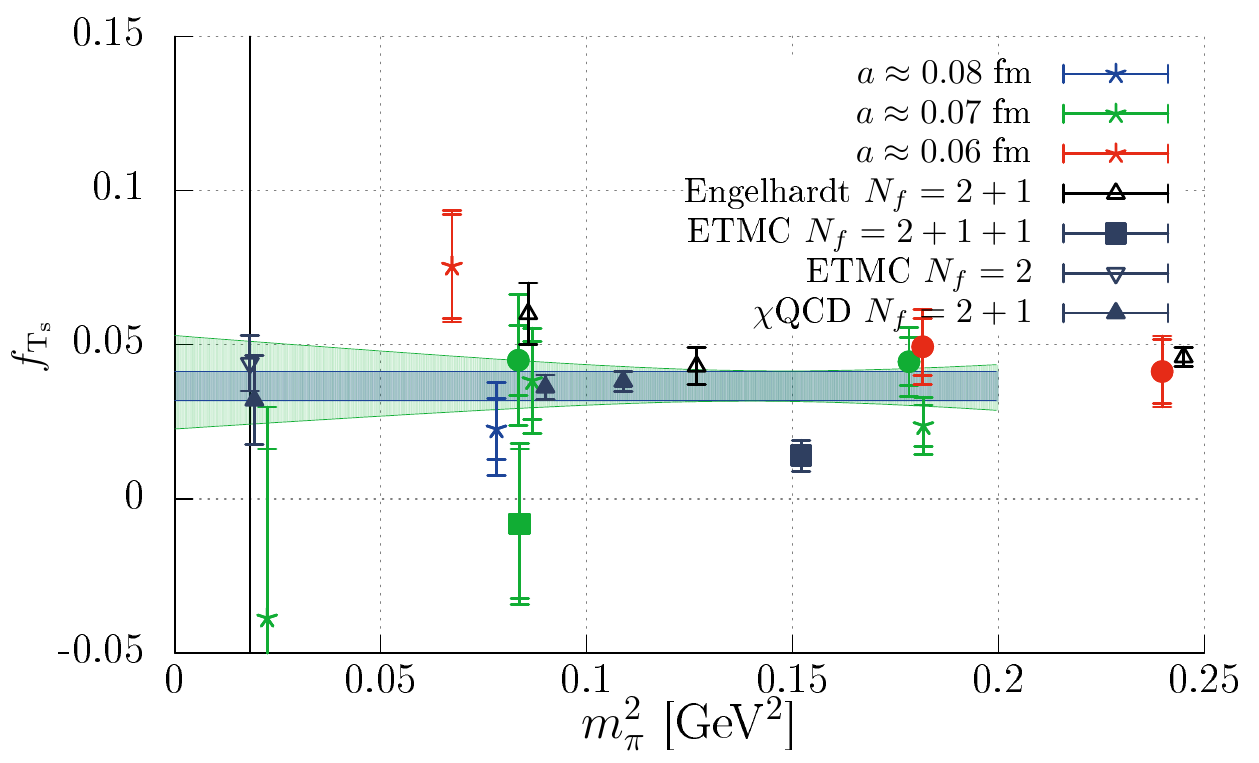}
\includegraphics[width=.48\textwidth,clip=]{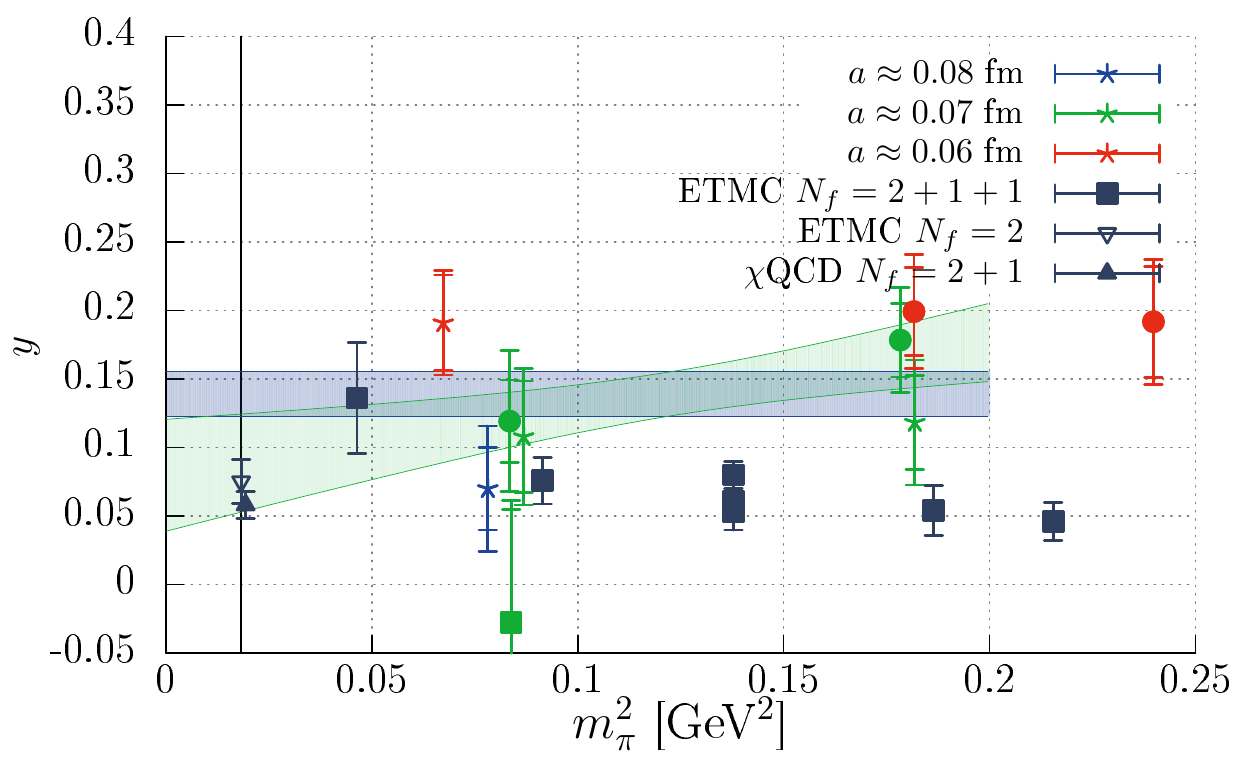}
}
\caption{The strange quark mass fraction $f_{T_s}$~(left) and $y$
  ratio~(right) as a function of $m_\pi^2$. Recent determinations by
  ETMC~\cite{Dinter:2012tt,Alexandrou:2013nda,Abdel-Rehim:2016won},
  Engelhardt~\cite{Engelhardt:2012gd} and $\chi$QCD~\cite{Yang:2015uis}
  (unitary values only) are also shown.  The results are displayed
  as in Fig.~\ref{fig:nucsigma}. For both quantities, the blue~(green)
  shaded regions indicate constant~(constant plus linear in $m_\pi^2$) fits with $\pm
  1\sigma$ error band for $m_\pi\lesssim 420$~MeV.}
\label{fig:fts}
\end{figure*}

To test whether our statistically more precise results at
$m_\pi\gtrsim 260$~MeV are consistent with the near physical point
value, we perform a phenomenological fit to $\frac{\sigma^\infty_{\pi
    N}}{(m^\infty_\pi)^2}(m_\pi^{\rm phys})^2$ based on
Eq.~(\ref{eq:nucmpidep}) of the form (a) $a-bm_\pi^\infty$, with $a$
and $b$ determined from the fit and (b)
$a-bm_\pi^\infty+c(m_\pi^\infty)^2\ln(m^\infty_\pi/\lambda)$, setting
$\lambda=1$~GeV and with $b$ fixed using
$g_A/F_\pi=13.88(29)$~GeV$^{-1}$~\cite{Bali:2014nma}. The fit range is
the same throughout, $m_\pi\lesssim 420$~MeV~(including the $150$~MeV
point), which gives us roughly three pion mass values, see
Fig.~\ref{fig:nucsigma}~(right). Higher orders in the expansion are
needed to include the $m_\pi\sim 500$~MeV data point.
Figure~\ref{fig:nucsigma}~(right) shows both fits give consistent
results at the physical point, only slightly below the central value
for $m_\pi\sim 150$~MeV, with $a=0.037(3)$~GeV,
$b=0.031(9)$ and $\chi^2/d.o.f.=1.0$ for fit (a) and
$a=0.039(3)$~GeV, $c=-0.33(2)$~GeV$^{-1}$ and $\chi^2/d.o.f.=1.1$ for
fit (b). The slope from fit (a) is significantly smaller than the ChPT
expectation of $ 9 g_A^2 (m_\pi^{\rm phys})^2 /(64\pi F_\pi^2)\sim
0.16$. We comment more on the application of ChPT to
$\sigma_{\pi N}$ and $m_N$ in Section~\ref{comparison}.  The spread in
the results at $m_\pi\sim 290$~MeV due to volume and lattice spacing
dependence, and similarly at $m_\pi\sim 420$~MeV is less than the
total uncertainty of the near physical point result. This observation,
together with the insignificant remaining extrapolation, motivates us
to quote $\frac{\sigma^\infty_{\pi N}}{(m^\infty_\pi)^2}(m_\pi^{\rm
  phys})^2$ for ensemble VIII, given in Table~\ref{tab:phys}, as our
final, more conservative, result at the physical point including all
systematics.

One can also extract the individual light quark sigma terms,
$\sigma_{q=u,d}$ and the non-singlet combination
$\sigma_0$~(Eq.~(\ref{eq:nonsinglet})) for the proton. Note that in
the isospin symmetric limit that we use for the neutron:
$\sigma_u^{n}=\sigma_d^p$, $\sigma_d^{n}=\sigma_u^p$. Corrections to
this limit are discussed for instance in
Refs.~\cite{Crivellin:2013ipa,Gonzalez-Alonso:2013ura}. We apply the same finite volume
corrections to $\sigma_0$ as for $\sigma_{\pi N}$, since the strange
contribution is sub-leading, while for $\sigma_q$ we correct in
proportion to the fraction $\sigma_q/\sigma_{\pi N}$. The final
results in all cases, given in Table~\ref{tab:phys}, are taken from
ensemble VIII after rescaling with $(m_\pi^{\rm phys}/m_\pi)^2$. The
quark fractions $f_{T_{q=u,d}}$ are found by dividing the light quark
sigma terms by the nucleon mass in the isospin limit\footnote{We
  remove the electromagnetic and quark mass effects for the nucleon
  using the charged hyperon splitting: $m_N=m_{\rm
    neutron}+\frac{1}{4}(m_{\Sigma^+}-m_{\Sigma^-})$.}
$m_N=938.6$~MeV. Note that the central value for $\sigma_0$ evaluated
in this way is larger than $\sigma_{\pi N}$. While the opposite should
be the case this is not significant considering the size of the error.
The wrong ordering of the central values of $\sigma_0$ and
$\sigma_{\pi N}$ is due to the fact that the central value for
$\sigma_s$ comes out negative at $m_\pi=150$~MeV, see
Fig.~\ref{fig:fts}, with a very large error. At heavier quark masses
the expected ordering is respected.

The strange quark content of the nucleon is encoded in
$\sigma_s$~($f_{T_s}$) and the $y$ ratio. The large cancellations
under renormalization, mentioned previously, mean our values are not
so precise.  Figure~\ref{fig:fts} shows that there is a fairly large spread
in our results, although this does not depend significantly on the
pion mass, lattice spacing or volume.  Due to the large uncertainty on
the near physical point ensemble we opt to extrapolate the
$m_\pi\lesssim 420$~MeV results to $m_\pi^{\rm phys}$ using (a) a fit
to a constant and (b) a fit including a constant plus linear term in
$m_\pi^2$. The central values and errors of the final results in
Table~\ref{tab:phys} are computed using the average and half of the
difference, respectively, of the maximum and minimum values at
$m_\pi^{\rm phys}$ obtained considering the $\pm 1\sigma$ error bands
of both fits.  

Apart from the ETMC~\cite{Dinter:2012tt} $N_f=2+1+1$ result, which is
somewhat low, other recent determinations of $f_{T_s}$ displayed in
Fig.~\ref{fig:fts} are in agreement with our fits, including those at
the physical point. Note that, due to the symmetry properties of the
twisted mass~(at maximal twist) and overlap actions used by ETMC and
$\chi$QCD, respectively, there is no mixing of quark flavours for the
scalar current and $\sigma_s$~($f_{T_s}$) is only multiplicatively
renormalized leading to reduced uncertainty in their results. The use of
domain wall fermions~(Engelhardt~\cite{Engelhardt:2012gd}) is
similarly advantageous.  For the $y$ ratio the
ETMC~\cite{Alexandrou:2013nda} $N_f=2+1+1$ results give $y\sim 0.05$ for
$m_\pi\gtrsim 300$~MeV. The ratio increases as the pion mass reduces
and they obtain $y=0.173(50)$ on extrapolation to $m_\pi^{\rm phys}$.
This is higher than the physical point determinations of the ETMC at
$N_f=2$~\cite{Abdel-Rehim:2016won} and $\chi$QCD for
$N_f=2+1$~\cite{Yang:2015uis}. Our results are generally higher but
given the large errors the difference is not significant.

\begin{figure}
\centerline{
\includegraphics[width=.45\textwidth,clip=]{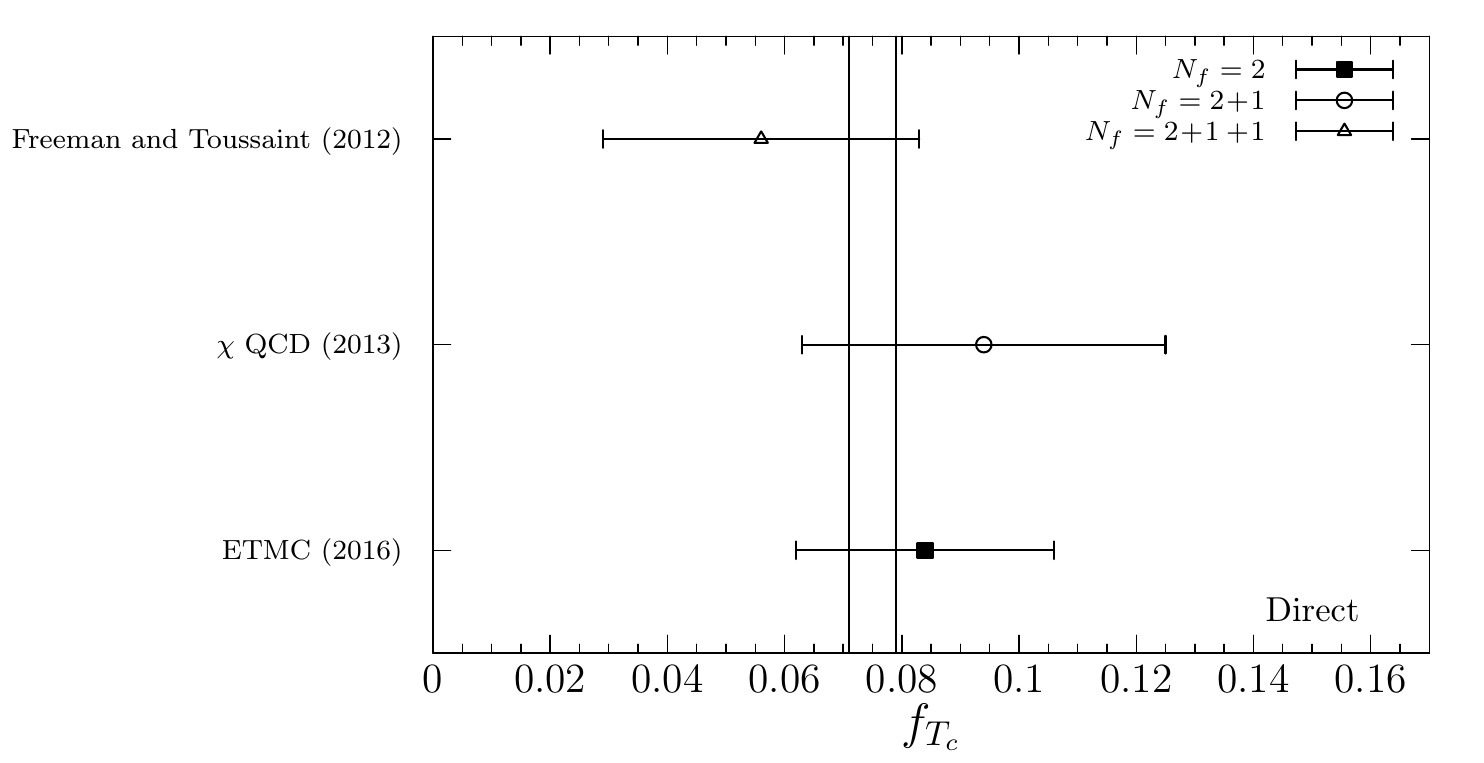}
}
\caption{Direct determinations of $f_{T_c}$ by Freeman and
  Toussaint~\cite{Freeman:2012ry}, $\chi$QCD~\cite{Gong:2013vja} and
  ETMC~\cite{Abdel-Rehim:2016won} compared to our indirect result
  indicated by the black lines, see Table~\ref{tab:phys}.}
\label{fig:sumftc}
\end{figure}

\begin{figure*}
\centerline{
\includegraphics[width=.48\textwidth,clip=]{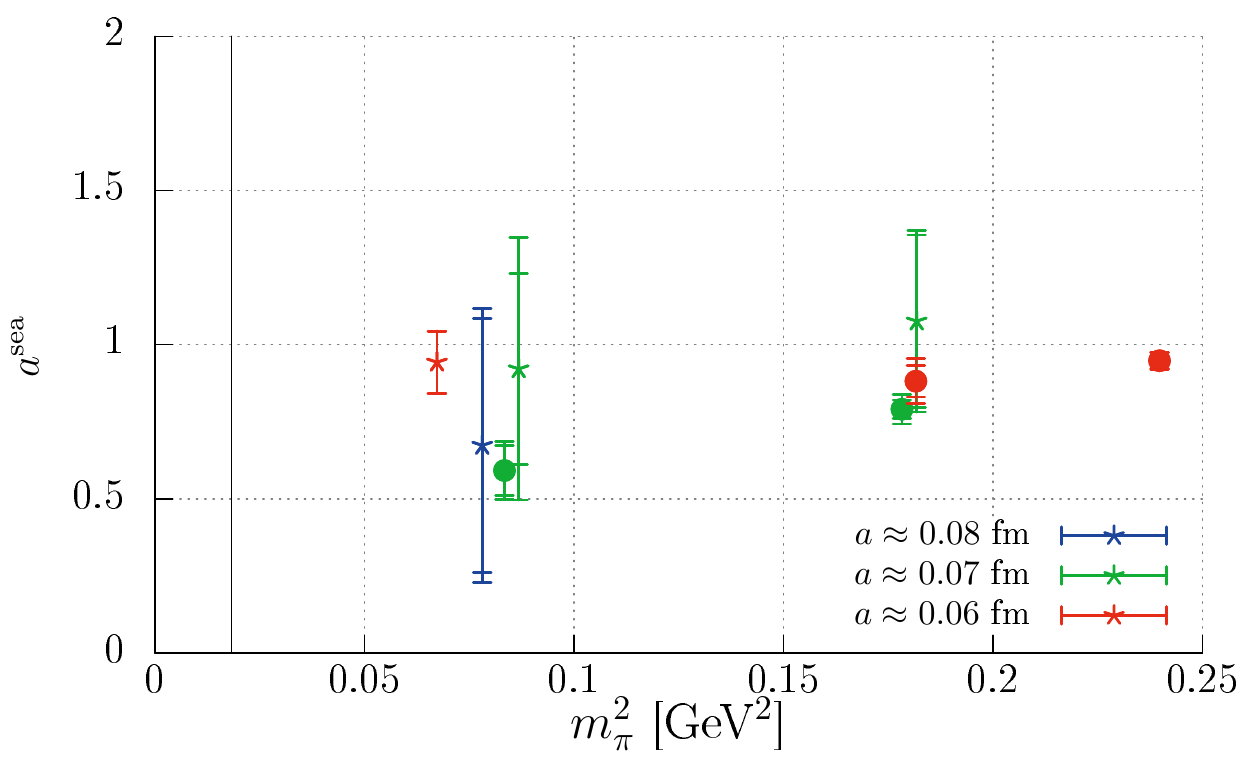}
\includegraphics[width=.48\textwidth,clip=]{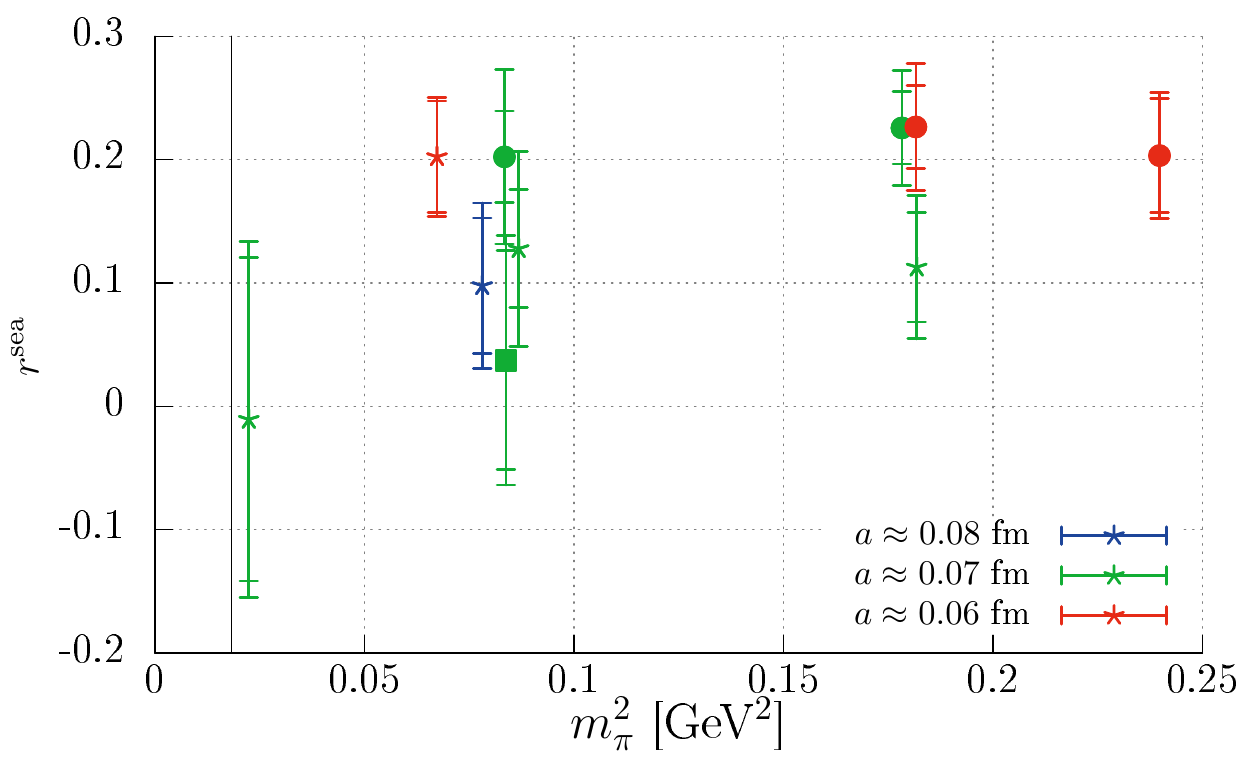}
}
\caption{ The strange to~(light) sea quark matrix elements~(left) and
  the ratio of sea to total light quark matrix elements~(right) vs.\
  $m_\pi^2$. The results are displayed as in Fig.~\ref{fig:nucsigma}.}
\label{fig:nucr}
\end{figure*}

Also of interest are the $c$, $b$ and $t$ quark fractions as these are
non-negligible due to the large quark masses accompanying the scalar matrix
element. As mentioned in the Introduction, in the heavy quark limit
the heavy quark fractions can be expressed in terms of
$f_{T_G}=1-\sum_{q=u,d,s} f_{T_q}$, to leading order in $1/m_h$ and
$\alpha$~\cite{Shifman:1978zn}, see Eq.~(\ref{eq:sigheavy}). Beyond
leading order in $\alpha$ the relation between $f_{T_G}$ and the light
quark fractions and also between $f_{T_G}$ and $f_{T_{c,b,t}}$ is
modified.  The relevant $\alpha^3$ matching expressions from a theory
with $N_f$ light quarks to one with an additional heavy quark are
given in Refs.~\cite{Chetyrkin:1997un,Hill:2014yxa}.
We utilize the full result for $f_{T_c}$, for which the strong
coupling at the relevant scale $m_c$ is largest, while for $f_{T_b}$
and $f_{T_t}$ we truncate after $\mathcal{O}(\alpha)$, arriving at the
values given in Table~\ref{tab:phys}. The perturbative error is taken
to be half the difference with the leading order value, i.e. $(2/27)
f_{T_G}$.  This is included in the total uncertainty quoted in
Table~\ref{tab:phys}. Perturbative matching of $N_f=3$ to $N_f=4$ QCD
in the heavy quark approximation at the scale $m_c$ may be considered
unreliable since neither $\alpha_{\MS}(m_c)\approx 0.39$ nor
$\Lambda/m_c$ are particularly small parameters in this case. However,
the first non-perturbative matching results are very
encouraging~\cite{Bruno:2014ufa}.  Direct determination of these
fractions is difficult, due to the large statistical uncertainty and
systematics involved, such as discretisation effects. The recent
results for $f_{T_c}$, shown in Fig.~\ref{fig:sumftc}, are consistent
with our value $f_{T_c}=0.075(4)$.

\begin{figure*}
\centerline{
\includegraphics[width=.48\textwidth,clip=]{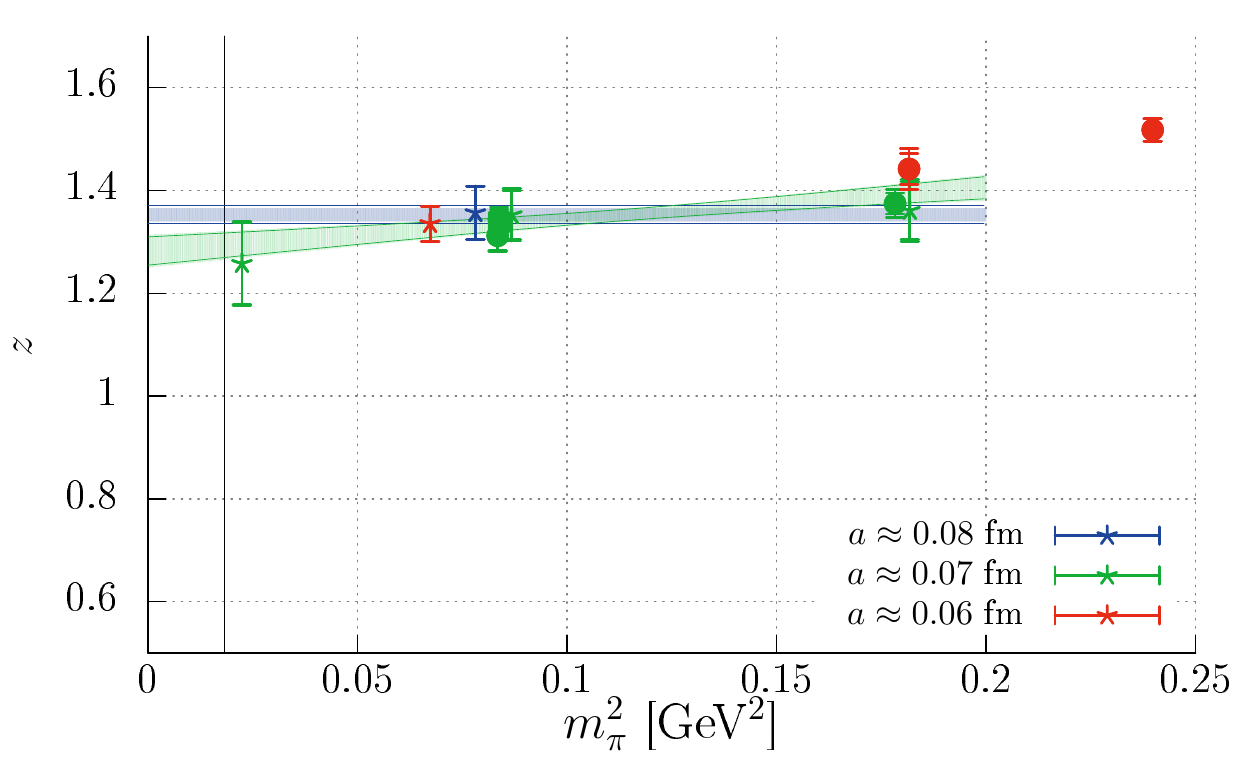}
\includegraphics[width=.48\textwidth,clip=]{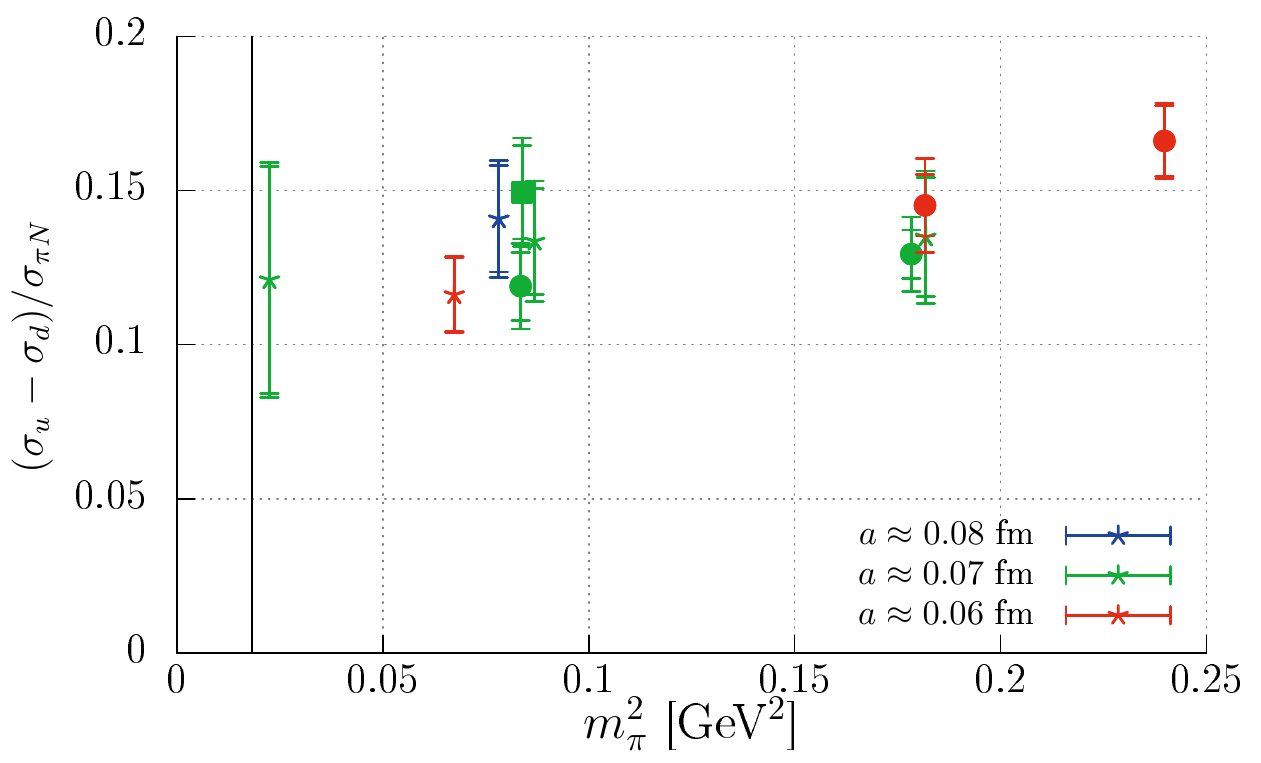}}
\caption{ (Left)  The $z$ ratio as a function of
  $m_\pi^2$. Constant~(blue shaded region) and constant plus linear
  in $m_\pi^2$ fits~(green) are shown with $\pm 1\sigma$ error band for
  $m_\pi\lesssim 420$~MeV. (Right) The $(\sigma_u-\sigma_d)/\sigma_{\pi N}$
  fraction. The results are displayed as in
  Fig.~\ref{fig:nucsigma}.}
\label{fig:nuczed}
\end{figure*}

In total, the quarks represent $\sum_{q=u,d,s,c,b,t} f_{T_q}
=0.291(15)$ or 273~MeV of the nucleon mass. The mass decomposition,
Eq.~(\ref{eq:massd}), now reads for the
$N_f=6$ theory:
\begin{align}
m_N \approx \left(0.29\, m_N\right)_{m} + \left(0.53\, m_N\right)_{\rm
  kin} + \left(0.18\, m_N\right)_{\rm a}.\label{eq:finalmsdecomp}
\end{align}
For the first term, $\sim 35$~MeV is due to the light
quarks~($\sigma_{\pi N}$) and roughly the same amount comes from the
strange quark~($\sigma_s$), the rest is due in almost equal parts from
the charm, bottom and top quarks. Comparing Eq.~(\ref{eq:finalmsdecomp})
with the $N_f=2$ theory, Eq.~(\ref{eq:massdecomp1}), the anomaly
contribution is relatively unchanged, while the kinetic term is
decreased to compensate for the larger quark mass term.

Note that at a low energy scale
the heavy quark contributions are indistinguishable from the kinetic
part: the matching to the $N_f+1$ theory was performed, assuming that
the nucleon mass is not affected by the existence of, e.g., the top
quark. Nevertheless, the Higgs~(where in Eq.~\eqref{eq:fn}
$\alpha_q\propto m_q$) at small recoil will couple to this fraction of
the nucleon mass, including the contributions from all the heavy
flavours. Since the heavy flavour scalar matrix elements alone are
very small, $\langle \bar{h}h\rangle\propto 1/m_h$, for hypothetical
particles with couplings that are insensitive to the quark mass, these
terms would be negligible. In this case the scalar couplings rather
than the sigma terms are relevant and we find for the proton
\begin{align}
& g_S^u=5.2(1.0),\,\,\,\,\,\,\,\, g_S^d=4.1(0.8), \,\,\,\,\,\,\,\, g_S^{u+d}=9.3(1.8), \nonumber\\
& g_S=g_S^{u-d}=1.0(2) \,\,\,\,\,\,\,\, \mathrm{and}\,\,\,\,\,\,\,\, g_S^s=0.35(15),\label{eq:gsresults}
\end{align}
in the $\overline{\rm MS}$ scheme at 2~GeV. The couplings were
extracted in the same way as for the sigma terms: $g_S^u$, $g_S^d$ and
$g_S^{u+d}$ are the results on ensemble VIII at $m_\pi=150$~MeV and
the value for $g_S^s$ is determined considering both a constant and
linear extrapolation in $m_\pi^2$ for $m_\pi \lesssim 420$~MeV. For $g_S$, see
Ref.~\cite{Bali:2014gha}. We expect $g_S^u$ and $g_S^d$ to be less
sensitive to isospin breaking effects than $\sigma_u$ and
$\sigma_d$~(that are approximately proportional to the quark masses
$m_u$ and $m_d$, respectively).

One can decompose the sigma terms further and
compare sea and valence quark contributions.  The ratios $a^{\rm
  sea}$ and $r^{\rm sea}$, shown in Fig.~\ref{fig:nucr}, indicate that
the sea is approximately SU${}_F$(3) symmetric, while the light quark
sea accounts for less than $30\%$ of the total light quark
contribution. Again, there is a fairly large spread in our results but
no significant dependence on pion mass, volume or lattice
spacing. Furthermore, one can look at isospin asymmetry in the form of
the $z$ ratio and the $\sigma_u-\sigma_d$ difference~(as a ratio with
$\sigma_{\pi N}=\sigma_u+\sigma_d$), both given in
Fig.~\ref{fig:nuczed}. Here, the results are more precise and the
insensitivity to the simulation parameters, in particular, the pion
mass, is clear. As discussed in Section~\ref{intro}, $z$ in
combination with $\sigma_{\pi N}$ and $\sigma_0$ is often used in the
literature to predict $f_{T_{q=u,d,s}}$. Fits to a constant and
constant plus a term linear in $m_\pi^2$ in the range $m_\pi\lesssim
420$~MeV give values for $z$ at the physical point consistent with the
results from ensemble VIII.  In keeping with the analysis for
$\sigma_{\pi N}$ and $\sigma_0$, the latter values are used for $z$ at
$m_{\pi}^{\rm phys}$.  We find $z=1.258(81)$, which is $3\sigma$ below
the expectation of $1.49$~\cite{Cheng:1988im} from the SU(3) flavour symmetry
breaking of octet baryon masses. Similarly,
$(\sigma_u-\sigma_d)/(\sigma_u+\sigma_d)=0.12(4)$ is significantly
below $1/3$, obtained from simple quark counting. The physical point
results for all quantities discussed above are displayed in
Table~\ref{tab:phys}.

\section{Comparison with other recent determinations of $\sigma_{\pi N}$ and $f_{T_s}$}
\label{comparison}
\begin{figure*}
\centerline{
\includegraphics[width=.5\textwidth,clip=]{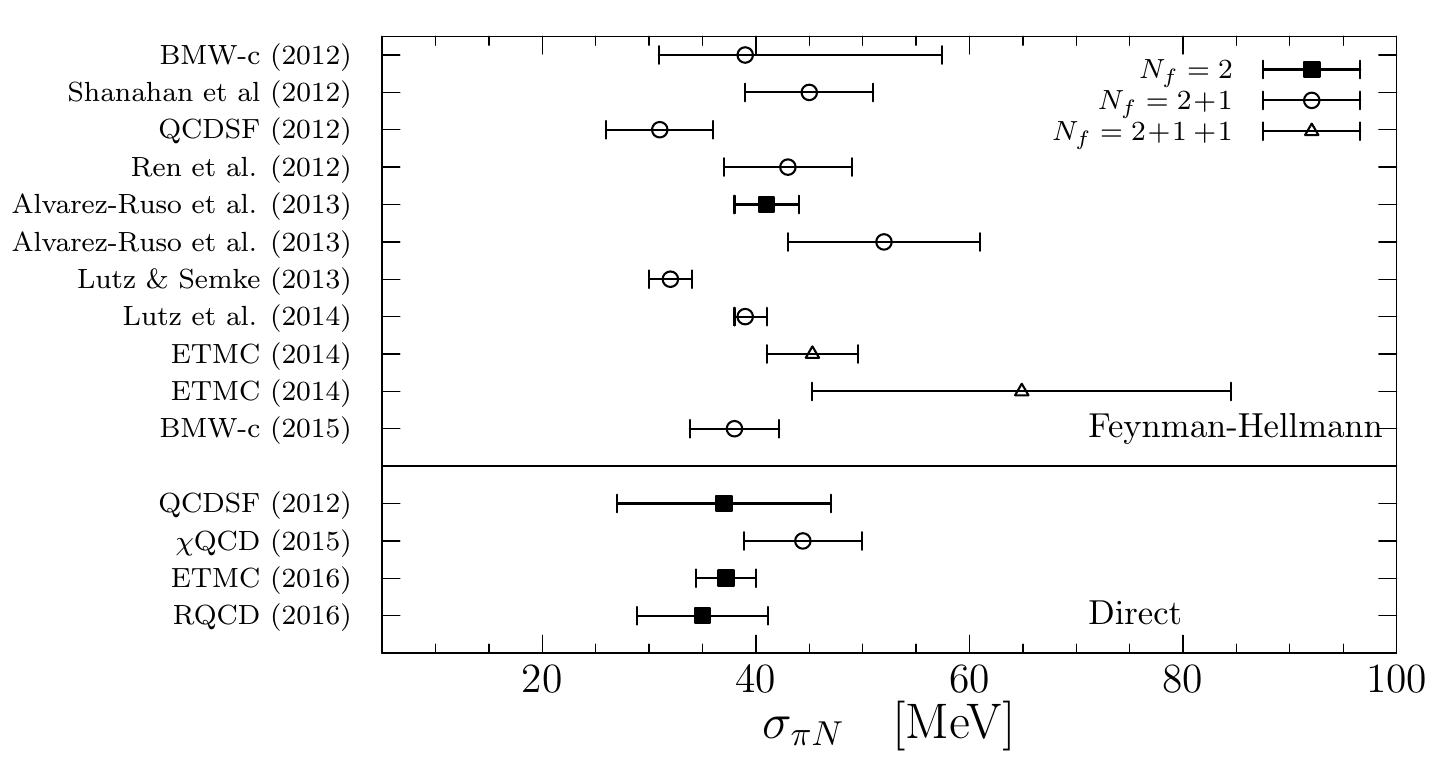}
\includegraphics[width=.52\textwidth,clip=]{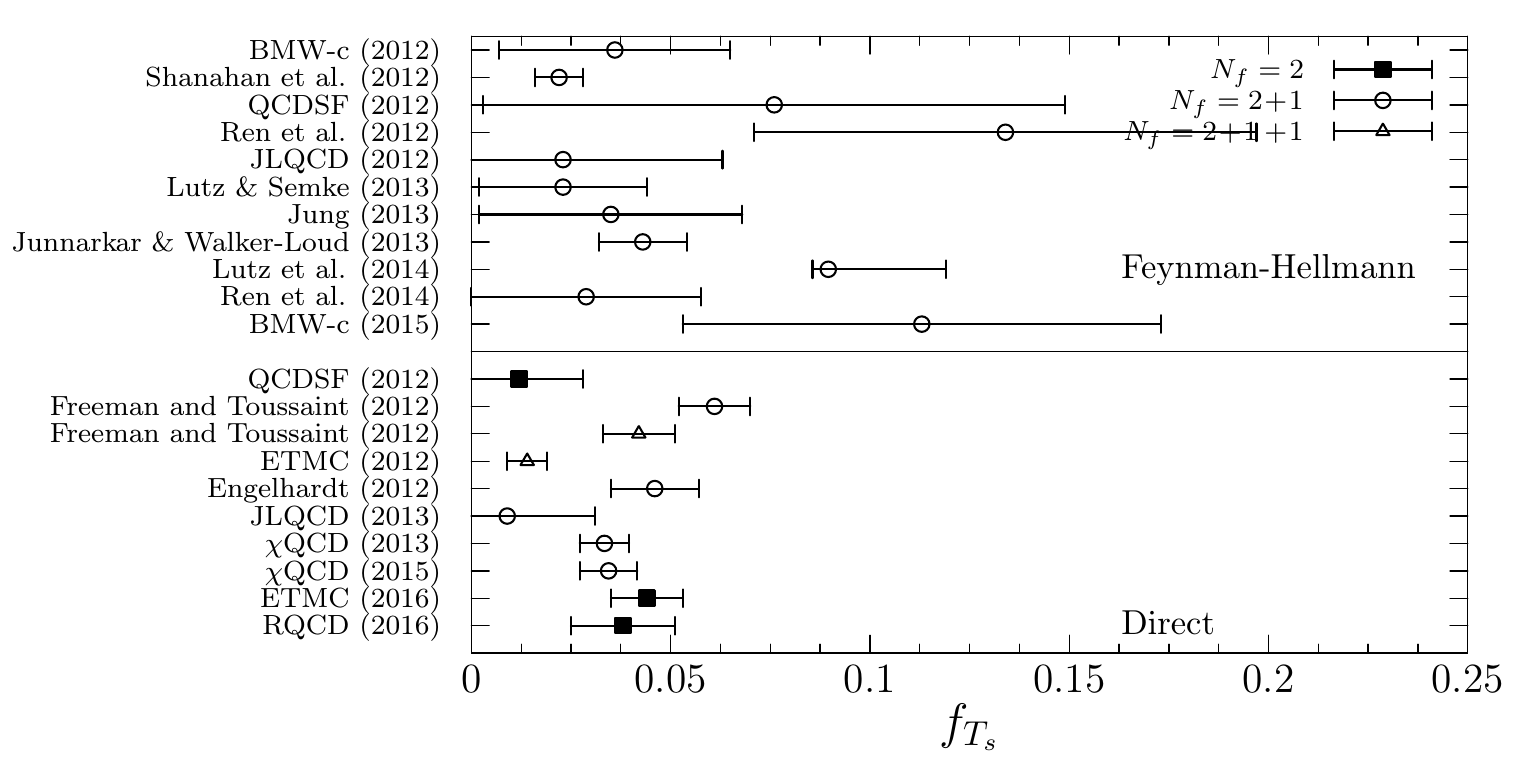}
}
\caption{Summary of recent lattice determinations of $\sigma_{\pi
    N}$~(left) and of the quark fraction $f_{T_s}$~(right).  For
  $\sigma_{\pi N}$ obtained using the Feynman-Hellmann theorem we have
  included the work of {BMW-c}~\cite{Durr:2011mp, Durr:2015dna},
  Shanahan et al.~\cite{Shanahan:2012wh}, QCDSF~\cite{Horsley:2011wr},
  Ren et al.~\cite{Ren:2012aj}, Alvarez-Ruso et
  al.~\cite{Alvarez-Ruso:2013fza}, Lutz and Semke~\cite{Lutz:2013kq},
  Lutz et al.~\cite{Lutz:2014oxa} and
  ETMC~\cite{Alexandrou:2014sha,Alexandrou:2014hsa}.  Direct
  determinations are displayed from
  QCDSF~\cite{Bali:2011ks,Bali:2012qs}, $\chi$QCD~\cite{Yang:2015uis}
  and ETMC~\cite{Abdel-Rehim:2016won}. For $f_{T_s}$, the
  Feynman-Hellmann results shown are from
  {BMW-c}~\cite{Durr:2011mp,Durr:2015dna}, Shanahan et
  al.~\cite{Shanahan:2012wh}, QCDSF~\cite{Horsley:2011wr}, Ren et
  al.~\cite{Ren:2012aj, Ren:2014vea}, JLQCD~\cite{Oksuzian:2012rzb},
  Lutz and Semke~\cite{Lutz:2013kq}, Jung~\cite{Jung:2012rz},
  Junnarkar and Walker-Loud~\cite{Junnarkar:2013ac} and Lutz et
  al.~\cite{Lutz:2014oxa}.  The direct evaluations were performed by
  QCDSF~\cite{Bali:2011ks}, Freeman and
  Toussaint~\cite{Freeman:2012ry},
  ETMC~\cite{Dinter:2012tt,Abdel-Rehim:2016won},
  Engelhardt~\cite{Engelhardt:2012gd}, JLQCD~\cite{Oksuzian:2012rzb}
  and $\chi$QCD~\cite{Gong:2013vja,Yang:2015uis}. RQCD refers to the
  present article.}
\label{fig:sumfts}
\end{figure*}

In Fig.~\ref{fig:sumfts} we compare our results and other direct
determinations of $\sigma_{\pi N}$ and $f_{T_s}$ with those extracted
via the Feynman-Hellmann theorem. The latter indirect evaluations need to
determine the slope of the nucleon mass at the physical point in terms
of the light and the strange quark masses. This requires simulations
which ideally include quark masses which are varied around the
physical values. For light quarks this is usually missing due to the
computational cost while for strange quarks the mass is normally kept
fixed as the physical point is approached. A notable exception is the
recent work of {BMW-c}~\cite{Durr:2015dna}.  These problems are
reflected in the larger variation in the results compared to the
direct methods, in particular, for $f_{T_s}$. As remarked above, the
direct evaluations are consistent and favour small values for
$\sigma_{\pi N}\sim 35-45$~MeV and $f_{T_s}\lesssim 0.05$. 

Alternative approaches involve the analysis of pion-nucleon scattering
data. Results for $\sigma_{\pi N}$ include, for example, $45(8)$~MeV
from Gasser et al.~\cite{Gasser:1990ce}, $64(7)$~MeV from Pavan et
al.~\cite{Pavan:2001wz}, $59(7)$~MeV from Alarcon et
al.~\cite{Alarcon:2011zs} and $52(7)$~MeV from Chen et
al.~\cite{Chen:2012nx} and most recently $59.1(3.5)$~MeV from
Hoferichter et al.~\cite{Hoferichter:2015dsa,Hoferichter:2015hva}~(see
also references in \cite{Hoferichter:2015hva}).  As can be seen from
Fig.~\ref{fig:sumfts}, $\sigma_{\pi N}\sim 60$~MeV is somewhat above
the direct lattice results. In the Roy Steiner analysis of scattering
data presented in Ref.~\cite{Hoferichter:2015hva} not only the scalar
formfactor and its slope near the Cheng-Dashen (CD) point are
determined but also ChPT low energy constants are obtained by matching
the ChPT expressions to the sub-threshold
parameters~\cite{Hoferichter:2015tha}. From the slope and the
formfactor at the CD point the sigma term is estimated neglecting
corrections that are formally of order $m_{\pi}^2/m_N^2$. The low
energy constants extracted from $\pi N$ scattering data enable a
detailed comparison with lattice results, also away from the physical
point. In view of approximations made in some of the above
analyses, the convergence of ChPT expansions at or near the physical
pion mass, clearly, is of great interest.

\begin{figure*}
\centerline{
\includegraphics[width=.5\textwidth,clip=]{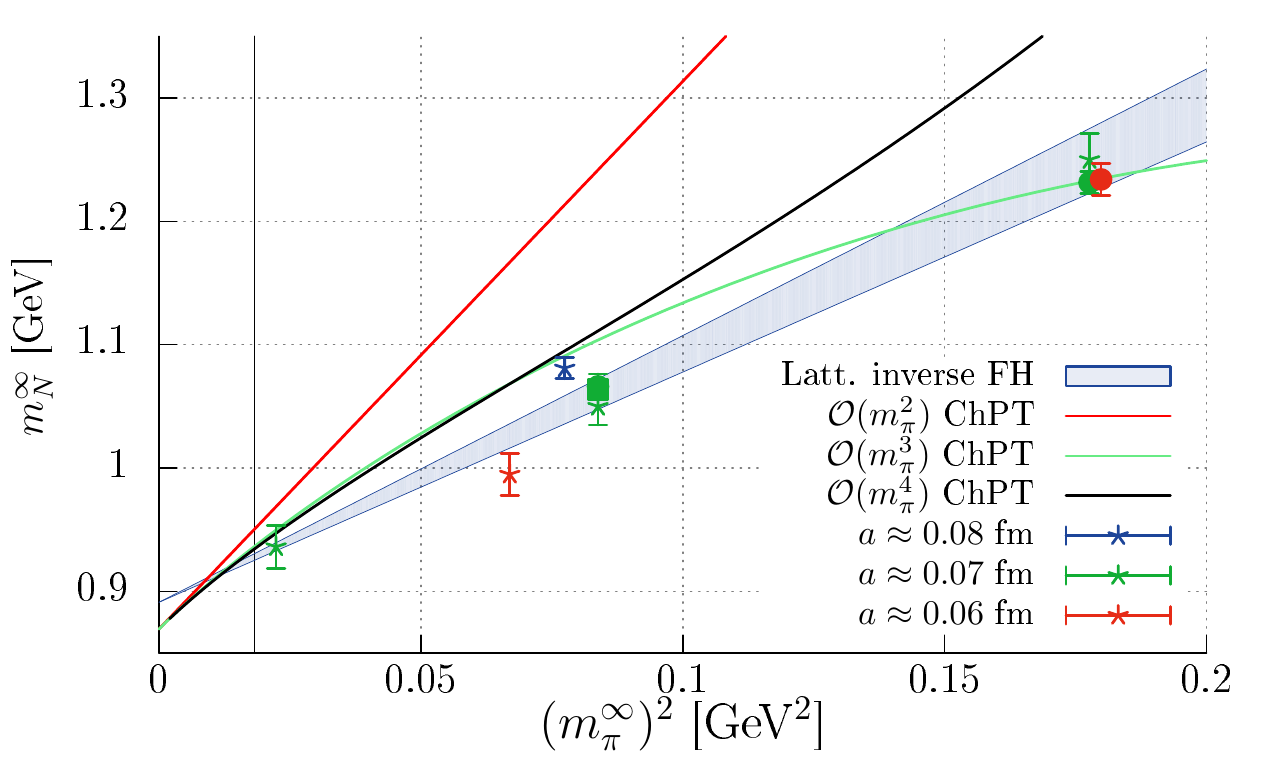}
\includegraphics[width=.5\textwidth,clip=]{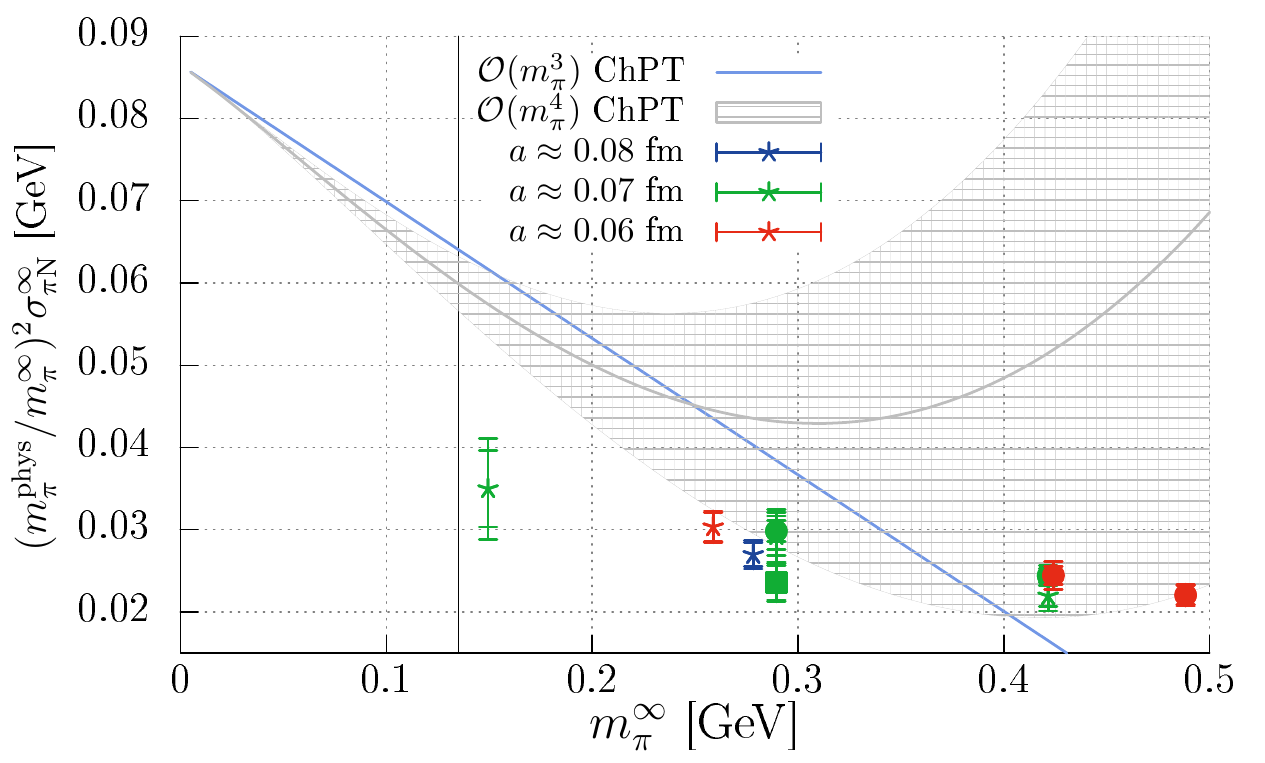}
}
\caption{(Left) Comparison of the prediction for the dependence of the
  nucleon mass on the pion mass provided by our determinations of
  $\sigma_{\pi N}$ via the Feynman-Hellmann theorem~(inverse FH, the
  blue shaded region indicates $\pm 1\sigma$ in the parametrization)
  with the nucleon masses extracted from the simulation, see the
  text. Also shown as coloured lines are the central values of the
  heavy baryon ChPT predictions for $m_N$ up to $\mathcal{O}(m_\pi^4)$
  obtained using the next-to-next-to-next-to-leading order~(NNNLO) low
  energy constants given in Ref.~\cite{Hoferichter:2015hva}. (Right) A
  similar comparison between the pion-nucleon sigma term in the
  combination $(m^{\rm phys}_\pi/m_\pi^\infty)^2\sigma^\infty_{\pi N}$
  and ChPT predictions. For $\mathcal{O}(m_\pi^3)$ ChPT only the
  central value is shown, while for $\mathcal{O}(m_\pi^4)$ the
  uncertainty due to the error on $e_1$~(see Eq.~(\ref{eq:hofsig})) is
  indicated by the grey shaded region. Note that we use $m_\pi^{\rm
    phys}=139.57$~MeV in the ChPT expressions~(only) in keeping with
  Ref.~\cite{Hoferichter:2015hva}.}
\label{fig:cmpchpt}
\end{figure*}

First we carry out a consistency check of our $\sigma_{\pi N}$ data,
shown in Fig~\ref{fig:cmpchpt}: we compare our nucleon masses
(corrected for finite size effects, incorporating systematic errors,
in the same way as for the sigma term) with the expectation obtained
by integrating the phenomenological $(a-bm_\pi)m_{\pi}^2$
parametrization of our $\sigma_{\pi N}$ values~(see
Section~\ref{nucleonsig}) as a function of $m_{\pi}^2$ (inverse
Feynman-Hellmann method). The integration constant is adjusted so that
the curve goes through the central value of the smallest pion mass
point.  Modulo the coarsest lattice point~($\beta=5.2$, $a=0.08$~fm)
the parametrization describes the nucleon mass behaviour very
well. However, it is clear from the figure that from a global fit to
the nucleon mass data alone it would have been difficult to obtain the
slope at the physical pion mass reliably in a parametrization
independent way, unless data at smaller than physical pion masses were
available.  Note that such a Feynman-Hellmann study from {BMW-c} found
$\sigma_{\pi N}=38(3)(3)$~MeV~\cite{Durr:2015dna}, in agreement with
our direct evaluation.

For comparison we superimpose the 
heavy baryon ChPT expression used in Ref.~\cite{Hoferichter:2015hva}, truncating
this at different orders in $m_{\pi}^n$,
\begin{widetext}
\begin{align}
m_N  = & m_N^0 -4c_1m_\pi^2-\frac{3 g_A^2 m_\pi^3}{32\pi F_\pi^2}- \frac{3}{32\pi^2 F_\pi^2 m_N^0}
\left[g_A^2+m_N^0\left(-8c_1+c_2+4c_3\right)\right]
 m_\pi^4\ln \left(\frac{m_\pi}{m_N^0}\right) \nonumber\\&\qquad 
+\left[e_1-\frac{3}{128\pi^2F_\pi^2m_N^0}\left(2g_A^2-c_2m_N^0\right)\right]m_\pi^4 + \mathcal{O}\left(m_\pi^5\right)\,,\label{eq:hofmn}
\end{align}
\end{widetext}
using their set of low energy constants: $c_1=-1.11(3)$~GeV$^{-1}$,
$c_2=3.13(3)$~GeV$^{-1}$, $c_3=-5.61(6)$~GeV$^{-1}$.  The value
$e_1=11.7(4.6)$~GeV$^{-3}$ is adjusted to reproduce $\sigma_{\pi
  N}=59.1(3.5)$~MeV at the charged physical pion mass
$m_\pi=139.57$~MeV and we use their determination $m_N^0=0.8695$~GeV,
while $g_A=1.2723$ and $F_\pi=0.0922$~GeV are taken from experiment. The
$\mathcal{O}(m_{\pi}^4)$ curve in Fig.~\ref{fig:cmpchpt}~(left)
corresponds to that shown in Fig.~28 of
Ref.~\cite{Hoferichter:2015hva}. At least above the physical pion
mass, there is disagreement between lattice data and ChPT using this
set of low energy constants, also varying these (including $e_1$)
within error bands. Note that the parametrization of our data shown in
Fig.~\ref{fig:cmpchpt} gives $c_1=-0.51(4)$~GeV$^{-1}$. Fitting to the
sigma term data,\footnote{The lower limit can be obtained by
  extrapolating from the $m_\pi\sim 150$~MeV point using $\sigma_{\pi
    N} = -4c_1m_\pi^2-9 g_A^2 m^3_\pi/(64\pi F_\pi^2)$, and ignoring higher order
corrections that are in the opposite direction.} it is
not possible to achieve a $c_1$ value of less than $-0.8$~GeV$^{-1}$.

In Fig.~\ref{fig:cmpchpt}~(right) the parametrization used in
Ref.~\cite{Hoferichter:2015hva},
\begin{widetext}
\begin{align}
\sigma_{\pi N}  = & -4c_1m_\pi^2-\frac{9 g_A^2 m^3_\pi}{64\pi F_\pi^2}- \frac{3}{64\pi^2 F_\pi^2 m_N^0}
\left[g_A^2+m_N^0\left(-8c_1+c_2+4c_3\right)\right]
 m_\pi^4\left[4\ln \left(\frac{m_\pi}{m_N^0}\right)+1\right]\nonumber\\
&\qquad + 2\left[e_1-\frac{3}{128\pi^2F_\pi^2m_N^0}\left(2g_A^2-c_2m_N^0\right)+\frac{c_1\left(\bar{l}_3-1\right)}{16\pi^2F_\pi^2}\right]m_\pi^4 + \mathcal{O}\left(m_\pi^5\right),\label{eq:hofsig}
\end{align}
\end{widetext}
where $\bar{l}_3=3.41(41)$, is directly compared with our results.
Unlike the nucleon mass, the sigma term is very sensitive to the value of $e_1$,
as indicated by the shaded region in the figure. Clearly, the
convergence of ChPT appears to be even inferior to that for the nucleon
mass, at least for $m_\pi \ge m_{\pi}^{\rm phys}$.
Discussions on the application of ChPT to the determination of $\sigma_{\pi N}$
can be found, for example, in Ref.~\cite{Leutwyler:2015jga}.

\section{Conclusions}
\label{conc}
In summary, we performed a high statistics study of the pion and
nucleon sigma terms with $N_f=2$ dynamical non-perturbatively improved
clover fermions for pion masses ranging from $m_\pi\sim 500$~MeV down
to close to the physical point. The set of ensembles available enabled
a study of volume dependence~($Lm_\pi=3.42-6.71$) and, for the
nucleon, also of lattice spacing effects~($a=0.06-0.08$~fm) for
$m_\pi\gtrsim 260$~MeV. Finite volume corrections derived from ChPT
turn out to be small, in particular, close to the physical point, and
any remaining volume dependence could be ascribed to
statistics. Similarly, for nucleon observables discretisation effects
were not discernible, although leading $\mathcal{O}(a)$ terms are
expected for most quantities and our lattice spacings only vary over a
limited range.  Extrapolations of the nucleon sigma term data using
simple forms for the chiral behaviour gave consistent results with
those obtained at the near physical point and we used these $m_\pi\sim
150$~MeV values to form the final results for quantities dominated by
light quarks, such as $\sigma_{\pi N}$ and $\sigma_{q=u,d}$ for the
proton in the isospin symmetric theory, summarized in
Table~\ref{tab:phys}. For the scalar couplings, that are expected to
be less sensitive to isospin breaking effects, see Eq.~(\ref{eq:gsresults}).
For the strange quark sigma terms and ratios,
$\sigma_s$, $f_{T_s}$ and $y$, our results are not so precise due to
large cancellations under renormalization for Wilson type fermions at
our moderate lattice spacings and we quote final values obtained by
extrapolation of the very mild quark mass dependence from
$m_\pi\gtrsim 260$~MeV to the physical point.

A careful analysis procedure was implemented to extract the connected
and disconnected quark line scalar matrix elements to ensure excited
state contamination is minimized. In particular, for the pion this
involved removing multi-pion states which propagate around the
boundary and can contribute significantly if the temporal extent of
the lattice is not very large. This improvement enabled us to show the
GMOR expectation $\sigma_\pi=m_\pi/2$ is valid up to
$m_\pi\sim 420$~MeV consistent with the GMOR behaviour of the pion mass as
a function of the renormalized quark mass over the same range. The
improvement technique may also be useful in the evaluation of other pion
matrix elements. 

Our study shows that with $\sigma_{\pi N}=35(6)$~MeV, the light
quarks contribute very little to the mass of the nucleon. About 30$\%$
of this 3--4$\%$ fraction is due to light sea quarks. The strange quark
contribution, $\sigma_s=35(12)$~MeV is similarly small. Appealing
to the heavy quark limit, we utilized the $\alpha^3$ perturbative
matching results~\cite{Chetyrkin:1997un,Hill:2014yxa} between a theory
of $N_f$ light quarks and one containing an additional heavy quark in
order to evaluate the mass contributions from the charm, bottom and
top quarks.  These contributions are significantly larger than for the
light and strange quarks due to the large quark masses in the
combinations $m_h\langle \bar{h}h\rangle$. Overall, the quarks
contribute about 29$\%$ of the mass, with the kinetic energies of
the quarks and gluons and the anomaly accounting for $53\%$ and
$18\%$, respectively, see Eq.~(\ref{eq:finalmsdecomp}). In
Table~\ref{tab:phys} we also provide values for $\sigma_0$ and
$z$. Estimates of these quantities and of $\sigma_{\pi N}$ from
(non-lattice) approaches are sometimes used in the
literature to predict $f_{T_{q=u,d,s}}$.

Good agreement was found for most quantities with other direct
determinations involving different quark actions, pion masses, numbers
of dynamical flavours, lattice spacings and volumes, in particular for
$\sigma_{\pi N}$ and $f_{T_s}$, displayed in Fig.~\ref{fig:sumfts},
determined around the physical point. These determinations favour
small values for both quantities compared to, for example,
$\sigma_{\pi N}=59.1(3.5)$ MeV from Hoferichter et
al.~\cite{Hoferichter:2015dsa,Hoferichter:2015hva} from a dispersive analysis of
pion-nucleon scattering data. The pion-nucleon sigma term
gives the slope of the nucleon mass as a function of $m_\pi^2$ via the
Feynman-Hellmann theorem. We showed our values for the sigma terms
describe the nucleon mass data up to $m_\pi\sim 420$~MeV, providing a
consistency check of the results.  In contrast, the heavy baryon ChPT
expansion did not seem to be well controlled above the physical
point. Direct lattice calculation is the most
theoretically clean approach to evaluate the sigma terms. Improvements
in techniques have led to an increase in the statistical precision for
$\sigma_{\pi N}$ and $f_{T_s}$ determined in this way and this must be
accompanied by a thorough investigation of the systematics.
Future calculations will involve $N_f=2+1$ simulations on
CLS ensembles~\cite{Bruno:2014jqa} with open boundaries to remove the
uncertainty of omitting the strange quark in the sea~(although this is
not expected to be a dominant effect) and to achieve smaller lattice
spacings for which the cancellations in $f_{T_s}$ under renormalization are less
severe. In addition, discretisation and finite volume effects will be
addressed systematically also for small pion masses.

\acknowledgments

We thank Martin Hoferichter for communicating to us the set of
constants used in Fig. 28 of Ref.~\cite{Hoferichter:2015hva} and for
discussion. We also thank Jose Manuel Alarcon for comments. The
ensembles were generated primarily on the QPACE
computer~\cite{Baier:2009yq,Nakamura:2011cd}, which was built as part
of the DFG (SFB/TRR 55) project.  The authors gratefully acknowledge
the Gauss Centre for Supercomputing
e.V.~(\url{http://www.gauss-centre.eu}) for granting computer time on
SuperMUC at the Leibniz Supercomputing Centre~(LRZ,
\url{http://www.lrz.de}) for this project.  The
BQCD~\cite{Nakamura:2010qh} and CHROMA~\cite{Edwards:2004sx} software
packages were used, along with the locally deflated domain
decomposition solver implementation of openQCD~\cite{luscher3}.  Part
of the analysis was also performed on the iDataCool cluster in
Regensburg.  Support was provided by the DFG (SFB/TRR 55) and the EU
(ITN STRONGnet).  We thank Rainer Schiel for generating some of the
data used in this article and Benjamin Gl\"a{\ss}le for software
support.

\appendix

\section{Conventions}
\label{app:definitions}
We work in Euclidean space-time throughout. Our continuum partition function is
defined as
\begin{align}
Z &= \int [dA][d\bar{q}][dq] e^{-\int_{V_4}\! d^4x \mathscr{L}(x)}\,,
\end{align}
where $q$ stands for the quark flavours of the theory. Suppressing the
flavour index, the Lagrangian reads
\begin{equation}
\mathscr{L}=\frac14F_{\mu\nu}F_{\mu\nu}+\bar{q}\left(D_{\mu}\gamma_{\mu}+m_q\right)q,
\end{equation}
where $D_{\mu}=\partial_{\mu}+igA_{\mu}$, $A_{\mu}=A_{\mu}^at^a$, $F_{\mu\nu}=-\frac{i}{g}[D_{\mu},D_{\nu}]$ and
 $F_{\mu\nu}F_{\mu\nu}=F^2=F_{\mu\nu}^aF_{\mu\nu}^a$.
This gives the energy-momentum tensor~\cite{Freedman:1974gs,Freedman:1974ze,Caracciolo:1989pt}
\begin{equation}
T_{\mu\nu}=F_{\mu\rho}F_{\nu\rho}-\frac14\delta_{\mu\nu}F^2+\frac14
\bar{q}\overleftrightarrow{D}_{\{\mu}\gamma_{\nu\}}q,
\end{equation}
where $\overleftrightarrow{D}_{\mu}=\overrightarrow{D}_{\mu}-\overleftarrow{D}_{\mu}$.
We define the $\beta$ function and the quark mass anomalous dimension $\gamma$-function as
\begin{align}
\beta(\alpha)&=\frac{d\alpha(\mu)}{d\ln\mu}=-2\alpha\left[\beta_0\frac{\alpha}{4\pi}+
\beta_1\left(\frac{\alpha}{4\pi}\right)^2+\cdots\right],\\
\gamma_m(\alpha)&=\frac{d\ln m(\mu)}{d\ln\mu}=-8\pi\left[\gamma_0\frac{\alpha}{4\pi}+
\gamma_1\left(\frac{\alpha}{4\pi}\right)^2+\cdots\right],
\end{align}
respectively. In these conventions
\begin{equation}
\beta_0=11-\frac{2}{3}N_f\,,\quad\gamma_0=1\,.
\end{equation}
The (classical plus anomalous) trace of the energy momentum tensor,
i.e. the interaction measure, can be obtained as the logarithmic
derivative of the free energy density with respect to a scale
$M$~\cite{Coleman:1970je,Chanowitz:1972da,Crewther:1972kn,Nielsen:1977sy,Adler:1976zt},
\begin{equation}
\label{eqtmunu}
T_{\mu\mu}=\frac{1}{V_4}\frac{d\ln Z}{d\ln M}=
\frac{\beta(\alpha)}{\alpha}\frac14F^2+\left[\gamma_m(\alpha)-1\right]m_q\bar{q}q.
\end{equation}
Note that the covariant derivative is independent of the coupling.
Rescaling $gA_{\mu}\mapsto A_{\mu}$ makes this explicit. As
$F^2/4=-(16\pi\alpha)^{-1}[D_{\mu},D_{\nu}][D_{\mu},D_{\nu}]$ with
$\alpha=g^2/(4\pi )$, the derivative of the gluon kinetic term gives
$-\beta(\alpha)/(4\alpha) F^2$.  The anomalous quark mass dimension is
obtained from applying the Leibniz rule to the derivative of the combination
$(m/M)\left(M\int\!d^4x\,\bar{q}q\right)$.

We decompose $T_{\mu\nu}=\overline{T}_{\mu\nu}+\hat{T}_{\mu\nu}$, where
$\overline{T}$ is traceless and
\begin{equation}
\hat{T}_{\mu\nu}=\frac14\delta_{\mu\nu}T_{\rho\rho}.
\end{equation}
With $F^2=2(\mathbf{E}^2+\mathbf{B}^2)$, $F_{4\mu}F_{4\mu}=\mathbf{E}^2$ and
using the equations of motion for the quark fields,
this gives
\begin{equation}
\overline{T}_{44}=\frac12\left(\mathbf{E}^2-\mathbf{B}^2\right)-\bar{q}\boldsymbol{D}\cdot
\boldsymbol{\gamma}q-\frac{3}{4}m_q\bar{q}{q},
\end{equation}
where $-\frac14 m_q\bar{q}q$ is the classical contribution to
$\hat{T}_{44}$. Note that $-T_{44}$ is the energy density.

Within Eq.~\eqref{eqtmunu} the combinations
\begin{equation}
m_q\bar{q}q\,,\quad
\frac{\beta(\alpha)}{4\alpha}F^2+\gamma_m(\alpha)m_q\bar{q}q,\label{eq:last}
\end{equation}
taken between physical states, are both renormalization group
invariants~(RGI), however, the second term is discontinuous at flavour
thresholds.  Note that this term, multiplied by $-8/\beta_0$ gives the
combination whose vacuum expectation value is known as the RGI
definition of the non-perturbative gluon
condensate~\cite{Tarrach:1981bi}.  The scale independence of the two
contributions shown in Eq.~(\ref{eq:last}) enables, within the heavy quark approximation, the
matching of a theory of $N_f$ quark flavours at a scale $M<m_h$ to a
theory of $N_f$ light flavours plus one heavy flavour of mass $m_h$ at
a scale $M>m_h$~\cite{Shifman:1978zn}.

\section{Spectral decomposition of the pion two- and three-point functions}
\label{app:spectral}
In order to motivate our method for reducing excited state
contributions and the subsequent choice of fit forms we start with the
transfer matrix expressions for $C_{{\rm 2pt}}$ and $C_{{\rm 3pt}}$ in
Eqs.~(\ref{eq:twopt0}) and~(\ref{eq:threept0}), respectively, with periodic boundary
conditions:
\begin{widetext}
\begin{align}
C_{{\rm 2pt}}(t_{\mathrm{f}},0)  = &
 \frac{1}{Z(T)} \mathrm{Tr}\left[
e^{-(T-t_{\rm f})\hat{H}}{\cal H}(0)e^{-t_{\rm f}\hat{H}} {\overline{\cal H}}(0)\right]
 = \sum_{n,m} \langle n |{\cal H}(0)|m \rangle\langle m |{\cal \overline{H}}(0)|n \rangle
e^{-(T-t_{\rm f})E_n}e^{-t_{\rm f}E_m},\label{eq:twoptops}\\
C_{{\rm 3pt}}(t_{\mathrm{f}},t,0) =  &
 \frac{1}{Z(T)} \mathrm{Tr}\left[
e^{-(T-t_{\rm f})\hat{H}}{\cal H}(0)e^{-(t_{\rm f}-t)\hat{H}}S(0)e^{-t\hat{H}}{\overline{\cal H}}(0)\right] \nonumber\\ & 
- \frac{1}{Z(T)} \mathrm{Tr}\left[
e^{-(T-t)\hat{H}}S(0)e^{-t\hat{H}}\right] \frac{1}{Z(T)} \mathrm{Tr}\left[
e^{-(T-t_{\rm f})\hat{H}}{\cal H}(0)e^{-t_{\rm f}\hat{H}} {\overline{\cal H}}(0)\right]\\
 = & \sum_{k,n,m} \langle n |{\cal H}(0)|m \rangle\langle m |S(0)|k \rangle\langle k |{\cal \overline{H}}(0)|n \rangle 
 e^{-(T-t_{\rm f})E_n}e^{-(t_{\rm f}-t)E_m}e^{-tE_k}\nonumber\\
&- \left[\sum_{n} \langle n |S(0)|n \rangle  e^{-TE_n}\right]\left[\sum_{n,m} \langle n |{\cal H}(0)|m \rangle\langle m |{\cal \overline{H}}(0)|n \rangle
e^{-(T-t_{\rm f})E_n}e^{-t_{\rm f}E_m}\right],\label{eq:threeptops}
\end{align}
\end{widetext}
where $\hat{H}$ is the lattice Hamiltonian and $Z(T) = \langle 0|
e^{-T\hat{H}}|0\rangle$ the partition
function.\footnote{In principle, the spectral decomposition of the
  partition function $Z(T) = \sum_n \langle n| e^{-T\hat{H}} |n\rangle =
  \sum_n e^{-E_n T}$ should also be considered, however, we are always
  interested in ratios of correlation functions where this factor
  drops out at leading order.} For convenience, we assume that the source time
$t_{\rm i}=0$ and ${\cal H}(t) = \sum_{\vec{x}} {\cal
  H}(\vec{x},t)$. For an interpolator ${\cal H}=\bar{u}\gamma_5 d$
with pseudoscalar quantum numbers, $J^P=0^-$, the overlap matrix
$\langle n |{\cal H}(0)|m \rangle$ can link any~(single- or
multi-particle) states $|n\rangle$ and $|m\rangle$ with $J=0$ if and
only if the states have opposite parity and $\Delta I = 1$. Similarly,
the matrix element $\langle n |S(0)|m \rangle$ for the scalar operator
$S=\bar{q}q$ is non-zero for $n$ and $m$ with the same parity, $J$,
isospin and strangeness. We denote the even states, $|0\rangle,
|2\rangle, |4\rangle, \ldots$ and the odd states, $|1\rangle,
|3\rangle, |5\rangle \ldots$, where $|0\rangle$ represents the vacuum
and $|1\rangle$ the ground state pion.  Since the lowest lying single-particle
$0^+$ state\footnote{This is the $\sigma/f_0(500)$.} is
heavier in mass than $2m_\pi$ and the radially excited pion lies above
1~GeV then $|n\rangle$ can be thought of as an $n$-pion multi-particle
state for small $n$.  Considering only $n\le 2$ to begin with, the
spectral decompositions are given by
\begin{widetext}
\begin{align}
C_{{\rm 2pt}}(t_{\mathrm{f}},0) &  = 
|Z_{01}|^2 e^{-t_{\rm f} E_1}\left\{1+e^{-(T-2t_{\rm f}) E_1} + \frac{|Z_{12}|^2}{|Z_{01}|^2} \left[e^{-(T-t_{\rm f})E_2} + e^{-t_{\rm f} E_2}e^{-(T-2t_{\rm f}) E_1} \right]\right\},\label{eq:spectwopt}\\
C_{{\rm 3pt}}(t_{\mathrm{f}},t,0) & = 
|Z_{01}|^2 e^{-t_{\rm f} E_1} \left\{\langle 1|S|1\rangle_{\rm sub} +\frac{|Z_{12}|^2}{|Z_{01}|^2}\left( \langle 1|S|1\rangle_{\rm sub} e^{-(T-t_{\rm f})E_2}+\langle 2|S|2\rangle_{\rm sub} e^{-(T-2t_{\rm f})E_1}e^{-t_{\rm f}E_2}\right)\right. \nonumber\\&+ \left.\frac{Z^*_{01}Z_{21}}{|Z_{01}|^2}\langle 0|S|2\rangle e^{-(T-2t_{\rm f})E_1}\left( e^{-tE_2}+ e^{-(t_{\rm f}-t)E_2}\right)\right. \nonumber\\&
- \left(\langle 1|S|1\rangle e^{-TE_1}+ \langle 2|S|2\rangle e^{-TE_2}\right) 
\left.\left(1+e^{-(T-2t_{\rm f}) E_1}+ \frac{|Z_{21}|^2}{|Z_{01}|^2} \left[e^{-t_{\rm f} E_1}e^{-(T-t_{\rm f})E_2}+ e^{-t_{\rm f} E_2}e^{-(T-t_{\rm f})E_1}\right]\right)\right\},
\label{eq:threept1}
\end{align}
\end{widetext}
for $T>t_{\rm f}>t>0$, where $\langle n|S|n\rangle_{\rm sub} = \langle
n|S|n\rangle - \langle 0|S|0\rangle$, the overlap $Z_{nm}=
Z_{mn}^*=\langle n|{\cal \overline{H}}(0)|m\rangle$ and $E_n\approx
n E_1$ is the energy of state $|n\rangle$. Note that the
expressions above are relevant for correlators generated with the same
source and sink interpolator, for example, smeared-smeared~(SS) two-
and three-point functions. Corrections to ground state dominance
involve terms arising from a forward propagating pion state together
with a scalar (two-pion)
state propagating backward around the temporal boundary and vice versa.
Depending on the size of the overlaps and
matrix elements, some of the terms in Eq.~(\ref{eq:threept1}) can
be large for $t_{\rm f} \gtrsim T/2$, in particular since
$E_2$ is rather small, for example, $2am_\pi\approx 0.11$ for ensemble
VIII in Table~\ref{tab:res}.

Contributions involving an odd parity state propagating across the
boundary in the backward direction can be removed by constructing 
correlation functions from quark propagators with different boundary
conditions in time.  For example, the two-point function with the
spectral decomposition of Eq.~(\ref{eq:spectwopt}) is computed using
\begin{equation}
C_{{\rm 2pt}}(t_{\mathrm{f}},0) = \sum_{\vec{x}} \mathrm{Tr}\left[(M^{-1})^\dagger(\vec{x},t_{\rm f};\vec{0},0)M^{-1}(\vec{x},t_{\rm f};\vec{0},0)\right],
\end{equation}
where both propagators, $M^{-1}(\vec{x},t_{\rm f};\vec{0},0)$, have
anti-periodic boundary conditions~(AP) in time imposed. If instead one
of the propagators has periodic boundary conditions~(P), then the
two-point function for this AP-P combination will change sign when
crossing the temporal boundary. This choice corresponds to the
H-boundary condition of Ref.~\cite{Kim:2003xt} that had been used in
earlier studies of nucleon excited states~\cite{Sasaki:2001nf}. Such
boundary effects were first discussed in
Ref.~\cite{Martinelli:1982bm}. Returning to Eq.~(\ref{eq:twoptops})
and separating the terms into two sums gives:
\begin{widetext}
\begin{align}
&C_{{\rm 2pt}}(t_{\mathrm{f}},0) 
 = \left(\sum_{n\,\, {\rm even},\,m\,\, {\rm odd}}+\sum_{n\,\, {\rm odd},\,m\,\, {\rm even}}\right) \langle n |{\cal H}(0)|m \rangle\langle m |{\cal \overline{H}}(0)|n \rangle
e^{-(T-t_{\rm f})E_n}e^{-t_{\rm f}E_m}.
\end{align}
The AP-P two-point function, $C_{{\rm 2pt}}^{\rm AP-P}$, will have a minus
sign for the second sum relative to $C_{{\rm 2pt}}^{\rm AP-AP}$. Taking the
average of these, we obtain the forward propagating odd parity states
only:
\begin{align}
C^{\rm improv}_{{\rm 2pt}}(t_{\mathrm{f}},0) = \frac{1}{2}\left[C^{\rm AP-P}_{{\rm 2pt}}(t_{\mathrm{f}},0)+C^{\rm AP-AP}_{{\rm 2pt}}(t_{\mathrm{f}},0)\right] &  = 
|Z_{01}|^2 e^{-t_{\rm f} E_1}\left[1 + \frac{|Z_{12}|^2}{|Z_{01}|^2} e^{-(T-t_{\rm f})E_2} +\ldots\right].\label{eq:improv2}
\end{align}
The same effect can be achieved for the three-point function by
combining both AP and P quark propagators:
\begin{align}
 C^{\rm improv}_{{\rm 3pt}} (t_{\mathrm{f}},t,0) & =   \frac{1}{2}\left[C^{AP-P}_{{\rm 3pt}}(t_{\mathrm{f}},t,0)+ C^{AP-AP}_{{\rm 3pt}}(t_{\mathrm{f}},t,0)\right].
\end{align}
For the disconnected part this corresponds to
\begin{equation}
C^{\rm
    improv,dis}_{{\rm 3pt}}(t_{\mathrm{f}},t,0) = \langle C^{\rm
    improv,c}_{{\rm 2pt}}(t_{\mathrm{f}},0) L^c(t)\rangle_c, 
\end{equation} 
cf. Eq.~(\ref{eq:dis}), where the loop is constructed from a
propagator with AP boundary conditions, $(M^{-1})^{\rm AP}$, while for
the connected part,
\begin{align}
C^{\rm improv,conn}_{{\rm 3pt}}(t_{\mathrm{f}},t,0) = &\frac{1}{2}\left\{\langle
  \mathrm{Tr}\left[(M^{-1})^{\rm P \dagger}(t_{\rm f},0)(M^{-1})^{\rm
      AP}(t_{\rm f};t)(M^{-1})^{\rm AP}(t;0)\right] \rangle\right.\nonumber\\ 
&\left. + \langle
  \mathrm{Tr}\left[(M^{-1})^{\rm AP \dagger}(t_{\rm f},0)(M^{-1})^{\rm
      AP}(t_{\rm f};t)(M^{-1})^{\rm AP}(t;0)\right] \rangle\right\}.
\end{align}
The improved three-point function has the spectral decomposition
\begin{align}
 C^{\rm improv}_{{\rm 3pt}} (t_{\mathrm{f}},t,0) & =  
|Z_{01}|^2 e^{-t_{\rm f} E_1} \left[1 + \frac{|Z_{12}|^2}{|Z_{01}|^2} e^{-(T-t_{\rm f})E_2}+\ldots\right] \left[ \langle 1|S|1\rangle_{\rm sub} - \langle 1|S|1\rangle e^{-TE_1} - \langle 2|S|2\rangle e^{-TE_2} - \ldots \right], \label{eq:threept2}\\
& \approx  |Z_{01}|^2 e^{-t_{\rm f} E_1} \left[1 + \frac{|Z_{12}|^2}{|Z_{01}|^2} e^{-(T-t_{\rm f})E_2}+\ldots\right] \langle 1|S|1\rangle_{\rm sub}. \label{eq:threept2b}
\end{align}
In the last step we neglect the terms with factors, $e^{-TE_1}$ and
$e^{-TE_2}$, which are $e^{-TE_1}<0.03$ and $e^{-TE_2}<0.001$,
respectively, for the ensembles in Table~\ref{tab:sim}. These limits
are calculated using $E_2=2m_\pi$ and the smallest value for $T m_\pi
\sim 3.5$~(obtained from ensemble VIII). Note that such terms can be
significant in finite temperature studies~\cite{Umeda:2007hy},
where, however, the use of AP boundary conditions is mandatory.

In some cases in our study the improved three-point functions still contain
significant contributions from the next state~(the forward
propagating $|3\rangle$ state).  Including the appropriate terms, we have
\begin{align}
 C^{\rm improv}_{{\rm 2pt}} (t_{\mathrm{f}},0) & = |Z_{01}|^2 e^{-t_{\rm f} E_1}\left[1 +  \frac{|Z_{21}|^2}{|Z_{01}|^2}e^{-(T-t_{\rm f})E_2} + \frac{|Z_{03}|^2}{|Z_{01}|^2}e^{-t_{\rm f}\Delta E} +\ldots \right]\label{eq:improv3} \\
& \approx |Z_{01}|^2 e^{-t_{\rm f} E_1} \left[1 + \frac{|Z_{03}|^2}{|Z_{01}|^2}e^{-t_{\rm f}\Delta E} \right],\label{eq:improv4} \\
 C^{\rm improv}_{{\rm 3pt}} (t_{\mathrm{f}},t,0) &
 =  |Z_{01}|^2 e^{-t_{\rm f} E_1}\left[\langle 1|S|1\rangle_{\rm sub}\left(1+ \frac{|Z_{21}|^2}{|Z_{01}|^2} e^{-(T-t_{\rm f})E_2}\right)  \right. \nonumber\\
&\left. + \frac{|Z_{03}|^2}{|Z_{01}|^2}\langle 3|S|3\rangle_{\rm sub} e^{-t_{\rm f}\Delta E}+ 
\frac{Z^*_{10}Z_{30}}{|Z_{01}|^2}\langle 1|S|3\rangle \left( e^{-(t_{\rm f}-t)\Delta E} +  e^{-t\Delta E}\right) +\ldots \right], \\
& \approx  |Z_{01}|^2 e^{-t_{\rm f} E_1}\left[\langle 1|S|1\rangle_{\rm sub} + \frac{|Z_{03}|^2}{|Z_{01}|^2}\langle 3|S|3\rangle_{\rm sub} e^{-t_{\rm f}\Delta E} \right. 
\left. + \frac{Z^*_{10}Z_{30}}{|Z_{01}|^2}\langle 1|S|3\rangle\left(e^{-(t_{\rm f}-t)\Delta E}+ e^{-t\Delta E} \right)\right],
\label{eq:threept3}
\end{align}
in the limit $t_{\rm f} \ll T$, where $e^{-(T-t_{\rm f})E_2}\sim
0$. $\Delta E$ denotes the  difference $E_3-E_1$. We also compute the
ratio of the improved three-point and two-point functions. If the
excited state contribution  to $ C^{\rm improv}_{{\rm 2pt}}$
is small, the ratio has the time dependence
\begin{align}
R^{\rm improv}(t_{\mathrm{f}},t,0)
 \approx & \langle 1|S|1\rangle^{\rm dis}_{\rm sub} + \frac{|Z_{03}|^2}{|Z_{01}|^2}\left[\langle 3|S|3\rangle_{\rm sub}-\langle 1|S|1\rangle_{\rm sub}\right] e^{-t_{\rm f}\Delta E}
 + \frac{Z^*_{30}Z_{10}}{|Z_{01}|^2}\langle 3|S|1\rangle\left( e^{-(t_{\rm f}-t)\Delta E} + e^{-t\Delta E}\right), \label{eq:ratimprov3}
\end{align}
\end{widetext}
where terms with factors, $e^{-2t_{\rm f}\Delta E}$ and $e^{-(t_{\rm
    f}+t)\Delta E}$ and smaller are not included. For our data these
assumptions are reasonable as demonstrated in
Fig.~\ref{fig:pioneffmass} which shows the deviation of improved
two-point functions from ground state dominance for ensembles with
$m_\pi=289$~MeV and $m_{\pi}=150$~MeV. Excited state contributions are small
and drop below the noise for $t_{\rm f}\lesssim 10 a$.

The connected and disconnected contributions to the three-point
function are analysed individually. Equations~(\ref{eq:threept1}),
(\ref{eq:threept2b}), (\ref{eq:threept3}) and (\ref{eq:ratimprov3})
give the functional forms of the disconnected part, which includes the
subtraction of $\langle 0| S|0\rangle$.  For the connected part the
expressions are similar and can be obtained by replacing $\langle
n|S|n\rangle_{\rm sub}$ by $\langle n|S|n\rangle$ for $n=1,2,3$. Also in
Eq.~(\ref{eq:threept1}) the subtracted term in the last line is not
present. 

Finally, if different interpolators are employed at source and sink,
for example, connected or disconnected three-point
functions that are smeared at the source and local at the sink,
then one cannot simplify,
\begin{align}
& \frac{Z^*_{30}Z_{10}}{|Z_{01}|^2}\langle 3|S|1\rangle e^{-(t_{\rm f}-t)\Delta E} + \frac{Z^*_{10}Z_{30}}{|Z_{01}|^2}\langle 1|S|3\rangle  e^{-t\Delta E}  \nonumber\\
&=\frac{Z^*_{30}Z_{10}}{|Z_{01}|^2}\langle 3|S|1\rangle\left[e^{-(t_{\rm f}-t)\Delta E} + e^{-t\Delta E}\right],
\end{align}
and similarly in Eqs.~(\ref{eq:threept1}), (\ref{eq:threept2b}),
(\ref{eq:threept3}) and (\ref{eq:ratimprov3}). Accordingly, in this
case the functional forms must be modified to allow for different coefficients for
these pairs of terms.

\section{Finite volume corrections to the nucleon and pion sigma terms}
\label{app:chiralsigma}
For convenience we collect the expressions used for applying
finite volume corrections to the pion and nucleon sigma terms.  For
the pion we use NLO ChPT~\cite{Gasser:1986vb,Gasser:1987zq},
\begin{align}
m_\pi(L) &= m_\pi\left[1+\frac{2}{N_f}\frac{m_\pi^2}{16\pi^2F^2} I(\lambda)\right],
\end{align}
with
\begin{equation}
I(\lambda)=\sum_{\vec{n}}
\frac{K_1(\lambda |\vec{n}|)}{\lambda |\vec{n}|},
\end{equation}
where $\lambda=Lm_\pi$, $K_1$ is the modified Bessel function of the
second kind and $\vec{n}\neq \vec{0}$ is an integer valued vector.
Using the Feynman-Hellmann theorem and the GMOR
relation we have for
the finite volume pion sigma term, $\sigma_\pi(L) = 2\sigma_u(L)=
2\sigma_d(L)$,
\begin{align}
\sigma_\pi(L) & = \left. \frac{1}{m_\pi(L)}\frac{\partial m_\pi^2(L)}{\partial \ln m_u }\right|_{L\,\,\mathrm{fixed}} = 
\left. \sigma_\pi \frac{\partial m_\pi(L)}{\partial m_\pi}\right|_{L\,\,\mathrm{fixed}}\\
&=\sigma_\pi\left[1+\frac{2}{N_f}\frac{m_{\pi}^2}{16\pi^2F^2}\left(3I(\lambda)+\lambda\frac{dI(\lambda)}{d\lambda}\right)\right],
\end{align}
where $F$, $m_\pi$ and $\sigma_\pi$ are the pion decay constant, pion
mass and sigma term in the infinite volume limit, respectively. We can
then invert the equation above, truncating at $\mathcal{O}(m_\pi^2)$:
\begin{align}
\sigma_u & = \sigma_u(L)\left[1-\frac{2}{N_f}\frac{m_{\pi}^2}{16\pi^2F^2}\left(3I(\lambda)+\lambda\frac{dI(\lambda)}{d\lambda}\right)\right].\label{eq:infvolpi}
\end{align}

For the nucleon we again use NLO ChPT, see, for example, Ref.~\cite{AliKhan:2003cu}:
\begin{align}
m_N(L) & = m_N\left[1+\frac{3g_A^2m_\pi^2}{16\pi^2F^2}I_0(\lambda,m_N/m_\pi)\right],
\end{align}
where
\begin{align}
I_0(\lambda,m_N/m_\pi) & =\int_0^\infty\!\! dx\sum_{\vec{n}}K_0\left(\lambda|\vec{n}|\sqrt{\frac{m_N^2}{m_\pi^2}x^2+1-x}\right).
\end{align}
With the Feynman-Hellmann theorem and the GMOR relation,
\begin{align}
\sigma_{\pi N}(L) & = \left.\frac{\partial m_N(L)}{\partial \ln m_{\ell}} \right|_{L\,\,\mathrm{fixed}}\approx\left. m_\pi^2 \frac{\partial m_N(L)}{\partial m_\pi^2}\right|_{L\,\,\mathrm{fixed}},
\end{align}
this leads to
\begin{align}
\sigma_{\pi N}(L) & = 
\sigma_{\pi N}+\frac{3g_A^2m_\pi^2}{16\pi^2F^2}\left[(\sigma_{\pi N}+m_N)
I_0(\lambda,m_N/m_\pi)\right.\nonumber\\ & \left.\qquad+ m_N  I_1(\lambda,m_N/m_\pi)\right],
\end{align}
where
\begin{widetext}
\begin{align}
I_1(\lambda,m_N/m_\pi) & = \int_0^\infty dx \frac{x-1}{2\sqrt{\frac{m_N^2}{m_\pi^2}x^2+1-x}}\sum_{\vec{n}}\lambda |\vec{n}| K_1\left(\lambda|\vec{n}|\sqrt{\frac{m_N^2}{m_\pi^2}x^2+1-x}\right).
\end{align}
Inverting the above formula and truncating at $\mathcal{O}(m_\pi^2)$, we obtain
\begin{align}
\sigma_{\pi N} & = 
\sigma_{\pi N}(L)-\frac{3g_A^2m_\pi^2}{16\pi^2F^2}\left[(\sigma_{\pi N}(L)+m_N)
I_0(\lambda,m_N/m_\pi) + m_N  I_1(\lambda,m_N/m_\pi)\right].
\end{align}
\end{widetext}

For the corrections to both the pion and nucleon sigma terms we
estimate the error of the finite volume shifts to be half the
size of the correction applied. This is added in quadrature to
the statistical and the other systematic uncertainties.

The above formulae entail the pion mass in infinite volume. This is
obtained using the largest available volume for each $(\beta,\kappa)$
combination, the NNNLO analytic expressions of
Ref.~\cite{Colangelo:2005gd} and the low energy constants of
Refs.~\cite{Colangelo:2005gd,Aoki:2013ldr}, see
Ref.~\cite{Bali:2014nma} for details.

\bibliography{references}
\end{document}